\documentclass[11pt,a4paper]{article}
\pdfoutput=1
\usepackage{jheppub}

\usepackage{multirow, graphicx,amssymb,url,mathrsfs,amsmath}
\usepackage{wrapfig,boxedminipage,setspace,subfigure,epsfig}
\usepackage{amsxtra,amstext,latexsym,dsfont,amsfonts}
\usepackage{makeidx,pst-grad,pst-plot}
\usepackage{color,eucal}
\usepackage[titletoc]{appendix}

\newcommand{\be}{\begin{equation}}
\newcommand{\ee}{\end{equation}}
\newcommand{\ba}{\begin{array}}
\newcommand{\ea}{\end{array}}

\title{Small Fermi Surfaces and Strong Correlation Effects  \\  in Dirac Materials  with Holography} 

\author[a,b]{Yunseok Seo,} 
\author[a]{Geunho Song,}
\author[c,d]{Chanyong Park,}
\author[a]{and Sang-Jin Sin}

\emailAdd{yseo@hanyang.ac.kr}
\emailAdd{sgh8774@gmail.com}
\emailAdd{chanyong.park@apctp.org}
\emailAdd{sjsin@hanyang.ac.kr}

\affiliation[a]{ Department of Physics, Hanyang University, Seoul 133-791, Korea }
\affiliation[b]{ Research Institute for Natural Science, Hanyang University, Seoul 133-791, Korea }
\affiliation[c]{ Asia Pacific Center of Theoretical Physics, Pohang, 790-330, Korea }
\affiliation[d]{ Department of Physics, Postech, Pohang, 790-784, Korea}

\abstract{
Recent discovery of transport anomaly  in graphene  demonstrated that  a system known to be weakly interacting may become strongly correlated if  system parameter(s) can be tuned such that 
 fermi surface is sufficiently small. We study the strong correlation effects in the transport coefficients of Dirac materials doped with magnetic impurity under the magnetic field using holographic method.  
 The experimental data of magneto-conductivity are well fit by our theory,  however, 
  not much data are available for  other transports of Dirac material in such regime. Therefore, our  results on  
  heat transport, thermo-electric power and Nernst coefficients are left as predictions of holographic theory for generic Dirac materials  in the vicinity of charge neutral point with possible  surface gap. We give detailed look over each magneto-transport observable and  3Dplots  to guide  future experiments. 
}

\subheader{
          }

\begin{document}

\maketitle

\section{Introduction}
 
 Understanding  strongly correlated  electron systems has been a theoretical challenge  for several decades\cite{mott1937discussion}.
Typically, excitations of such system lose particle nature,  which invalidates   Fermi-liquid theory, leaving 
physicists helpless. On the other hand, such systems also exhibit mysteriously  rapid thermalization\cite{Sachdev:2011mz,zaanen2015holographic,Oh:2013qxn,Sin:2013yha}, which provides the hydrodynamic description\cite{HKMS,Lucas:2015sya}   near quantum critical point(QCP), where the system becomes universal: almost all details of the system are washed out. This is very analogous to the universality of a black  hole in the sense that it also  lose  all the information of its mother star apart from the criticality index and  a few conserved quantum numbers. 

The gauge-gravity duality\cite{Maldacena:1997re,Witten:1998qj,Gubser:1998bc} provided a 
mathematically rigorous example  and suggested  a natural setting 
to put the analogy on more quantitative framework, which  attracted much   interest  as a new paradigm for strongly interacting systems.
More recently,  large violation of Widermann-Frantz law was observed in graphene near charge neutral point, 
indicating that graphene is a strongly   interacting system\cite{pkim}  in some windows of parameters. 
The gauge-gravity principle applied with two currents, exhibited remarkable agreement  
with the  experimental data\cite{Seo:2016vks}, improving the hydrodynamic analysis\cite{Lucas:2015sya} of the same system.   

The fundamental reason for the appearance of the strong interaction in graphene is the 
{\it  smallness of the fermi sea}: the effective coupling in a system with a Dirac cone is  
\def\be{\begin{equation}}
\def\ee{\end{equation}}
\def\bea{\begin{array}}
\def\eea{\end{array}}
\be
\alpha_{eff}=\frac{e^{2}}{4\pi \epsilon \hbar c} \frac{c}{v_{F}} \sim 2.2/\epsilon_{r}, \label{aeff}
\ee
with ${v_{F}}$ fermi velocity. 
 If the fermi surface passes near the Dirac point, the tip of the cone, 
 electron hole pair creation from such small fermi surface is insufficient to screen the Coulomb interaction
 so that $\epsilon_{r} \sim 1$ and   $\alpha_{eff} >1$, making the system strongly interacting.  
 This argument works even when a gap is open as far as the fermi surface can be tuned to be small.  
 The Eq.(\ref{aeff}) also explains 
 why electron-electron Coulomb interaction is small in usual metal where fermi surface is large. 
 Since the above argument is  so simple and generic,  we expect that for any Dirac material, 
there should be a  regime of parameters where electrons are strongly correlated. 
The presence of Dirac cone  also provides  reasoning 
 why such system  has a QCP with dynamical exponent $z=1$ having Lorentz invariance.

The most well known Dirac material other than the graphene is the surface of a topological insulator(TI)\cite{hasan2010colloquium,qi2011topological}. The latter has an unpaired Dirac cone and strong spin-orbit coupling, and   as a consequence, it has a  variety of interesting  physics\cite{Yu61,PhysRevLett.102.216403,PhysRevB.78.195424} including weak anti-localization(WAL)\cite{bergmann1982weak}, quantized anomalous Hall effect\cite{Yu61}, Majorana fermion\cite{PhysRevLett.102.216403} and topological magneto-electric effect\cite{PhysRevB.78.195424}.  
Magnetic doping in TI can open  a gap in the  surface state  by  breaking the time reversal  symmetry\cite{liu2012crossover,zhang2012interplay,bao2013quantum},
and it is responsible  for the transition from WAL to weak localization(WL). 
For extremely low doping,  the  sharp horn of the magneto-conductivity curve  near zero magnetic field can be   
attributed to the particle nature of the basic excitations and indeed can be well 
described by  Hikami-Larkin-Nagaoka (HLN) function\cite{hikami1980spin}.
However, for intermediate doping  where the tendency of WAL and weak localization (WL) compete,   a satisfactory theory is still wanted \cite{lu2011competition,liu2012crossover,bao2013quantum}
\footnote{For extremely thin film case, there is a phenomenological description. In ref.\cite{liu2012crossover,lang2012competing},  
the authors assigned  weights  for two HLN functions of opposite sign by hand to fit the  data.  
In graphene case,   WAL-to-WL transition is  
better understood\cite{mccann2006weak,tikhonenko2009transition} in terms of inter-valley scattering versus spin-orbit interaction. The parameter there corresponds to  charge density which does not induces gap,  
while for TI case is the magnetic doping rate inducing surface gap. We believe that the physics involved is different.} .

\begin{figure}[ht!]
\centering
   {\includegraphics[width=8cm]{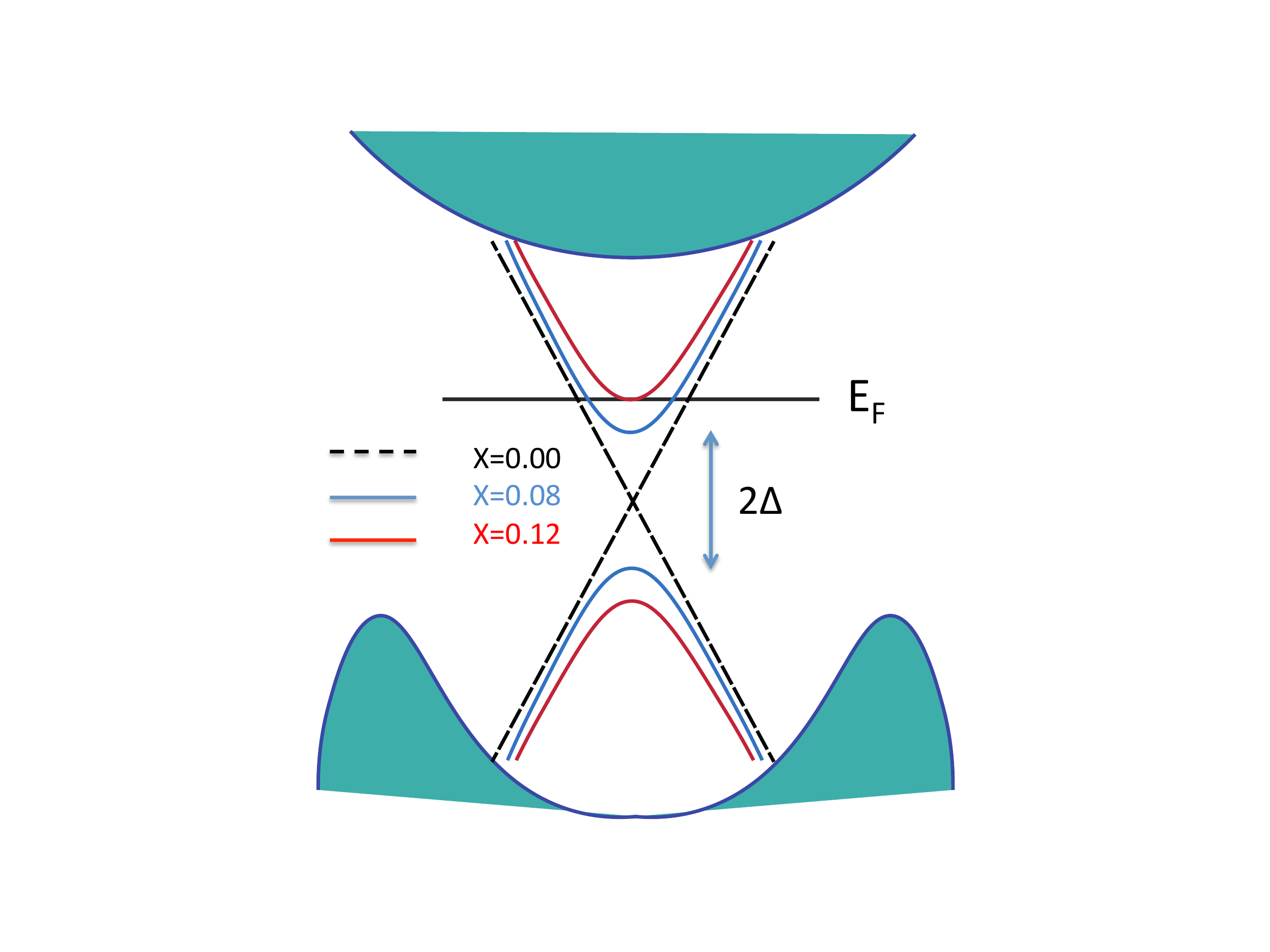} \label{}}
 \caption{ Evolution of density of state. As   we increase the doping and thereby the surface gap of the TI, 
  the fermi surface gets smaller. } \label{fig:gap}
\end{figure}
To understand why the transition regime is strongly interacting, look at the Figure \ref{fig:gap}. 
We start with  the case  where fermi surface is large at zero doping. 
Increasing the surface gap pushes up the dispersion curve, which makes the fermi sea smaller.
If gap is large enough but fermi surface  is not  small,  the transition from WAL to WL happens in a manner involving sharp peak in MC curve, demonstrating  the particle nature of the excitations. 
With more doping, the dispersion curve is pushed up more so that both large gap and small fermi surface  are achieved and the system achieve the transition with strongly interacting nature.    
Therefore electron system near such  transition region can be described by the holographic theory.  

In the previous paper \cite{Seo:2017oyh}, we compared magneto-conductivity calculated by the holographic calculation and  the data of Mn doped  Bi$_{2}$Se$_{3}$ and  Cr doped  Bi$_{2}$Te$_{3}$.  
We  showed that    the experimental data is well fit by the theoretical curves in the parameter island where 
fermi-surface is small. 
In this paper, we  study  all possible magneto-transport coefficients including thermal and  thermo-electric  
transports  of surface states of topological insulators in the regime of  strong correlation. 
We will give 3D plots of  each of them. 
Since not much data are available for heat transport  or thermo-electric transports of Dirac material in such regime, our  study can be regarded as predictions of holographic theory for generic Dirac materials  
in the vicinity of charge neutral point
\footnote{Our treatment can be applied for the case  with   surface gap as well as the case without gap as far as the system can be considered as a conductor. }.

\section{Gravity dual of the surface of TI  with magnetic doping}
Although our target is general Dirac material not just for Topological Insulator (TI), 
 we want to  setup holographic formalism to describe the surface of it,
  which is one of the most well studied   material with Dirac cone. 
  Phenomenologically, we  will be interested  in magneto-transport of TI surface as a consequence of 
 surface gap which is  generated by the  magnetic doping.  
 
\subsection{Holographic Formulation of  the surface state}
We setup the holographic model by a sequence of reasonings. 
 \begin{enumerate}
\item 
The key feature of Topological bulk band  is the presence of a surface {\it normalizable} zero-mode.   
It happens when the    bulk band is inverted  and  one known mechanism for band inversion is  large spin-orbit interaction. So considering  boundary is crucial  to discuss TI. 

\item
On the other hand, in Holographic theory, having both bulk and boundary of a physical system is very difficult, if not impossible, since the bulk of the physical system is already at the  boundary of AdS space. 
In this situation, we have to 'carefully' delete either bulk or boundary for holographic description, depending on one's 
goal. Our  main physical observable is low energy transport dynamics which happens only 
in the surface  while bulk is a boring insulator  as a gapped system. 

Therefore,  we delete the bulk part of TI  together with the question `how the system became topological insulator'. 
Our goal is to describe the surface phenomena knowing the system is  already a TI. 
This means that we consider the case when the interaction do not destroy the  inverted  band  structure of the bulk. 
In our case this is justified because we  get strongly interacting system only at the surface  
by  deforming the surface band only such that the fermi surface becomes small by use of the magnetic doping at the surface.

\item 
We focus   on   the consequence of the Dirac cone rather than the cause of the Dirac cone, 
 the latter being the question of bulk. 
Once we confine our attention to the surface, we can characterize it by the presence of single Dirac cone. 
The surface physics due to the Dirac cone is not much different from that of  graphene. 
Therefore, by treating the surface as a Dirac material, we already encoded the most important 
consequence of topological nature  of the bulk band.   
 
\item 
Our gravitational system  is a deformation of charged AdS blackhole, which is widely used one.  
Our point is that one has to ask  ``for what material is  such canonical gravity solution good?''  
Such  local Lorentz invariant gravitation solutions are  good only for Dirac materials, which, as a quantum critical system, can be characterized by the dynamical  exponent $z=1$. 

\item 
There is one essential difference between a genuine 2 dimensional Dirac material and 
surface of TI. 
It is the relative position of the Fermi level and the surface band. 
For the former (like pure graphene), fermi level  should pass the Dirac point at zero applied chemical potential. 
For the latter, it is not necessarily so.  See figure attached. This point was also emphasized by Witten in his lectures on 
Topological material \cite{Witten:2015aoa}. 
The position of fermi level is determined by the bulk physics, and there is no reason why 
it should pass the Dirac point of the surface band. This is the reason why the surface fermi energy at 
zero chemical potential  is off the Dirac point of the surface Dirac cone. 
See the   figure 1. This is what we mean  `delete carefully'.

 \item  The existence of the surface mode   is the primary consequence of topological band structure, and 
 whether  surface gap is open or not is secondary question.

\item 
We want to discuss the  transports of  magnetically doped surface of TI. 
The question is what  is the recipe  to describe the system. 
We need the interaction between the magnetic impurity and the surface electron and  
this is the central part for the formulation for the practical purpose. 
Our interaction term should be the minimal one that describes the interaction with magnetic impurity with the charge current  that breaks time reversal invariance.  
 \end{enumerate}

 \subsection{Setup and background solutions}
 While it is clear that we need metric, gauge field and scalars to care the energy-momentum, the current and impurities respectively, for TI, special care is necessary to encode strong spin-orbit coupling (SOC). 
The latter  induces  the band inversion which in turn induces massless  fermions at the boundary and the topological nature of the system. 
To  encode the effect of SOC in the presence of the magnetic impurities  breaking the time reversal symmetry (TRS),  
we introduce   a  coupling between the  impurity density and the instanton density. 
Such an interaction term  was first  introduced  in \cite{kkss} by us to discuss the SOC with  TRS  broken. 
 It is the leading order term that  can  take care of  gauge field coupling with impurity density 
 in a  TRS  breaking manner. 
 
With these preparations, our holographic model is defined by   the Einstein-Maxwell-scalar action  
 on an  asymptotically AdS$_4$ manifold $\mathcal{M}$,
 \begin{align}\label{action}
	2\kappa^2S&=\int_{\mathcal{M}} d^4x \sqrt{-g}\left\{ R+\frac{6}{L^2}-\frac{1}{4}F^2-\sum_{I,a=1,2}\frac{1}{2}(\partial\chi_I^{(a)})^2\right\}\cr
	&-\frac{q_{\chi}}{16}\int_{\mathcal{M}}\sum_{I}(\partial\chi_I^{(2)})^2\epsilon^{\mu \nu\rho\sigma}F_{\mu \nu} F_{\rho\sigma}+S_c,\\
	S_c&=-\int_{\partial\mathcal{M}}d^3x\sqrt{-\gamma}\left(2K+\frac{4}{L}+R[\gamma]-\sum_{I,a=1,2}\frac{L}{2}\nabla\chi_I^{(a)}\cdot\nabla\chi_I^{(a)}\right),  
\end{align} 
where $\kappa^2=8\pi G$ and $L$ is the AdS radius and we set $2\kappa^2=L=1.$ $S_c$ is the counter term for holographic renormalization. Here we introduce two scalar fields and only one scalar field contributes in the interaction term.  The equations of motion are 
\begin{align}
&\partial_{\mu}(\sqrt{-g} g^{\mu\nu}\sum_{a} \partial_{\nu} \chi_{I}^{(a)} ) +\frac{q_{\chi}}{8}\, \partial_{\mu} ( \epsilon^{\rho\sigma\lambda\gamma} F_{\rho\sigma} F_{\lambda\gamma} g^{\mu\nu}\partial_{\nu} \chi_{I}^{(2)})=0 \\
&\partial_{\mu} \left( \sqrt{-g} F^{\mu\nu} +\frac{q_{\chi}}{4} g^{\rho\sigma} \sum_{I} \partial_{\rho} \chi_{I} ^{(2)}\,\partial_{\sigma} \chi_{I} \epsilon^{\alpha\beta \mu \nu}F_{\alpha\beta}\right )=0 \\
&R_{\mu\nu}-\frac{1}{2} g_{\mu\nu}\left[ R+6-\frac{1}{4}F^2-\sum_{I,a}\frac{1}{2}(\partial\chi_I^{(a)})^2 \right] -\frac{1}{2} F^{\rho}_{\mu} F_{\rho\nu}-\frac{1}{2}\sum_{I,a}(\partial_{\mu}\chi_{I}^{(a)})(\partial_{\nu}\chi_{I}^{(a)})\cr
&-\frac{1}{\sqrt{-g}}\, \frac{q_{\chi}}{16} \sum_{I} (\partial_{\mu} \chi_{I}^{(2)})(\partial_{\nu}\chi_{I}^{(2)}) \epsilon^{\rho\sigma\lambda\gamma} F_{\rho\sigma}F_{\lambda \gamma} =0.
\end{align}
Assuming the solution of the equations of motion takes the form,
 \begin{align}\label{bgsol0}
 	&A=a(r)dt+\frac{1}{2}H(xdy-ydx),\nonumber\\
 	&\chi_I^{(1)}=\left(\begin{array}{c} \alpha \, x \\ \alpha \, y \end{array}  \right),\qquad \chi_I^{(2)}=\left(\begin{array}{c} \lambda \, x \\ \lambda \,y \end{array}  \right),\nonumber\\
 	&ds^2=-U(r)dt^2+\frac{dr^2}{U(r)}+r^2(dx^2+dy^2).
 \end{align}
 we can find the exact solution as follows:
 \begin{align}\label{bgsol}
 	&U(r)=r^2-\frac{\alpha^2+\lambda^2}{2}-\frac{m_0}{r}+\frac{q^2+H^2}{4r^2}+\frac{\lambda^4H^2q_{\chi}^2}{20r^6}-\frac{\lambda^2Hqq_{\chi}}{6r^4},\\
 	&a(r)=\mu-\frac{q}{r}+\frac{\lambda^2Hq_{\chi}}{3r^3},
 \end{align}
 where $\mu$ is the chemical potential, $q$ and $m_0$ are determined by the conditions $A_t(r_0)=U(r_0)=0$ at the black hole horizon$(r=r_0)$. $q$ is the conserved $U(1)$ charge and $\alpha$ and $\lambda$ is relevant to momentum relaxation which will be discussed later:
 \begin{align}\label{m0q}
 	q&=\mu r_0 +\frac{1}{3}\theta H \qquad \text{with} \quad \theta=\frac{\lambda^2q_{\chi}}{r_0^2}\\
 	m_0&=r_0^3\left(1+\frac{r_0^2\mu^2+H^2}{4r_0^4}-\frac{\alpha^2+\lambda^2}{2r_0^2}\right)+\frac{\theta^2H^2}{45r_0}.
 \end{align}

\subsection{Thermodynamics and magnetisation}
To obtain a thermodynamic potential for this black hole solution we compute the on-shell Euclidean action($S^E$) by analytically continuing to Euclidean time($\tau$) of which period is  the inverse temperature
$
t = - i \tau \,,    S^E = - i S_{ren} 
$
where $S^E$ is the Euclidean action. 
The temperature of the boundary system is identified by the Hawking temperature in the bulk,
\begin{align}\label{temperature}
4\pi T = U'(r_0)=3 r_0 -\frac{1}{4 r_0^3}\left[ H^2 +2 r_0^2 (\alpha^2 +\lambda^2) +(q -H \theta)^2   \right] \,,
\end{align}
and the entropy density is given by the area of the horizon,
$
 s= 4\pi r_0^2 \,.
$
One can directly check that $dT/dr_0 >0$ for positive $r_0$, therefore, temperature is monotonically increasing function of $r_0$. It implies that the entropy is monotonically increasing function of temperature.

From the Euclidean renormalized action, we can define the thermodynamic potential($\Omega$) and its density($\mathcal W$):
\begin{align} 
S^E  \equiv  \frac{\mathcal{V}_2}{T} \mathcal W \equiv \frac{\Omega}{T} \,,
\end{align}
where $\mathcal{V}_2 = \int  d x  d y$. Plugging the solution (\ref{bgsol}) into the Euclidean renormalised action, the potential density $\mathcal W$ can be expresses   as 
\begin{align} \label{W1}
\mathcal W  =  \frac{\Omega}{\mathcal{V}_2} =-m_0 +2 r_0^3+\frac{(q^2 -H^2)}{2 r_0} +\frac{q \theta H}{3 r_0} -\frac{3 \theta^2 H^2}{10 r_0}.
\end{align}
The boundary energy momentum tensor is given by 
\begin{align}\label{energy}
<T_{\mu\nu}>=\left(\begin{array}{ccc} 2 m_0 & 0&0\\ 0& m_0&0\\0&0&m_0 \end{array}\right), 
\end{align}
from which energy density $\varepsilon=2m_{0}$.
If we identify the boundary on-shell action to the negative pressure $-{\cal P}$ and combine background solutions, we get Smarr relation 
\begin{align}
\varepsilon +{\cal P} = s \, T + \mu\, q.
\end{align}
 
Variation of the potential density with respect to $H$, $r_0$, $\mu$ and $\beta$ gives
\begin{align}\label{dW1}
\delta {\mathcal W} &=-\left(- \frac{H}{r_0}+\frac{q \, \theta}{3 r_0} -\frac{\theta^2 H}{5 r_0}  \right) \delta H - s \delta T - q \delta \mu \cr
&-r_0 \delta(\alpha^2)-\left( r_0 +\frac{ q \, \theta H}{3 r_0 \lambda^2} -\frac{\theta^2 H^2}{5 r_0 \lambda^2} \right) \delta(\lambda^2) , 
\end{align}
where the variation of $r_0$ is replaced by that of temperature.
 If we define 
\begin{align}\label{Mtheta}
\tilde{M} &=- \frac{H}{r_0}+\frac{q \, \theta}{3 r_0} -\frac{\theta^2 H}{5 r_0} \cr
\Theta_{\alpha} &= r_0 \cr
\Theta_{\lambda} & = r_0 +\frac{ q \, \theta H}{3 r_0 \lambda^2} -\frac{\theta^2 H^2}{5 r_0 \lambda^2} ,
\end{align}
then (\ref{dW1}) becomes
\begin{align}\label{dW2}
\delta {\mathcal W} = - \tilde{M} \delta H - s \delta T -q \delta \mu -\Theta_{\alpha} \delta (\alpha^2) -\Theta_{\lambda} \delta (\lambda^2).
\end{align}
Combining the variation of the second line in (\ref{W1}) and (\ref{dW1}), we finally get the first law of thermodynamics;
\begin{align}\label{delE}
\delta \varepsilon =- \tilde{M} \delta H + T \delta s + \mu \delta q - \Theta_{\alpha} \delta (\alpha^2)-\Theta_{\lambda} \delta(\lambda^2).
\end{align}

Now, we can define $\tilde{M}$ as the    magnetization of the  2+1 dimensional system from the theromdynamic law. The magnetization  $\tilde{M}$ has finite value in the absence of the external magnetic filed. 
\begin{align}\label{M00}
\tilde{M}_0 \equiv \tilde{M}\Big|_{H=0} = \frac{q \theta}{3 r_0}= \frac{q \lambda^2 q_{\chi}}{3 r_0^3}.
\end{align}
We can interpret the boundary system as a ferro-magnetic material. The value of $\tilde{M}_{0}$ proportional to the charge density $q$ and  $\lambda^2$. The scalar field $\chi_{I}$ plays role of the impurity density which cause momentum relaxation. From this analogy, we can  identify $\lambda^2$ as magnetic impurity density and $\alpha^2$ as non-magnetic one.

The thermodynamic law (\ref{delE}), contains variation of impurity denisty
and the conjugate $\Theta_{\alpha,\lambda}$  can be interpreted as the energy dissipated per unit impurity.

The pressure also can be written as
\begin{align}\label{Txx}
{\cal P} &= <T_{xx} > + \tilde{M} H + \Theta_{\alpha} \alpha^2 +\Theta_{\lambda} \lambda^2 \cr
&= \frac{\varepsilon}{2} + \tilde{M} H + \Theta_{\alpha} \alpha^2 +\Theta_{\lambda} \lambda^2.
\end{align}

The energy magnetization density can be defined as the linear response of the system under the metric fluctuation $\delta g_{tx}^{(0)} = B_E \,y$ \cite{HKMS, Blake:2015ina}, that is, as the derivative of the on-shell action with respect to the $B_E$ with    temperature and   chemical potential  fixed. 
\begin{align}\label{ME}
\tilde{M}^{E} = \frac{1}{72 r_0^2} \left[ 18 q^2 \theta + 3 \theta H^2 (2+\theta^2) - 4 q\, H (9 + 4 \theta^2)  \right].
\end{align} 
This result  will be used   when we calculate the DC transport coefficient in next section. 
It is straightforward to show that (\ref{ME}) is reduced to  the energy magnetization of the dynonic black hole when we take $\theta \to 0$ limit.

We now discuss physical meaning of two impurities. As shown in (\ref{M00}), $\lambda^2$ plays role of the magnetic impurity density. On the other hand, in the absence of $q_{\chi}$, impurity term in the thermodynamic first law (\ref{delE}) becomes $r_0 (\alpha^2 +\lambda^2)$ which is same as the impurity density in non-magnetic theory\cite{Andrade:2013gsa}. Therefore, we can redefine the total impurity density(sum of magnetic and non-magnetic impurity density) and the ratio of the magnetic impurity density to the total impurity density as
\begin{align}
\beta^2 = \alpha^2 +\lambda^2, \quad
\gamma=\frac{\lambda^2}{\alpha^2 + \lambda^2}.
\end{align}
From now on, we will use $\beta^2$ and $\gamma$  instead of $\alpha$ and $\lambda$.

We finish this section with a comment on an issue on   magnetization.  
In 2+1 dimensional system, there is a serious issue on the physical reality of magnetization, 
which is measured by  looking  the total 
 magnetic induction $B$,  which is the sum of external field and its effect induced inside the matter.
 However, the mass dimension of $\tilde M$ and 
$H$ are different and therefore we can not add these to form $B$.  
The presence of the problem is independent of using holography and we analyze the problem in appendix.

\section{Magneto-transport coefficients}
\subsection{DC conductivities from horizon data}
In this section, we calculate DC transports for the system with finite magnetization from black hole horizon data \cite{Blake:2013bqa,Blake:2013owa,Donos:2014cya}. To do this, we turn on small fluctuations around background solution (\ref{bgsol});
\begin{align}
\delta G_{ti} &= -t U(r) \zeta_i +\delta g_{ti} (r) \cr
\delta G_{ri} &=  r^2 \delta g_{ri} \cr
\delta A_i  &= t (-E_i + \zeta_i a(r) ) + \delta a_i (r) ,
\end{align}
where $i =x$, $y$ and we also turn on the fluctuation of scalar fields $\delta \chi_{i}^{(a)} (r)$. With this ansatz equations of motion for fluctuation are time-independent.  The linearized equations for the fluctuations are
\begin{align}\label{eqFluc1}
&\sum_{a}{\delta \chi_{i}^{(a)}}''(r) +\left( \frac{2}{r} +\frac{U'(r)}{U(r)}  \right) \sum_{a} \delta \chi_{i}^{(a)} (r) -(\alpha +\lambda) \delta g_{ri}'(r)\cr
&-(\alpha +\lambda) \left(  \frac{2}{r} +\frac{U'(r)}{U(r)} \right) \delta g_{ri}(r) -\frac{(\alpha+\lambda) \zeta_{i}}{r^2 U(r)} = 0 ,
\end{align}
\begin{align}\label{eqFluc2}
&\left[ U(r) \delta a_{i}' + a_{t}'(r) \delta g_{ti}(r) +\epsilon_{ij} H U(r) \delta g_{ij}(r) \right]' + \epsilon_{ij} \left[ \frac{q_{\chi} \lambda^2}{r^2} E_{j} +\left(\frac{H}{r^2} -\frac{2 q_{\chi}\lambda^2 a_{t}(r)}{r^3}  \right) \zeta_{j}  \right] =0,
\end{align}
\begin{align}\label{eqFluc3}
&\delta g_{ti}''(r) -\frac{1}{2 r^4 U(r)}\left[ H^2 -12 r^4 +2 r^2 \beta^2+ 4 r^2 U(r) -r^4 a_{t}'(r)^2 +4 r^3 U'(r) +2 r^4 U''(r) \right] \delta g_{ti} (r)\cr
&+a_{t}'(r) \delta a_{i}(r) +\epsilon_{ij} H \left[a_t'(r) \delta g_{tj}(r) +\frac{a_{t} (r)}{r^2 U(r)} \zeta_{j} -\frac{E_{j}}{r^2 U(r)}    \right] =0 \cr
&\left[ \frac{H^2}{4 r^2} -3 r^2 +\frac{\beta^2}{2} -\frac{r^2 a_t'(r)^2}{4}+r U'(r) +\frac{r^2 U''(r)}{2} \right] \delta g_{ri} (x)-\frac{\lambda}{2}\left[1-\frac{H q_{\chi} a_t'(r)}{r^2}  \right] \delta {\chi_{i}^{(2)}}'(r)\cr
&-\frac{\alpha}{2}\delta \chi_{i}^{(1)}-\epsilon_{ij}\frac{H}{2r^2}\left[\delta a_{j}'(r) +\frac{a_t '(r) \delta g_{tj} (r)}{U(r)}  \right] +\frac{a_t'(r)}{2 U(r)} E_{i} -\left[ \frac{1}{r} +\frac{a_{t} (r) a_{t}'(r)}{2 U(r)} -\frac{U'(r)}{2 U(r)}  \right] \zeta_{i} =0.
\end{align}
(\ref{eqFluc1}) and (\ref{eqFluc2}) come from the equations from the scalars and the gauge field fluctuation and the last two equations (\ref{eqFluc3}) are the Einstein equations of the metric fluctuations.

To make the fluctuation equations to be regular at the horizon, we impose 
\begin{align}\label{nearhor}
\delta g_{ti} &\sim \delta g_{ti}^{(0)} + {\cal O}(r-r_0) \cr
\delta g_{ri} & \sim \frac{\delta g_{ti}^{(0)}}{r^2 U(r)} + {\cal O} (r-r0) \cr
\delta \chi_{i} & \sim {\cal O}(r-r_0) \cr
\delta a_{i} &\sim -\frac{E_{i}}{4\pi T} \log (r-r_0) +{\cal O}(r-r_0),
\end{align}
where the last line implies in-falling condition of the gauge field fluctuation at the horizon. We can easily see this in the Eddington-Finkelstein coordinate.

With the regularity condition (\ref{nearhor}), the last equation can be written in terms of the black hole data and external sources as
\begin{align}\label{eqhor}
&\frac{{\cal F}}{r_0^2} \delta g_{tx}^{(0)} -\frac{H {\cal G}}{r_0^2} \delta g_{ty}^{(0)} +{\cal G} E_x +H E_y + 4\pi r_0^2 T \zeta_x =0 \cr
&\frac{{\cal F}}{r_0^2} \delta g_{ty}^{(0)} +\frac{H {\cal G}}{r_0^2} \delta g_{tx}^{(0)} +{\cal G} E_y -H E_x + 4\pi r_0^2 T \zeta_y =0 , 
\end{align}
where
\begin{align}
{\cal F} &= r_0^2 \beta^2 +(1+\theta^2) H^2 -q\, \theta H \cr
{\cal G} &=q -\theta H. 
\end{align}
The solution of the algebraic equation (\ref{eqhor}) is
\begin{align}\label{horizondata}
\delta g_{ti}^{(0)} =-\frac{r_0^2}{{\cal F}^2 + H^2 {\cal G}} \left[ {\cal G}({\cal F}-H^2) E_i + H({\cal F}+{\cal G}^2) \epsilon_{ij} E_j + 4\pi r_0^2 T ({\cal F} \zeta_i + H {\cal G} \epsilon_{ij} \zeta_j )  \right].
\end{align}

Now, let's consider current defined by
\begin{align}\label{current0}
{\cal J}^{i} &=\sqrt{-g} F^{ir} \cr
&=-U(r) \delta a_{i}'(r) - a_t'(r) \delta g_{tx} (r) -\epsilon_{ij} \delta g_{rj}(r) \cr
{\cal Q}^{i} &=U(r)^2 \partial_r \left(\frac{\delta g_{ti}(r)}{U(r)}\right) - a_t (r) J^i \cr
&=U(r) \delta g_{ti}'(r)+(a_t (r) a_t'(r) -U'(r)) \delta g_{ti} (r) + a_t (r) U(r) \delta a_{i}'(r) +\epsilon_{ij} H a_t (r) U(r) \delta g_{rj} (r).\cr
\end{align}
One can easily show that these currents (\ref{current0}) become the electric current $J^{i}$ and the heat current $Q^{i} = <T^{ti}> - \mu J^{i}$ at the boundary($r\rightarrow \infty$). These quantities, however, can not be conserved along $r$ direction in general. The way to resolve this problem is to take derivative of both side of (\ref{current0}) with respect to $r$.  Substitute equations of motion (\ref{eqFluc1}), (\ref{eqFluc2}) and (\ref{eqFluc3}) to the right hand side and integrate it from black hole horizon to boundary. In summary;
\begin{align}\label{current1}
J_{total}^{i} &= {\cal J}(\infty) = {\cal J}(r_0) + \int_{r_0}^{\infty} d r \left(\partial_{r} {\cal J}(r) \Big|_{{\rm e.o.m}}\right) \cr
Q_{total}^{i} &={\cal Q}(\infty) ={\cal Q}(r_0)+ \int_{r_0}^{\infty} d r \left(\partial_{r} {\cal Q}(r) \Big|_{{\rm e.o.m}}\right),
\end{align}
here, we put subscript `total' to the current because it contains the magnetization current and the energy magnetization current contribution.
By imposing the background solution and the horizon expansion (\ref{nearhor}), we get
\begin{align}\label{current2}
J_{total}^{i} &= E_{x} +  \epsilon_{ij} \theta E_{j} -\frac{{\cal G}}{r_0^2}\delta g_{ti}^{(0)} -\frac{H}{r_0} \epsilon_{ij} \delta g_{tj}^{(0)}+M \epsilon_{ij} \zeta_{j} \cr
Q_{total}^{i} &= - 4\pi T \delta g_{ti}^{(0)} - M \epsilon_{ij} E_j - 2(M^{E} - \mu M) \epsilon_{ij} \zeta_j,
\end{align}
where $M$ and $M^E$ are the magnetization and the energy mangetization defined in the previous section. The last term in the electric current and the last two terms in the heat current correspond to the mangetization current and the energy magnetization current, which should not be taken into account to the electric and heat current. Subtracting these terms and combining with the horizon data (\ref{horizondata}), we get the electric and heat current in terms of external sources;
\begin{align}\label{current3}
J^{i} =& \frac{({\cal F}+{\cal G}^2)({\cal F}-H^2)}{{\cal F}^2 + H^2 {\cal G}^2} E_i  +\left[\theta +\frac{H {\cal G}(2 {\cal F} +{\cal G}^2 -H^2)}{{\cal F}^2 + H^2 {\cal G}^2}  \right] \epsilon_{ij} E_j \cr
&+ \frac{s T {\cal G}({\cal F}-H^2)}{{\cal F}^2 + H^2 {\cal G}^2} \zeta_i +\frac{s T H({\cal F}+{\cal G}^2)}{{\cal F}^2 + H^2 {\cal G}^2} \epsilon_{ij} \zeta_j \cr
Q^{i} =& \frac{s T {\cal G}({\cal F}-H^2)}{{\cal F}^2 + H^2 {\cal G}^2} E_{i} +\frac{s T H({\cal F} +{\cal G}^2)}{{\cal F}^2 + H^2 {\cal G}^2} \epsilon_{ij} E_{j} +\frac{s^2 T^2 {\cal F}}{{\cal F}^2 + H^2 {\cal G}^2} \zeta_{i} +\frac{s^2 T^2 H {\cal G}}{{\cal F}^2 + H^2 {\cal G}^2}\epsilon_{ij} \zeta_{j}  .
\end{align}
Now, the transport coefficients can be read off from
\begin{align}\label{transport}
\left(\begin{array}{c} J^{i} \\ Q^{i} \end{array}  \right)=\left(\begin{array}{cc} \sigma_{ij} & \alpha_{ij} T \\ \bar{\alpha}_{ij} T& \bar{\kappa}_{ij} T \end{array}\right) \left( \begin{array}{c} E_j \\ -\frac{\nabla_j T}{T} \end{array}   \right),
\end{align}
where the temperature gradient $-(\nabla_i T)/T = \zeta_{i}$ in previous expression (\ref{current3}). The results are summarized as;
\begin{align}\label{DC01}
\sigma_{ii} &= \frac{({\cal F}+{\cal G}^2)({\cal F}-H^2)}{{\cal F}^2 + H^2 {\cal G}^2} \cr
\sigma_{ij} &= \epsilon_{ij} \left[\theta +\frac{H {\cal G}(2 {\cal F} +{\cal G}^2 -H^2)}{{\cal F}^2 + H^2 {\cal G}^2}  \right] \cr
\alpha_{ii} &=\bar{\alpha}_{ii}= \frac{s  {\cal G}({\cal F}-H^2)}{{\cal F}^2 + H^2 {\cal G}^2} \cr
\alpha_{ij} &=\bar{\alpha}_{ij}=\epsilon_{ij}\frac{s  H({\cal F}+{\cal G}^2)}{{\cal F}^2 + H^2 {\cal G}^2} \cr
\bar{\kappa}_{ii} &=\frac{s^2 T {\cal F}}{{\cal F}^2 + H^2 {\cal G}^2} \cr
\bar{\kappa}_{ij} &=\epsilon_{ij} \frac{s^2 T H {\cal G}}{{\cal F}^2 + H^2 {\cal G}^2}.
\end{align}
The thermal conductivity when  electric currents are set to be zero is
\begin{align}
\kappa_{ij} &= \bar{\kappa}_{ij} - T \, (\bar{\alpha}\cdot \sigma^{-1} \cdot \alpha)_{ij},\label{kappa}
\end{align}
which  can be calculated to be 
\begin{align}
\kappa_{ii} &= \frac{s^2 T({\cal F}+{\cal G}^2-H^2 + 2 H {\cal G} \theta +{\cal F} \theta^2)}{({\cal F}+{\cal G}^2)^2+ H(H-2 {\cal G} \theta)(H^2 -2{\cal F} -{\cal G}^2)+({\cal F}^2 +H^2 {\cal G}^2)\theta^2} \cr
\kappa_{ij}&=\epsilon_{ij} \frac{s^2 T (H - {\cal G}\theta)({\cal G} + H \theta)}{({\cal F}+{\cal G}^2)^2+ H(H-2 {\cal G} \theta)(H^2 -2{\cal F} -{\cal G}^2)+({\cal F}^2 +H^2 {\cal G}^2)\theta^2}.
\end{align}

There are several interesting properties to the DC conductivities. 
The Onsaga's relation $\alpha_{ij}$=$\bar{\alpha}_{ij}$ can be easily checked. 
Also, notice that the off-diagonal components are anti-symmetric. 
These results are consistent with the DC transport coefficient of dyonic black hole\cite{Blake:2014yla,Blake:2015ina,Kim:2015wba} in $q_{\chi} \rightarrow 0$ or $\gamma \rightarrow 0$ limit, which gives some confidence for the validity of our results. 

The other interesting observables are the Seebeck effect and the Nernst effect. Under the external magnetic field, electric field can be generated by the longitudinal thermal gradient. The generation of the longitudinal and the transverse electric field  are called `Seebeck'  and `Nernst' effect respectively.  The Seebeck effect is used to check the Mott relation which is sensitive to the type of the interaction in material. Recently, people have found that the Mott relation near the charge neutrality point is broken   in the clean graphene\cite{Ghahari:2016aa}. The Nernst effect is known as the phenomena of the vortex liquid and  certain material shows large Nernst signal above critical temperature\cite{wang2006nernst,Anderson:2007aa}. The Seebeck coefficient($S$)   and the Nernst signal($N$) are defined as 
\begin{align}\label{SN01}
S = \left(\sigma^{-1} \cdot \alpha \right)_{xx}  , \quad 
N =-\left(\sigma^{-1} \cdot \alpha \right)_{yx}
\end{align}
respectively 
and can be calculated from (\ref{DC01}) : 
\begin{align}\label{TEPN}
S &= \frac{s({\cal F} +{\cal G}^2)({\cal G} + H \theta)}{({\cal F}+{\cal G}^2)^2+ H(H-2 {\cal G} \theta)(H^2 -2{\cal F} -{\cal G}^2)+({\cal F}^2 +H^2 {\cal G}^2)\theta^2} \cr
N &=\frac{s({\cal F}-H^2)(H - {\cal G} \theta)}{({\cal F}+{\cal G}^2)^2+ H(H-2 {\cal G} \theta)(H^2 -2{\cal F} -{\cal G}^2)+({\cal F}^2 +H^2 {\cal G}^2)\theta^2}.
\end{align}

 A technical remark is in order. Subtracting the contribution of the external magnetic field to the energy density amounts to  shifting on-shell action to satisfy thermodynamic relations. It corresponds to adding the finite counter term $\frac{3 H^2}{4 r_0^2}$ to the action. We take the same procedure to calculate DC transports. The final electric and heat current (\ref{current2}) would be written in terms of $\tilde{M}$ and $\tilde{M}^E$. After subtracting the magnetization current and the energy magnetization current contribution, we get same result of the DC transport coefficients (\ref{DC01}).

\subsection{Analysis of the transport coefficients}
In this section, we analyze the external parameter dependence of magneto-transport coefficients.  As shown in (\ref{DC01}), DC transport coefficients are complicated function of the external parameters and hence we need full numerical calculation. 
In the absence of the external field, it gives anomalous transport which comes from non-zero magnetization of the system, which  can be treated analytically. We will discuss several aspects of them. 

\subsubsection{Magnetotransport coefficients for  non-ferromagnetic case ($\mu =0$)}
In the presence of the external magnetic field, DC transport coefficients (\ref{DC01}) are complicated functions of other parameters.  In particular we can not solve $r_0$  in terms of others but we can calculate transport coefficients numerically.  In this section, we discuss transport coefficients in non-ferromagnetic case. As shown in (\ref{M00}), the magnetization at zero magnetic field proportional to the charge density or equivalently chemical potential. Therefore, we set $\mu=0$ to discuss magneto-transports in non-ferromagnetic material.

\begin{figure}[ht!]
\centering
    \subfigure[ ]
   {\includegraphics[width=5.5cm]{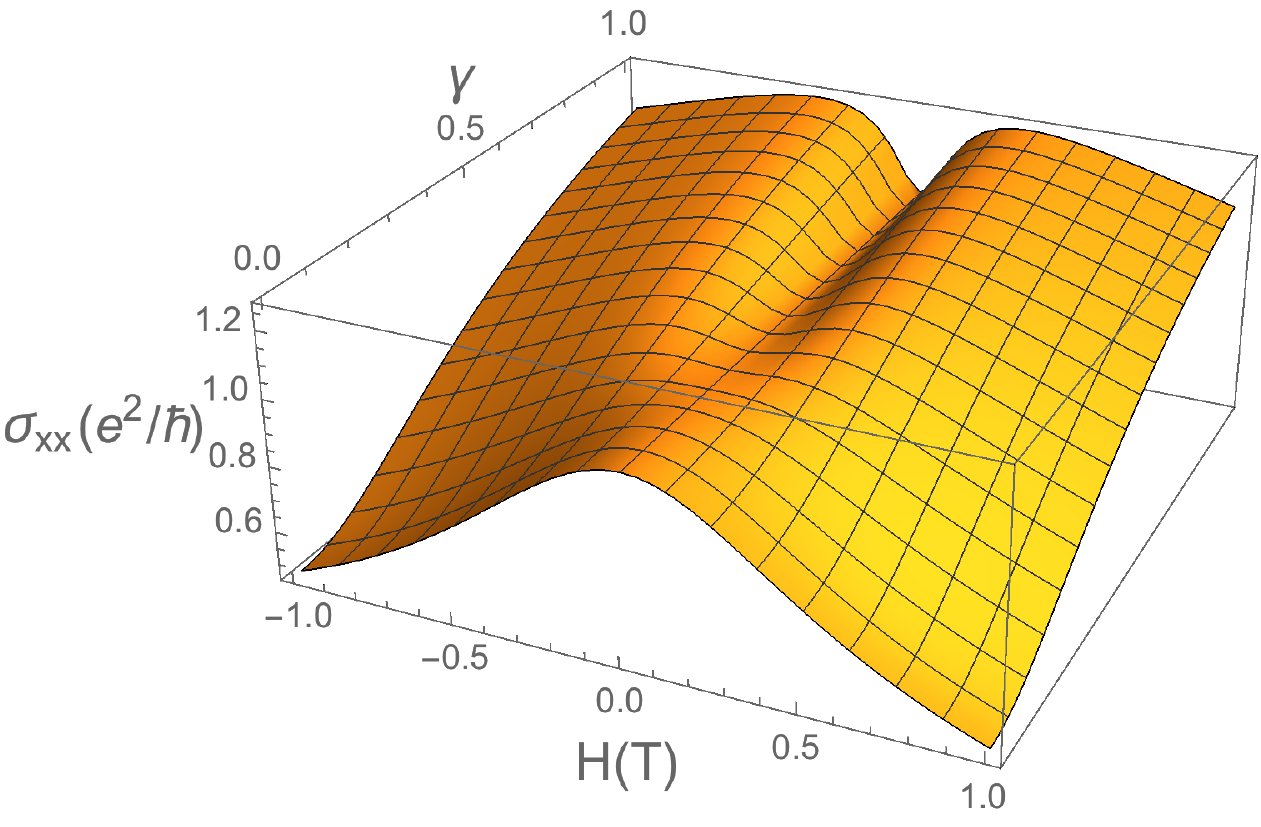} \label{}}
   \hspace{1cm}
       \subfigure[]
   {\includegraphics[width=5.5cm]{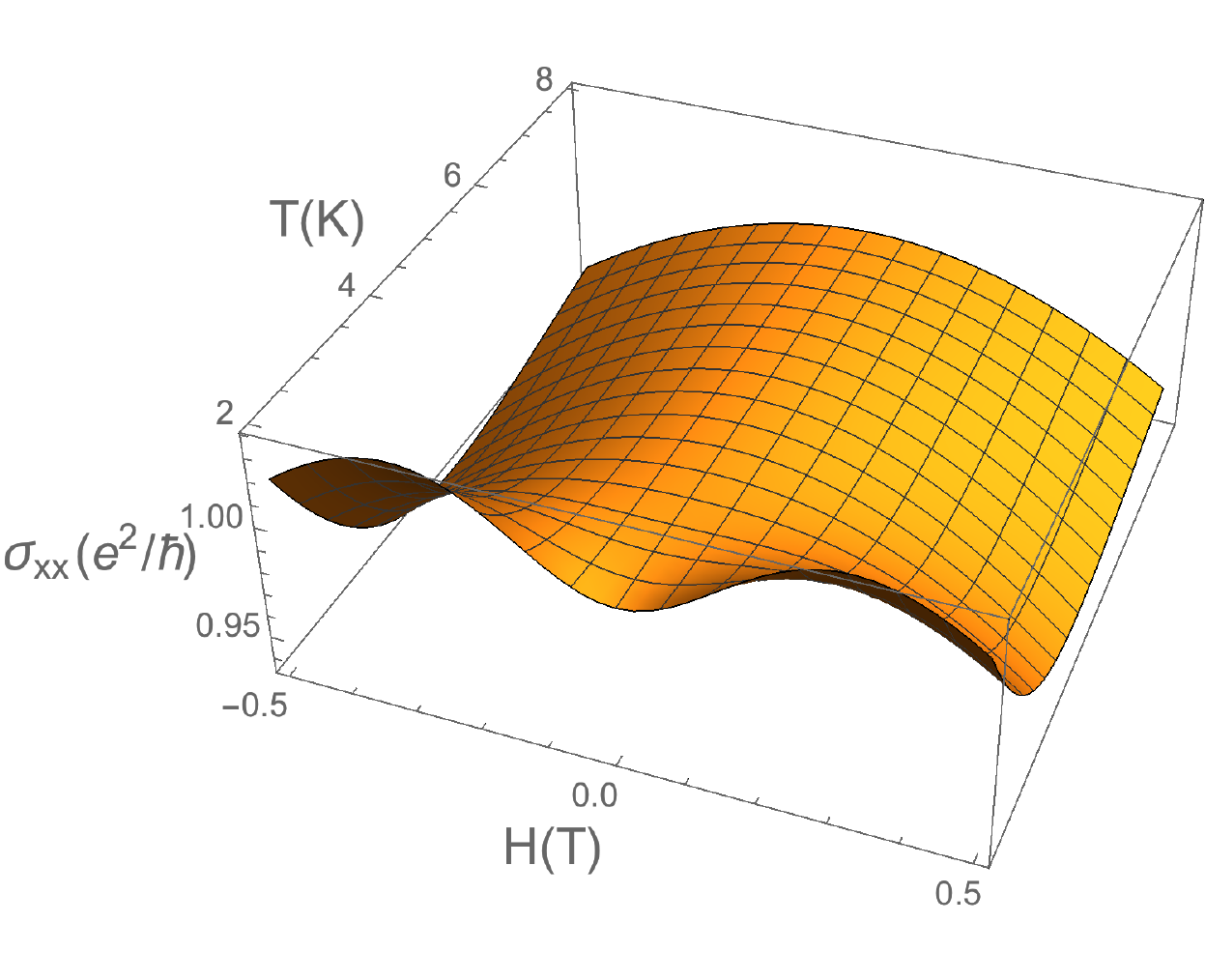} \label{}}
 \caption{Longitudinal conductivity $\sigma_{xx}$  
 (a) as a function of Magnetic doping ($\gamma$) and magnetic field ($B$)  at    $T=2K$, 
 (b) as a function of Temperature and magnetic field  at $q_{\chi} \gamma=0.7$. 
 In both case, we used  $\beta^2=\frac{2747}{(\mu m)^2}$ and  $v_{F} =7.5 \times 10^4 m/s$. 
 We use the same number in all the figures appearing in this section. } \label{fig:SxxHgH}
\end{figure}

Figure \ref{fig:SxxHgH} shows the magnetic doping and temperature dependence of the longitudinal conductivity.

As discuss in \cite{Seo:2017oyh}, there is transition between weak-antilocalization and weak localization as we change magnetic doping or temperature.  The longitudinal conductivity can be expressed at small magnetic field limit as
\begin{align}\label{MC01}
\sigma_{xx} \sim 1+ \frac{2}{  r_0^2 \beta^2} ({4\over9} \theta^2 -1) H^2 + {\cal O} (H^4).
\end{align}
At high temprature, $\theta$ behaves $\sim 1/T^2$ and the longitudinal conductivity alway has negative curvature at $H=0$. As temperature decreases, the sign of $H^2$ in (\ref{MC01}) is  flipped  at $\theta=3/2$. By combining with background solution, we get critical temperature given by the analytical expression 
\begin{align}\label{Tc}
T_C = \frac{\sqrt{3/2}}{2\pi} \, \frac{( q_{\chi} \gamma -1/4)}{\sqrt{q_{\chi} \gamma}} \, \beta.
\end{align}
If the value of $q_{\chi} \gamma$ smaller than $1/4$, then $T_C$ becomes negative and there is no transition to weak localization. The system has weak anti-localization in all temperature region. In the case of $q_{\chi} \gamma > 1/4$, the system shows weak  localization for  $T <T_C$ and transition to weak anti-localization appears at $T=T_C$.  
For high magnetic field, the longitudinal conductivity becomes
\begin{align}
\sigma_{xx} \Big|_{H \gg 1} \sim \frac{\beta^2}{H} + \frac{2}{3} \, \left( \frac{q_{\chi} \beta^2 \gamma}{H} \right)^2 + {\cal O} \left( \frac{1}{H^3} \right). 
\end{align}
 We see that total impurity term is dominant  over magnetic impurity term. 

\begin{figure}[ht!]
\centering
    \subfigure[ ]
   {\includegraphics[width=5.5cm]{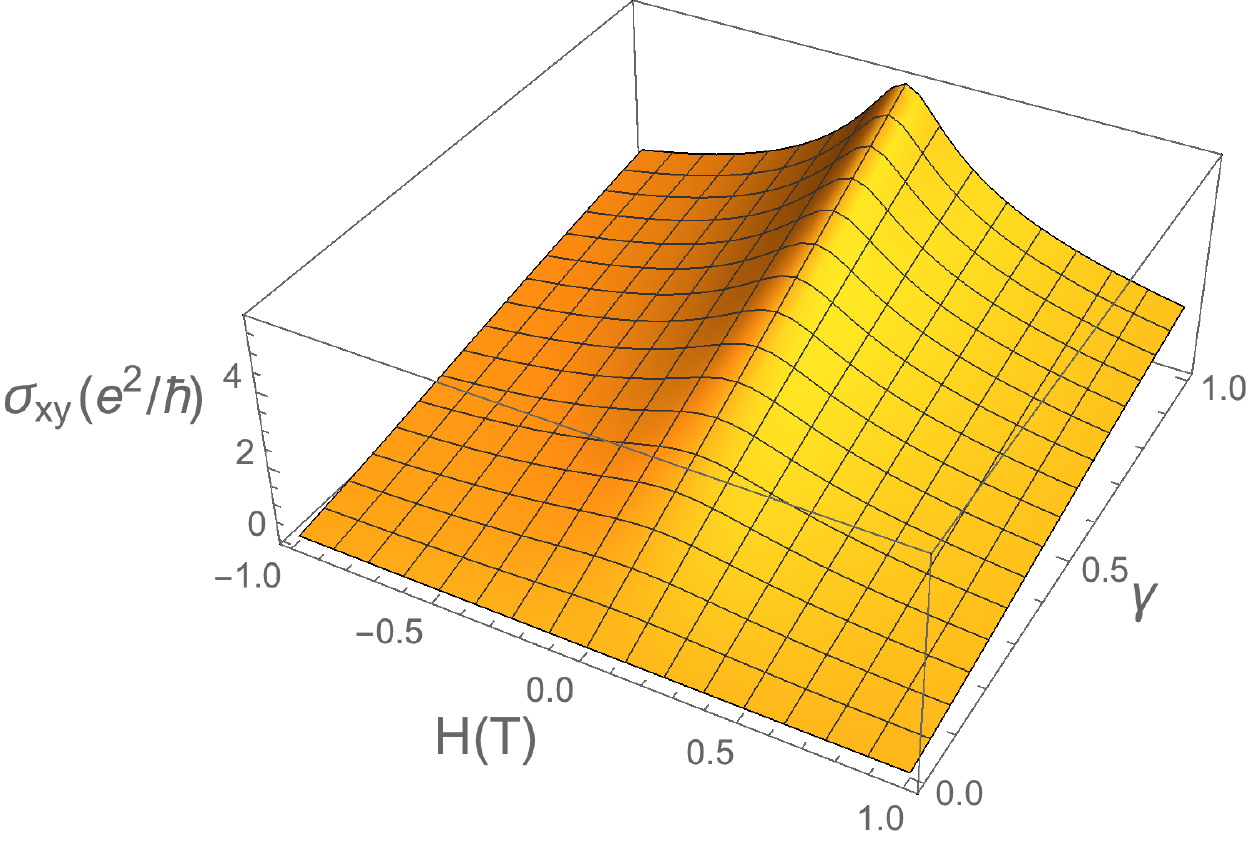} \label{}}
   \hspace{1cm}
       \subfigure[]
   {\includegraphics[width=5.5cm]{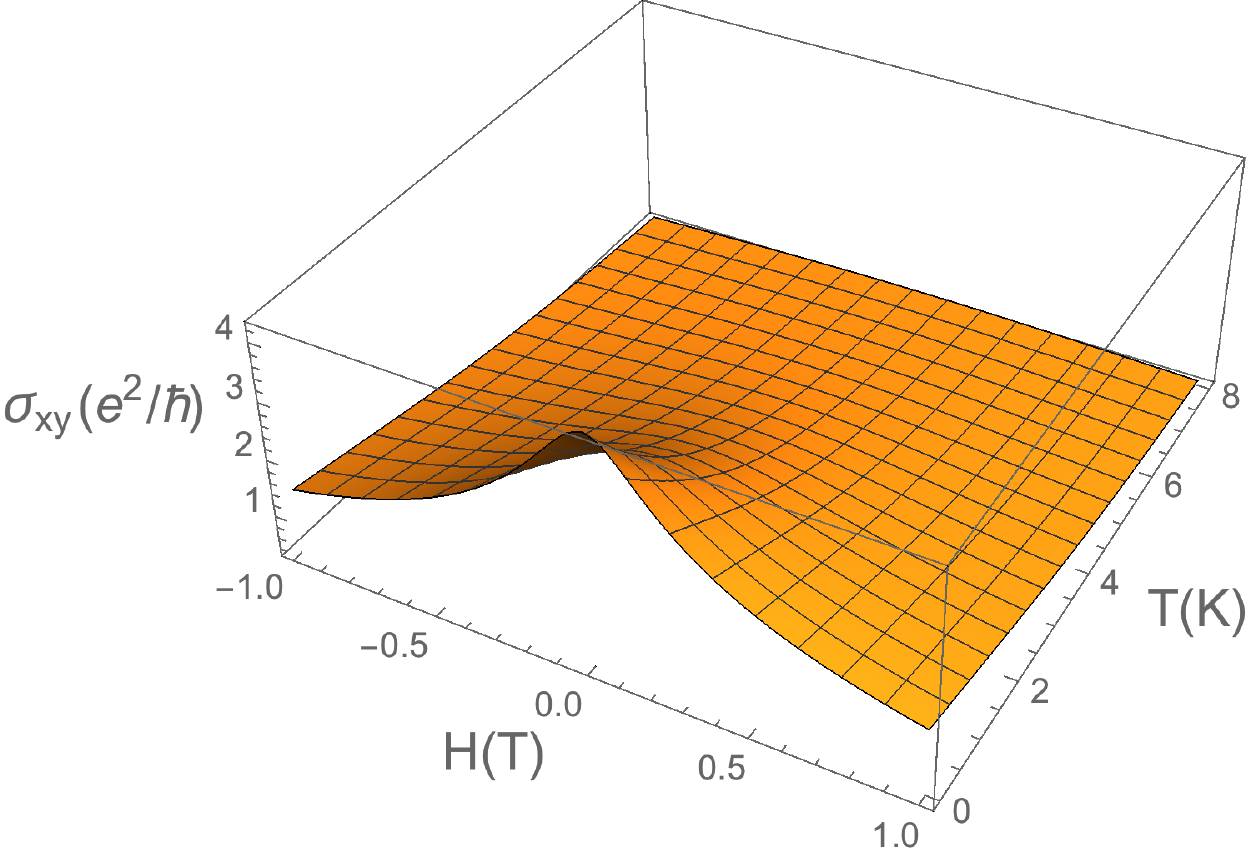} \label{}}
 \caption{Transverse conductivity $\sigma_{xy}$ (a) as a function of Magnetic doping and magnetic field    at $T=2K$ and  (b) as a function of Temperature and magnetic field   at $q_{\chi} \gamma=0.7$.   } \label{fig:SxyH}
\end{figure}

Figure \ref{fig:SxyH} shows transverse conductivity $\sigma_{xy}$  as a function of magnetic doping, temperature and magnetic field. 
Here one can see that  the $\sigma_{xy}$ has maximum at $H=0$ and the height is proportional to doping parameter and inverse temperature. We can understand this analytically since  it can be expressed as
\begin{align}
\sigma_{xy} \Big|_{H \ll 1} &\sim \theta - \frac{4 q_{\chi} \gamma}{3 r_0^4} H^2 + {\cal O}(H^3) \cr
\sigma_{xy} \Big|_{H \gg 1} &\sim \frac{q_{\chi} \beta^2 \gamma}{3 H} + {\cal O}\left( \frac{1}{H^3} \right),
\end{align}
here we use $r_0 \sim \sqrt{H}$ for $H \gg 1$.

\begin{figure}[ht!]
\centering
    \subfigure[ ]
   {\includegraphics[width=5.5cm]{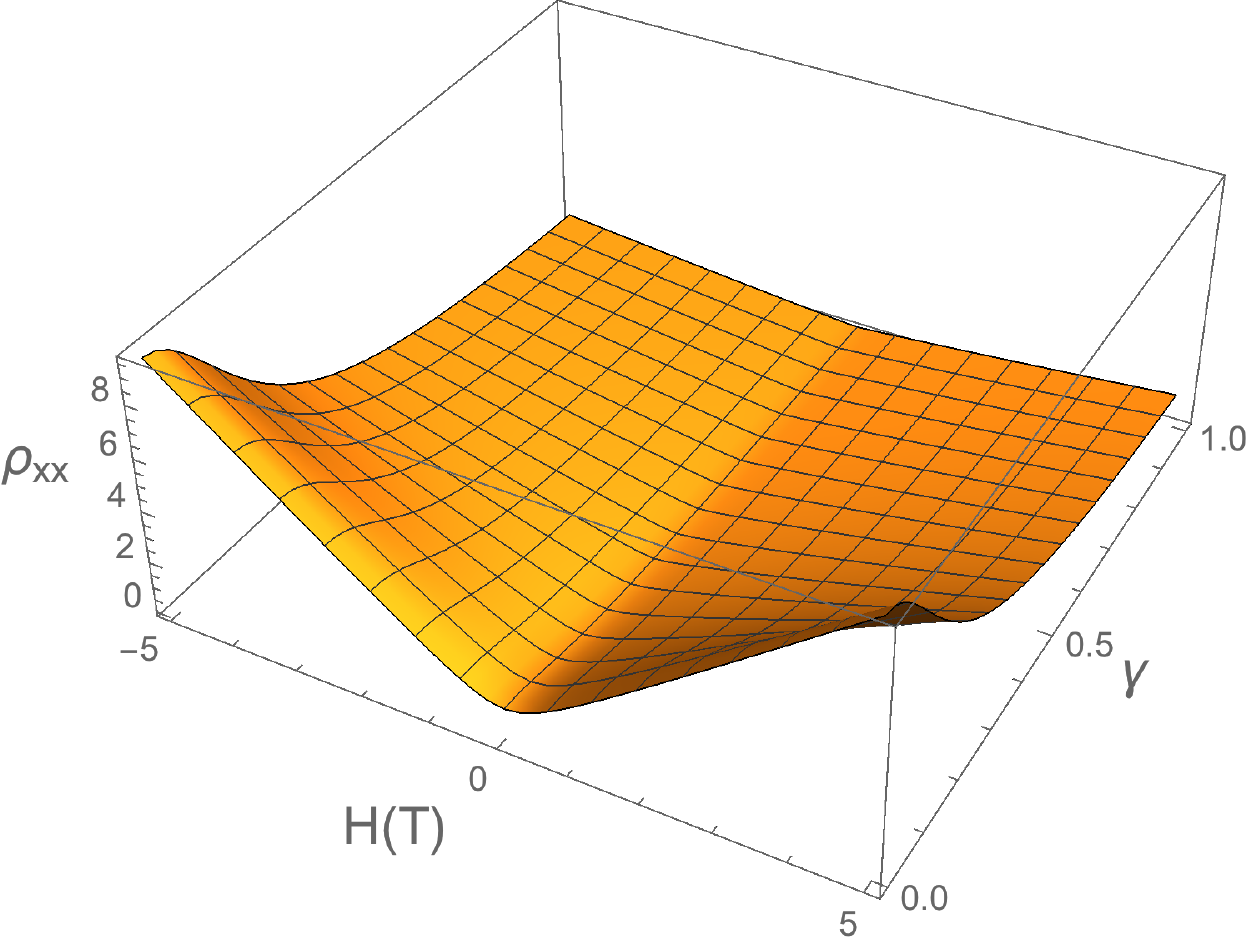} \label{}}
   \hspace{1cm}
       \subfigure[]
   {\includegraphics[width=5.5cm]{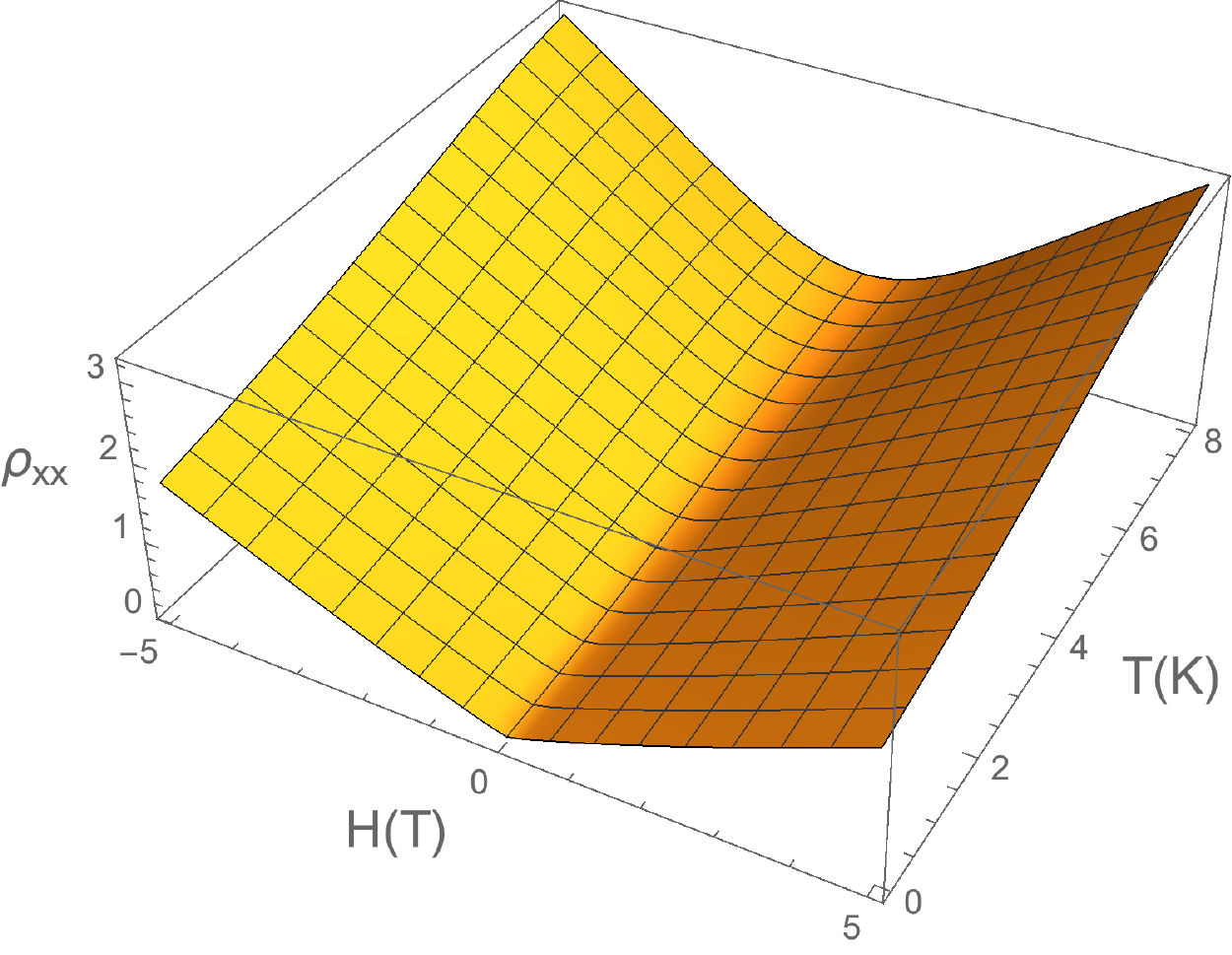} \label{}}
   
 \caption{  longitudinal resistivity  $\rho_{xx}$, (a)  as a function of Magnetic doping and magnetic field   at   $T=2K$ and (b)  as a function of Temperature and magnetic field     at $\gamma=0.5$. 
           } \label{fig:RxxH}
\end{figure}

\begin{figure}[ht!]
\centering
   {\includegraphics[width=6cm]{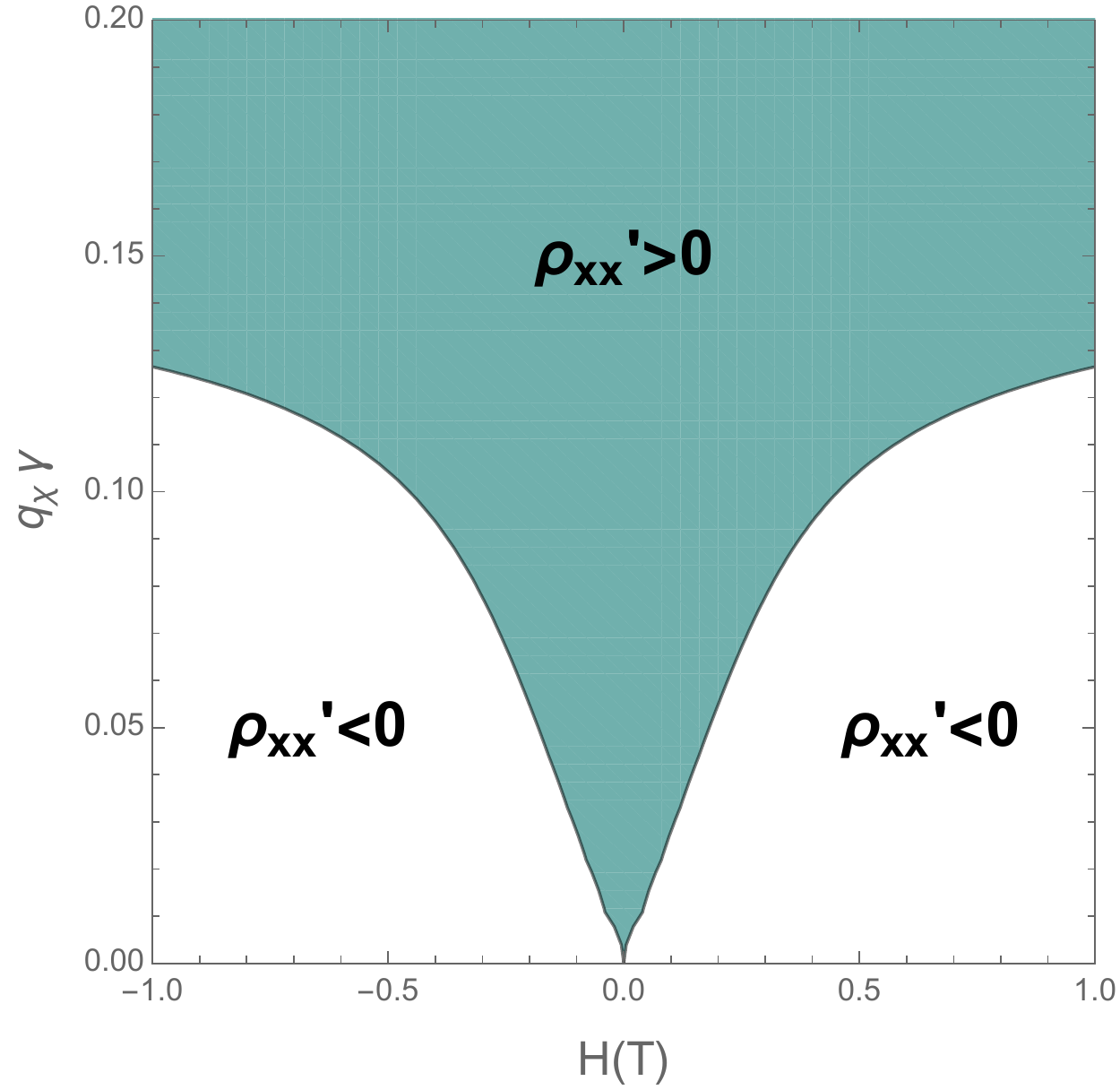} \label{}}
 \caption{(a) Metalicity, the sign of  $\partial \rho_{xx} /\partial T$,  as a function of Magnetic doping and magnetic field. Negative (white) regions indicate the instability.  } \label{fig:dRdT}
\end{figure}

The longitudinal resistivity is presented in Figure \ref{fig:RxxH}.
 Figure  \ref{fig:RxxH} (a) shows magnetic doping dependence of magneto-resistance.  One should notice that the signal of  transition to WL    does not appear in the resistivity contrary  to $\sigma_{xx}$ shown in Figure \ref{fig:SxxHgH} (a). 
 One can understand it in the following way. 
 The longitudinal resistivity comes from the inversion of the conductivity matrix as
\begin{align}
\rho_{xx} = \frac{ \sigma_{xx}}{\sigma_{xx}^2 + \sigma_{xy}^2},
\end{align}
where we use $\sigma_{xx}=\sigma_{yy}$ and $\sigma_{xy} = -\sigma_{yx}$.  
The relation $\rho_{xx} \sim 1/\sigma_{xx} $  holds only when    $\sigma_{xy}$ is sufficiently small. 
On the other hand, the weak localization in $\sigma_{xx}$  appears in  large doping region, where the transverse conductivity is large as one can see Figure \ref{fig:SxyH} (a).
 Therefore, transverse component of the conductivity affects the longitudinal magneto-resistance and  wash out the signal of weak localization.  So it is better to define the weak localization  by magneto-conductance  
 instead of magneto-resistance at least for strongly interacting system. 
  Figure  \ref{fig:RxxH} (b) shows magneto-resistivity as function of  $(B,T)$ at  a fixed doping rate.
Near zero temperature, we can define the metalicity as the sign of  $\partial \rho_{xx} /\partial T$, which is demonstrated at figure \ref{fig:dRdT}. 

\begin{figure}[ht!]
\centering
    \subfigure[ ]
   {\includegraphics[width=5.5cm]{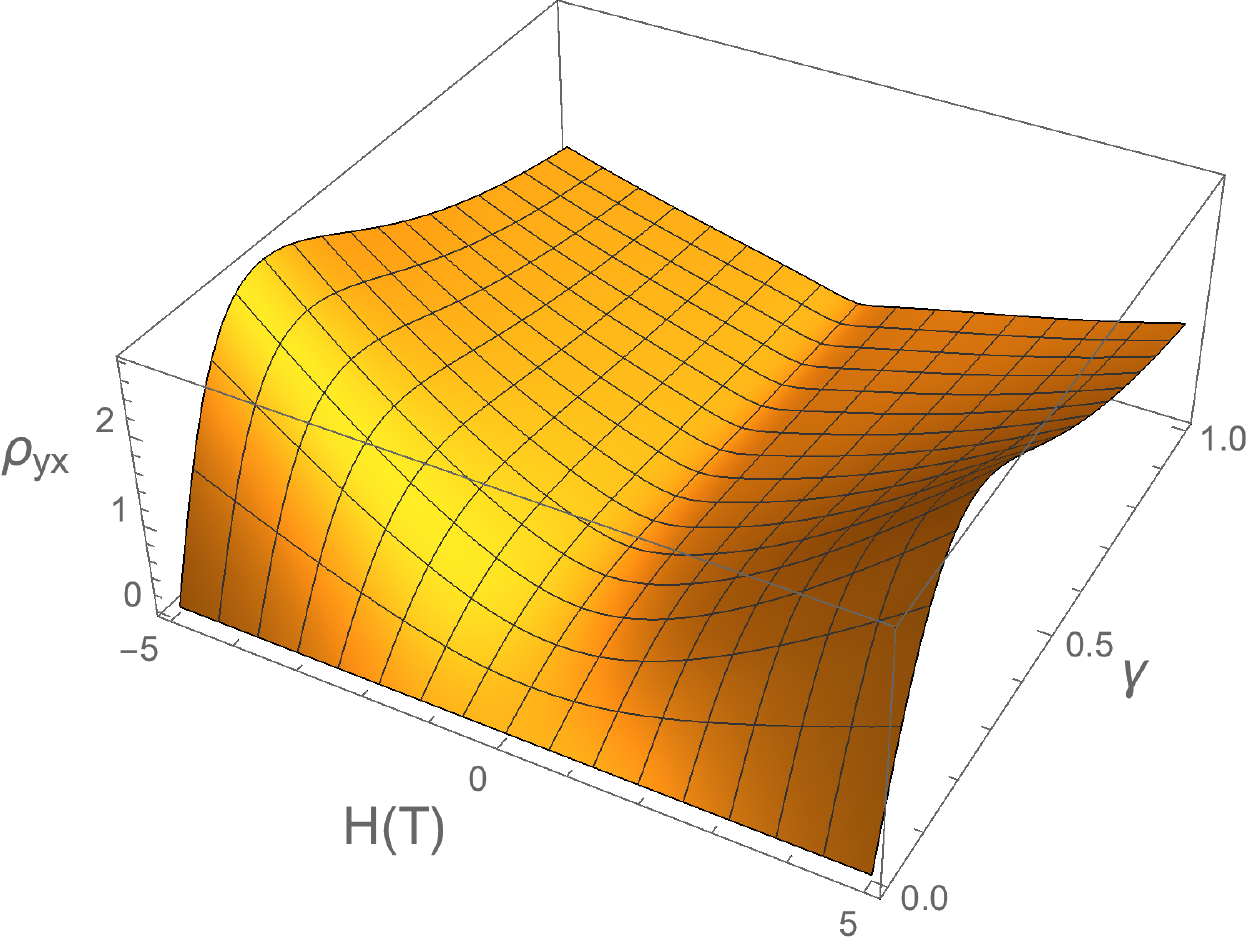} \label{}}
   \hspace{1cm}
       \subfigure[]
   {\includegraphics[width=5.5cm]{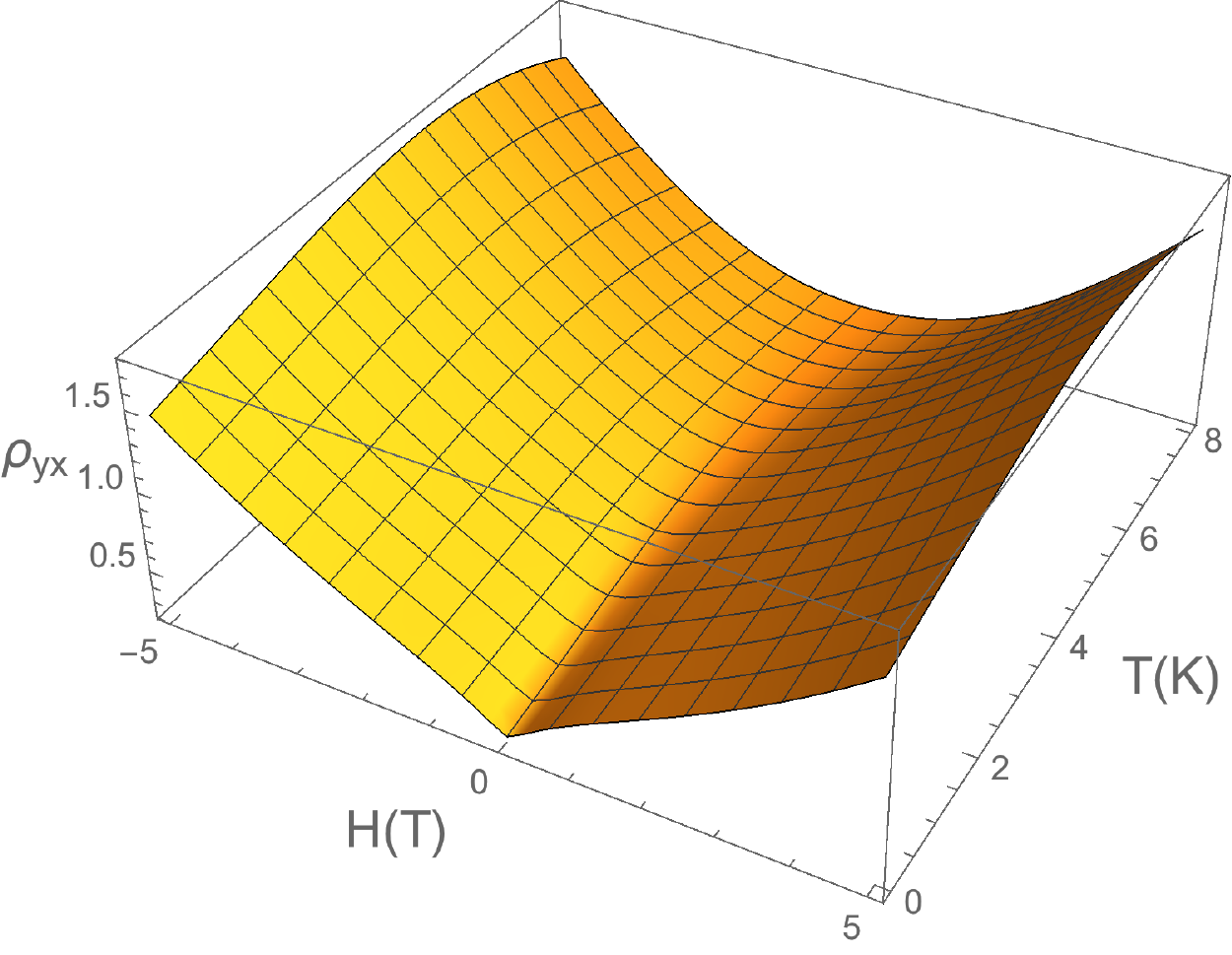} \label{}}
 \caption{ Transverse resistivity $\rho_{yx}$,  (a) as a function of Magnetic doping and magnetic field    at  $T=0.01$. (b)  as a function of Temperature and magnetic field    at $q_{\chi} \gamma=0.7$. 
           } \label{fig:RyxH}
\end{figure}

Due to the non vanishing $\theta$, the transverse resistivity  also has non-trivial behavior for other parameters. Near zero magnetic field and zero temperature, the transverse resistivity can be expressed as
\begin{align}
\rho_{yx} \Big|_{T \ll 1, H \ll 1} \sim\frac{ \tilde{\theta}  \left(4 \tilde{\theta} ^4+13 \tilde{\theta} ^2+3\right)}{3 \beta ^4 \left(\tilde{\theta} ^2+1\right)^2} H^2 +{\cal O} (H^4),
\end{align}
 where $\tilde{\theta} = 6  q_{\chi} \gamma$. On the other hand, the transverse resistivity linearly increases at large magnetic field limit as
 \begin{align}
 \rho_{yx} \Big|_{H \gg 1} \sim \frac{8 \tilde{\theta}}{45 \beta^2}\cdot H + \cdots.
 \end{align}
This behavior of the transverse resistivity is drawn in Figure \ref{fig:RyxH} (a). 

\begin{figure}[ht!]
\centering
    \subfigure[ ]
   {\includegraphics[width=5.5cm]{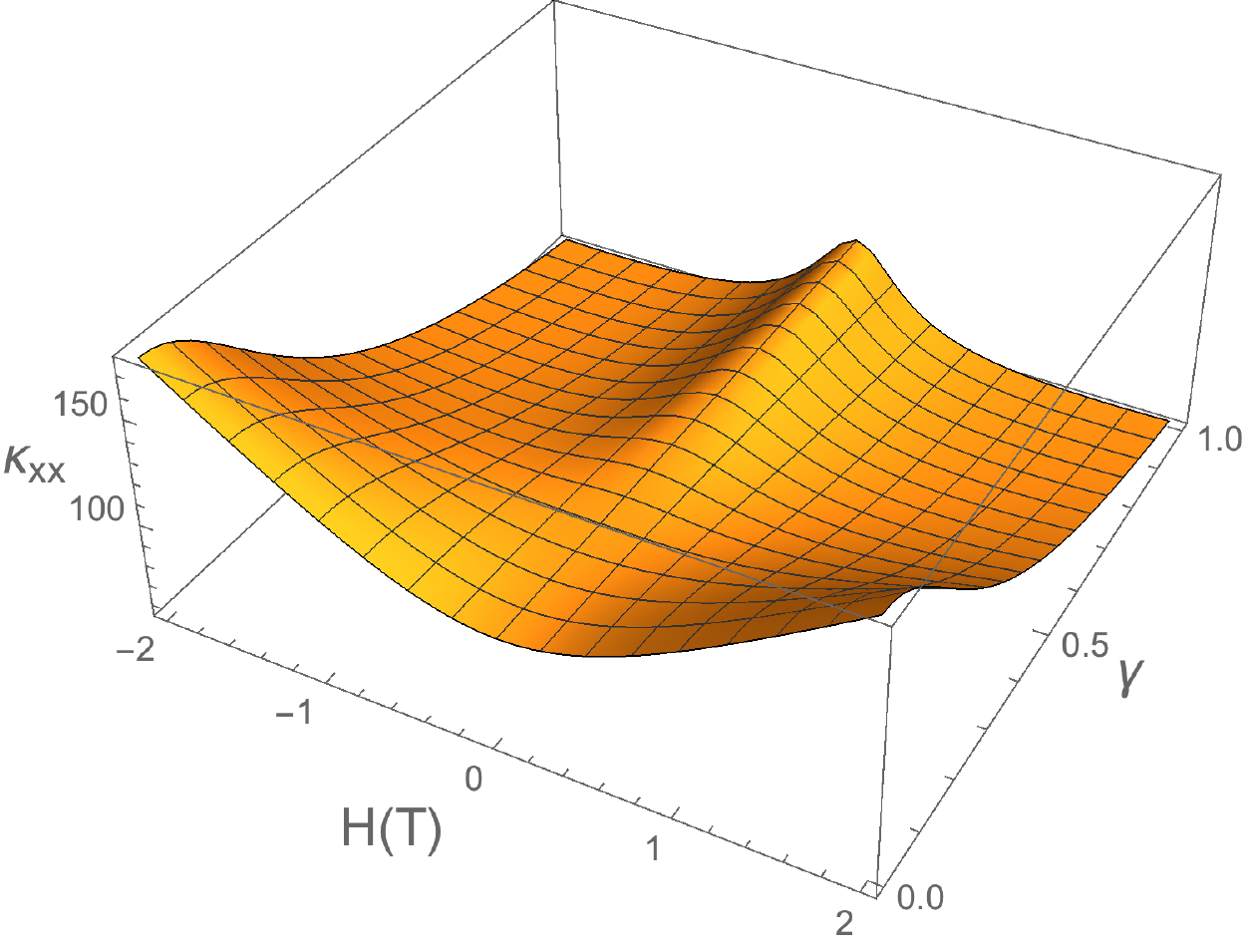} \label{}}
   \hspace{1cm}
       \subfigure[]
   {\includegraphics[width=5.5cm]{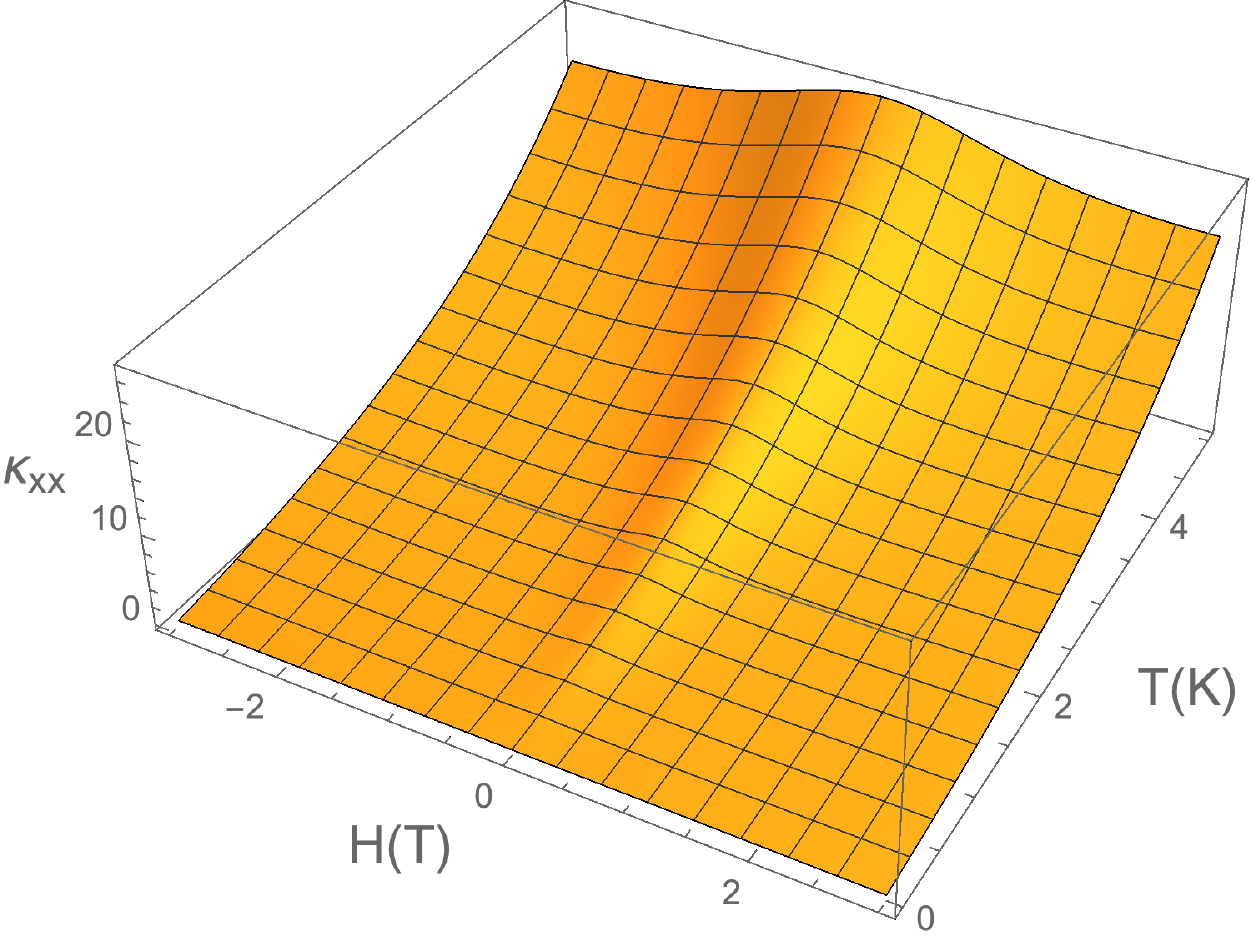} \label{ }}
   
 \caption{   Longitudinal thermal conductivity   $\kappa_{xx}$   (a) as a function of  Magnetic doping and magnetic field  at $T=2K$,  (b) as a function of Temperature and magnetic field  at $q_{\chi} \gamma=0.7$.             } \label{fig:KxxH}
\end{figure}

The impurity density dependence of the longitudinal thermal conductivity $\kappa_{xx}$ also has non trivial behavior. For low temperature limit, $\kappa_{xx}$ can be expanded as
\begin{align}\label{Kxx2}
\kappa_{xx} \Big|_{T \ll 1} \sim  -\frac{8 \pi^2 T (8 \tilde{\theta}^4+\tilde{\theta}^2 -9)}{9 \beta^4 (\tilde{\theta}^2 +1)} H^2 +{\cal O}(H^4),
\end{align}
where $\tilde{\theta} = 6  q_{\chi} \gamma$. Notice that  the numerator of (\ref{Kxx2}) changes sign as $\gamma$(or $\tilde{\theta}$) increases and it becomes zero at critical value $\tilde{\theta} =1$.  The impurity density dependence of $\kappa_{xx}$ is presented at Figure \ref{fig:KxxH} (a). For given impurity density, $\kappa_{xx}$ has power behavior in temperature as
\begin{align}
\kappa_{xx} \Big|_{T \ll 1} &\sim \frac{8 \pi^2 T}{3} +\cdots \cr
\kappa_{xx} \Big|_{T \gg 1} &\sim \frac{64 \pi^4 T^3}{9 \beta^2} + \cdots \cr
\end{align}
at zero magnetic field limit.

\begin{figure}[ht!]
\centering
    \subfigure[ ]
   {\includegraphics[width=5.5cm]{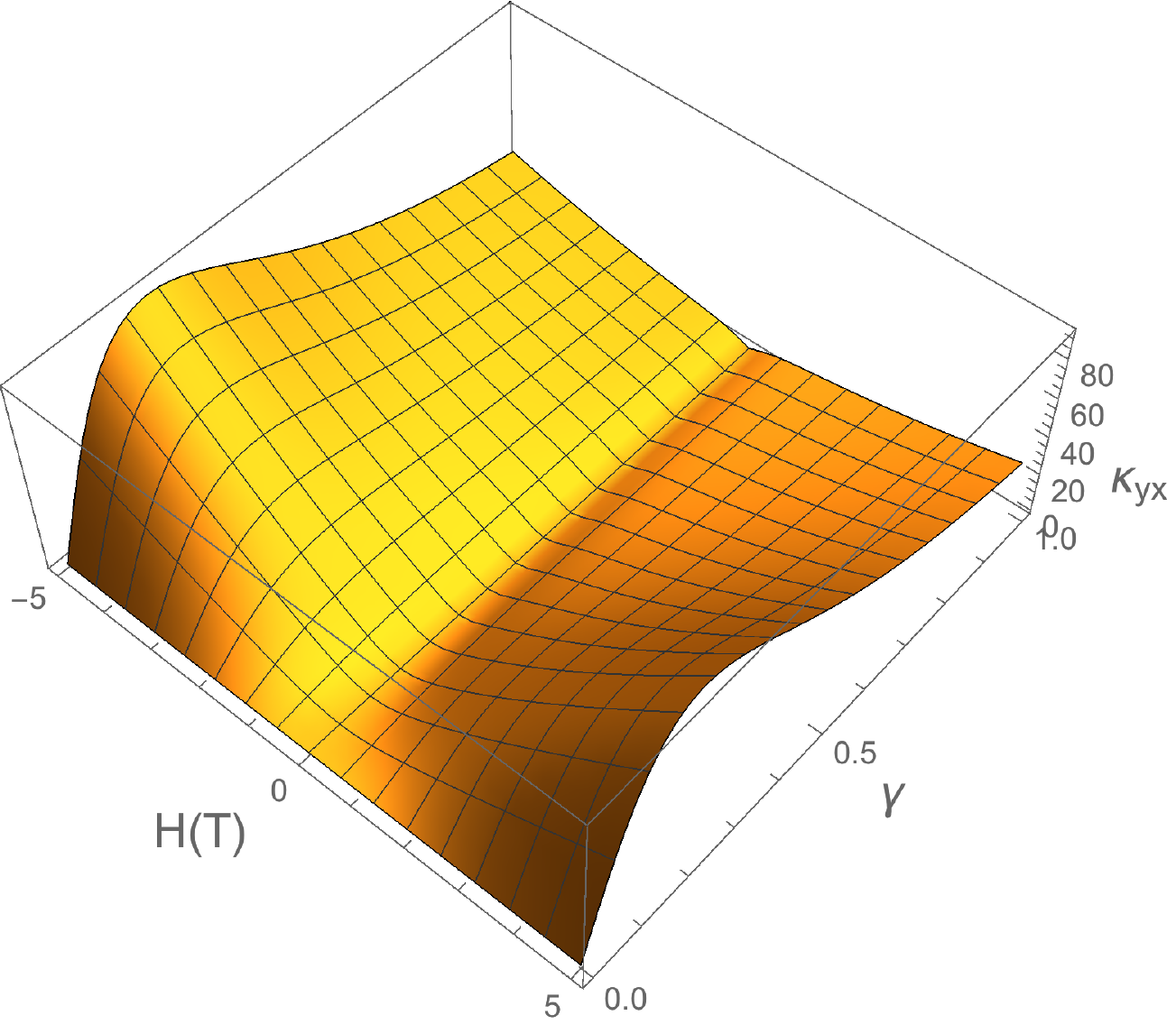} \label{}}
   \hspace{1cm}
       \subfigure[]
   {\includegraphics[width=5.5cm]{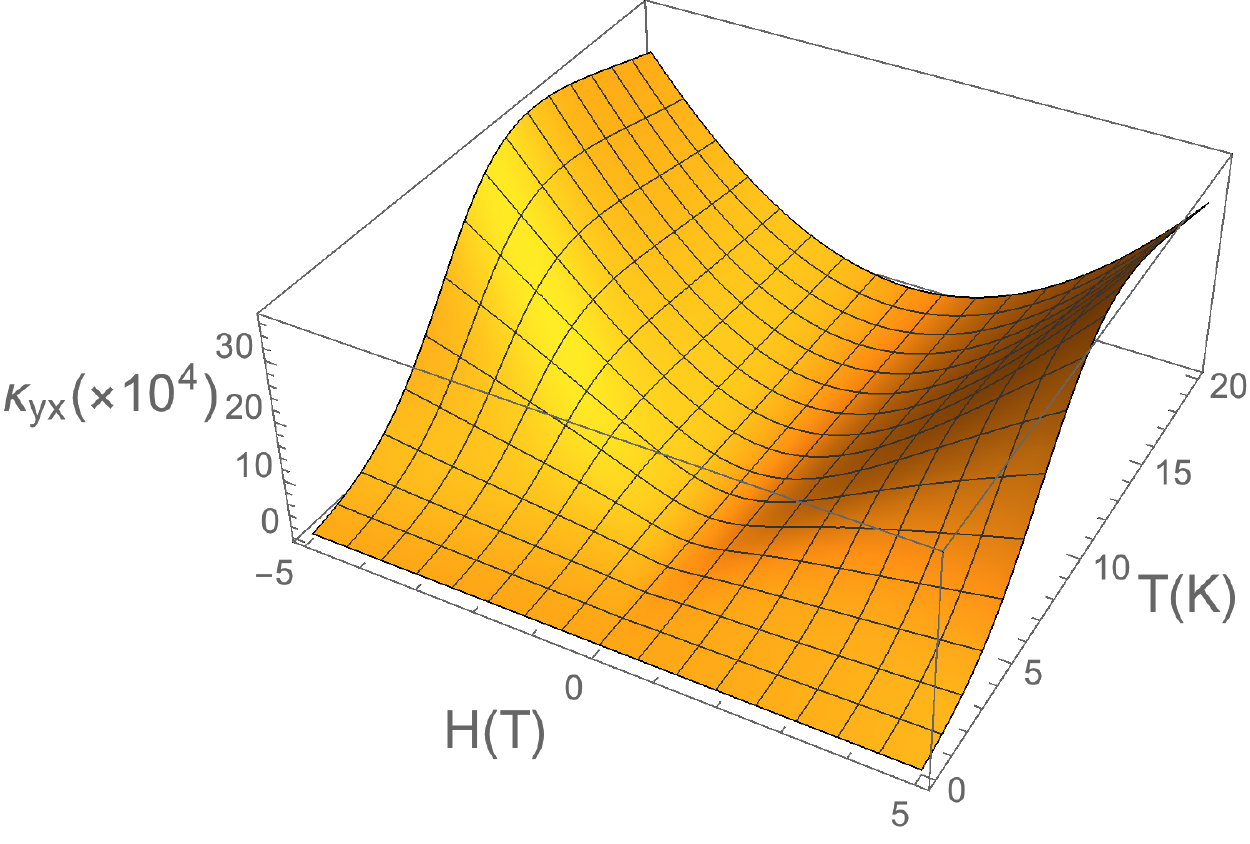} \label{}}
 \caption{Transverse thermal conductivity   $\kappa_{xy}$   (a) as a function of  Magnetic doping and magnetic field  at $T=2K$,  (b) as a function of Temperature and magnetic field  at $q_{\chi}$. 
           } \label{fig:KxyH}
\end{figure}

\begin{figure}[ht!]
\centering
    \subfigure[ ]
   {\includegraphics[width=5.5cm]{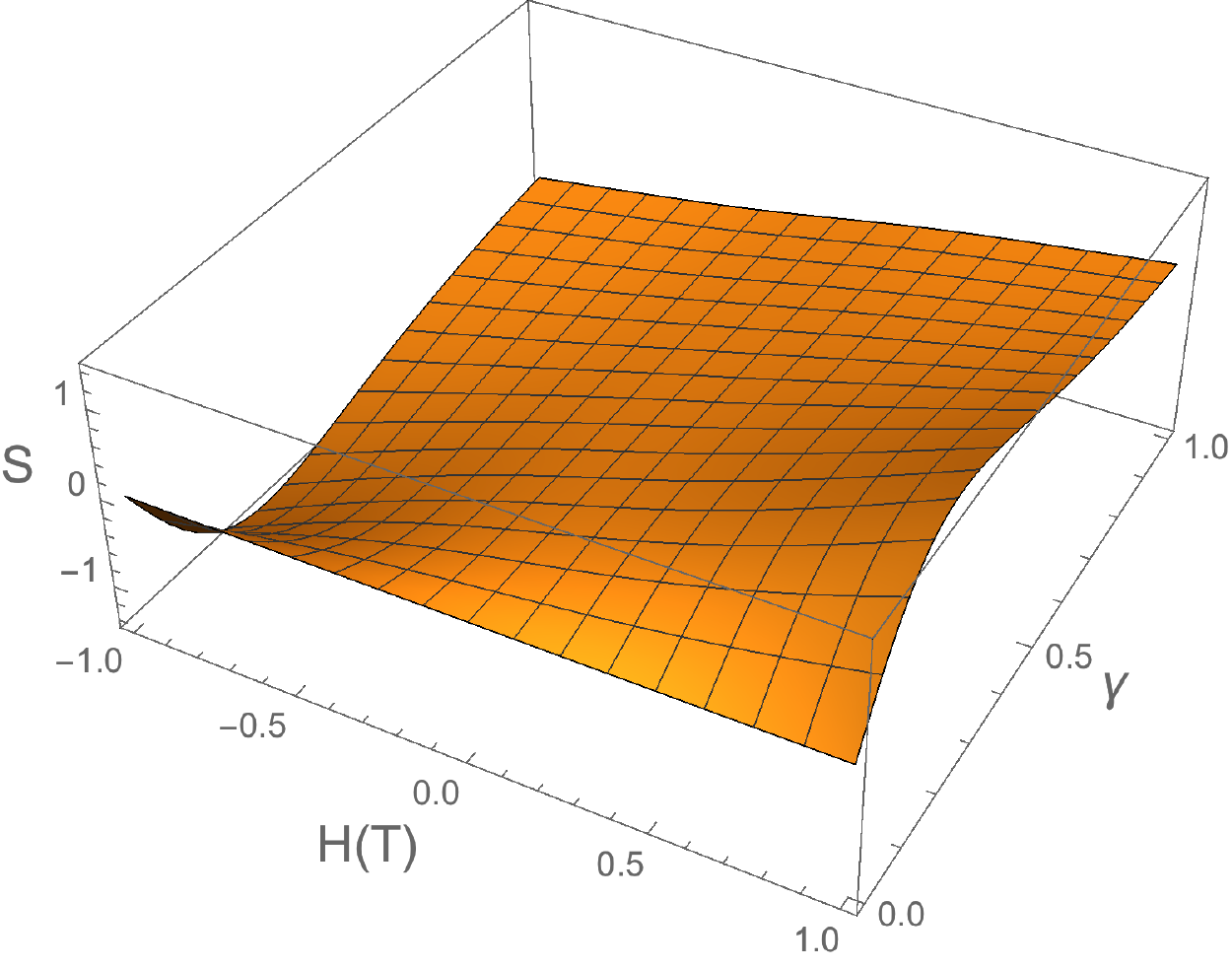} \label{}}
   \hspace{1cm}
       \subfigure[]
   {\includegraphics[width=5.5cm]{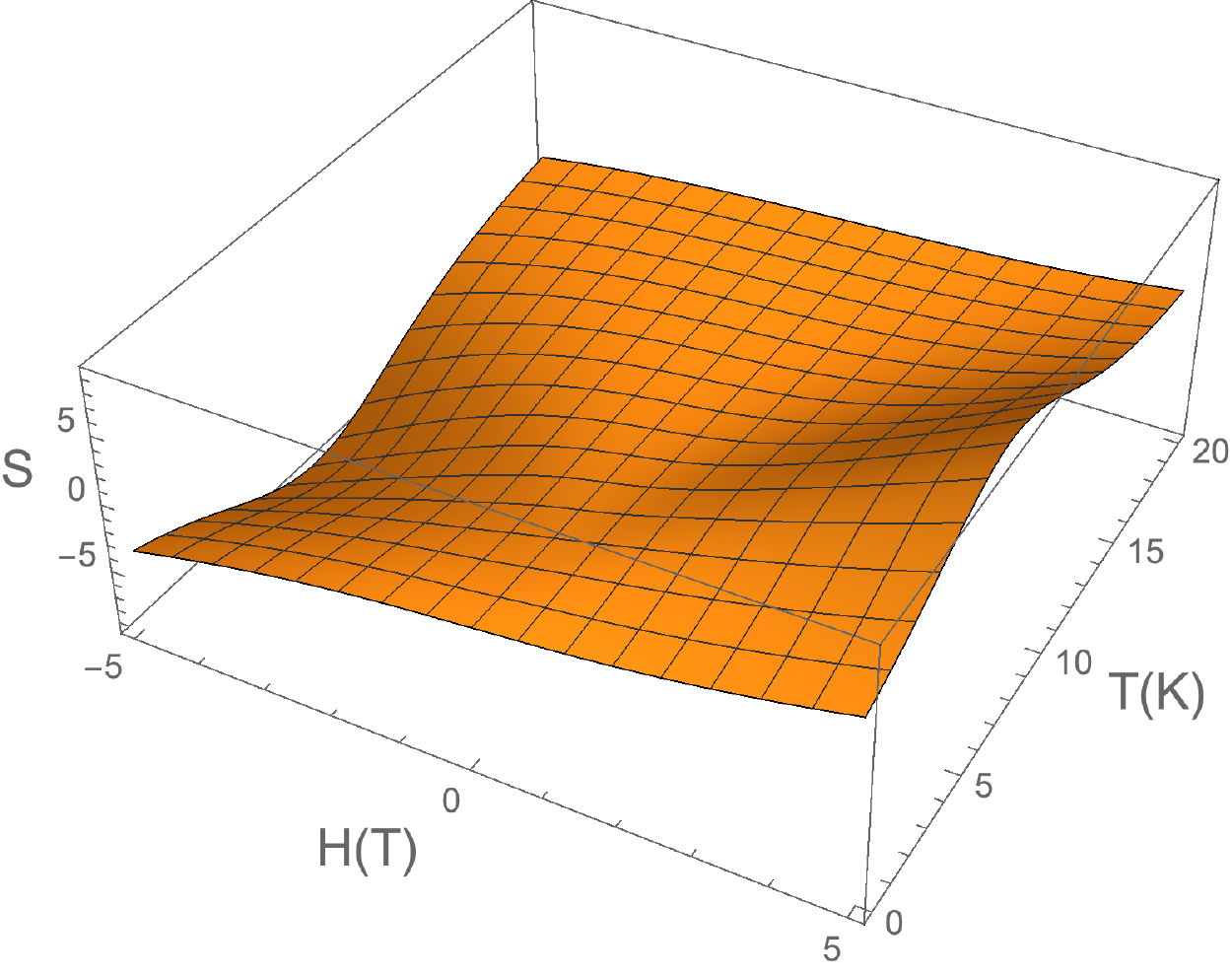} \label{}}

 \caption{
  Seebeck coefficient   $S $   (a) as a function of  Magnetic doping and magnetic field  at $T=2K$,  (b) as a function of Temperature and magnetic field  at $q_{\chi} \gamma=0.7$.          
  } \label{fig:SH}
\end{figure}

\begin{figure}[ht!]
\centering
    \subfigure[ ]
   {\includegraphics[width=5.5cm]{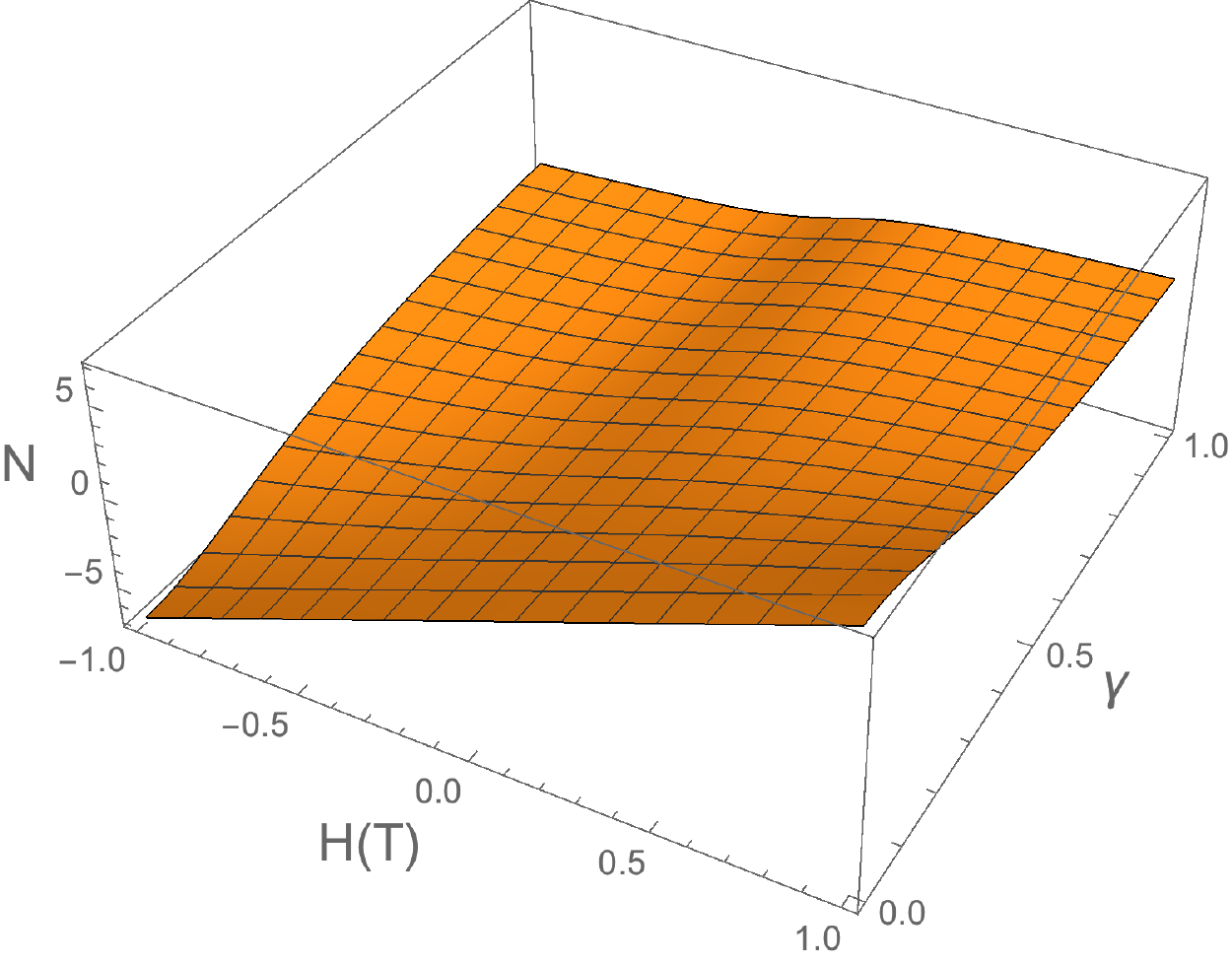} \label{}}
   \hspace{1cm}
       \subfigure[]
   {\includegraphics[width=5.5cm]{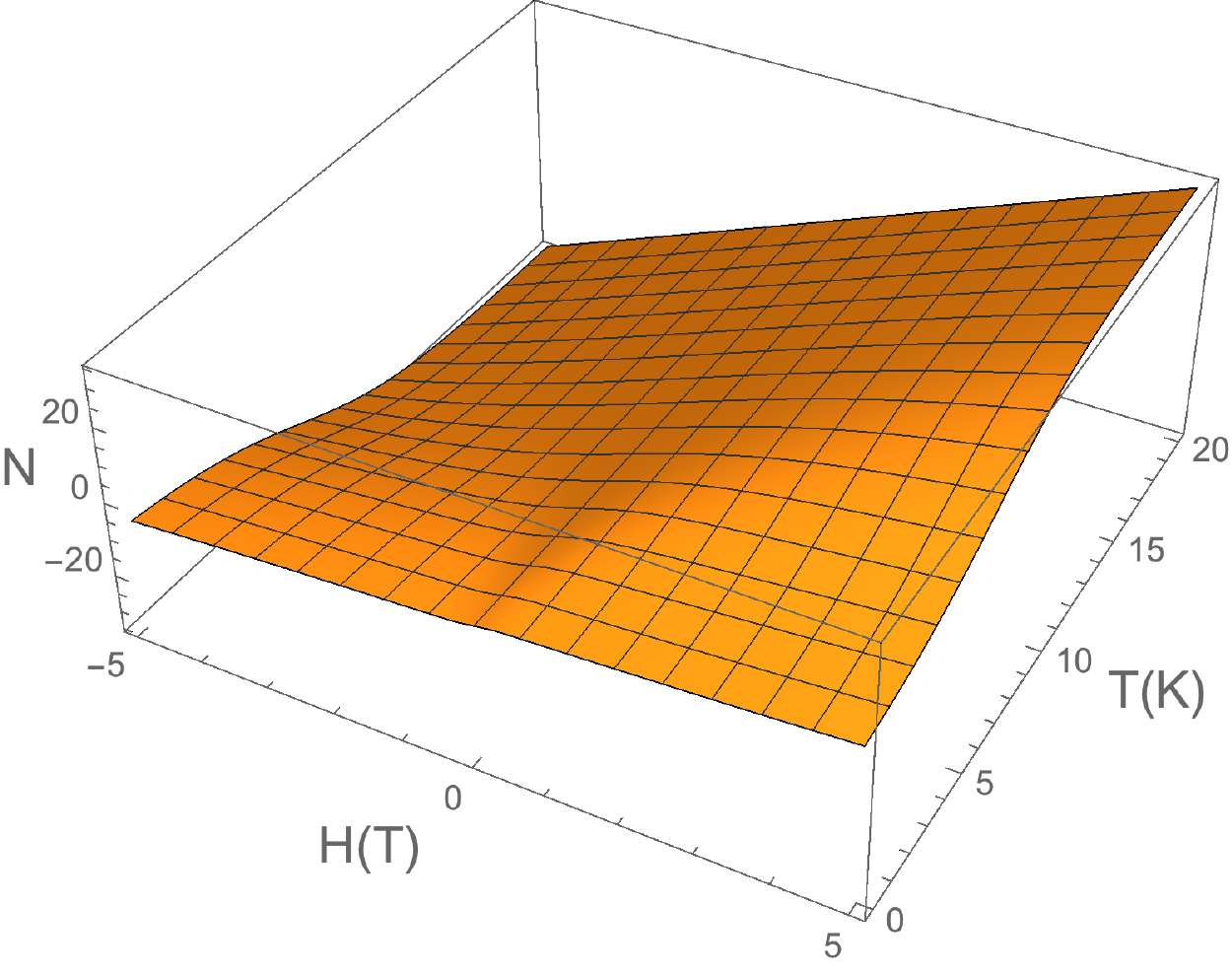} \label{}}
   
 \caption{  Nernst signal  $N $   (a) as a function of  Magnetic doping and magnetic field  at $T=2K$,  (b) as a function of Temperature and magnetic field  at $q_{\chi} \gamma=0.7$.          
   } \label{fig:NH}
\end{figure}

The Seebeck coefficient and the Nernst signal  (\ref{TEPN}) is expanded near zero magnetic field as
\begin{align}
S \Big|_{H \ll 1} &\sim \frac{4\pi \theta}{ 3\beta^2 (1+\theta^2)}\cdot H + \cdots \cr
N \Big|_{H \ll 1} &\sim \frac{4\pi (3+2 \theta^2)}{ 3\beta^2 (1+\theta^2)}\cdot H + \cdots.
\end{align}
The exact calculations for Seebeck coefficient are drawn in Figure \ref{fig:SH} and Figure \ref{fig:NH}. 

\subsubsection {Zero field limit and Anomalous Hall Transports}
 Taking the   $H \rightarrow 0$ limit, the transport coefficients (\ref{DC01}) become
\begin{align}\label{H0trans}
\sigma = \left( \begin{array}{cc} 1+\frac{q^2}{r_0^2 \beta^2} & \theta \\ -\theta & 1+\frac{q^2}{r_0^2 \beta^2} \end{array}   \right) , \quad 
\alpha = \left(\begin{array} {cc} \frac{4\pi q}{\beta^2} & 0 \\ 0 & \frac{4\pi q}{\beta^2}      \end{array}   \right) , \quad 
\bar{\kappa} = \left(\begin{array} {cc} \frac{4\pi s T}{\beta^2} & 0 \\ 0 & \frac{4 \pi s T}{\beta^2}   \end{array} \right).
\end{align}
Notice that in the absence of the external field,  only $\theta$  appears in  $\sigma_{xy}$. The other transport coefficients are the same as those of the RN-AdS black hole with momentum relaxation.  Notice that 
there are no   off-diagonal elements  of $\alpha$ and $\bar{\kappa}$. 
 In this zero magnetic field  limit, the black hole horizon radius $r_0$ is independent of $q_{\chi}$ and it behaves as 
\begin{align}\label{asymr0}
r_0 \Big|_{q \rightarrow \infty} \sim \frac{ 1}{3^{1/4} \sqrt{2}} \sqrt{q}+ {\cal O}(1/\sqrt{q}) , \quad\quad r_0 \Big|_{T \rightarrow \infty} \sim \frac{4\pi}{3} T +{\cal O}(1/T).
\end{align}

\begin{figure}[ht!]
\centering
    \subfigure[$\sigma_{xx}$ ]
   {\includegraphics[width=5.5cm]{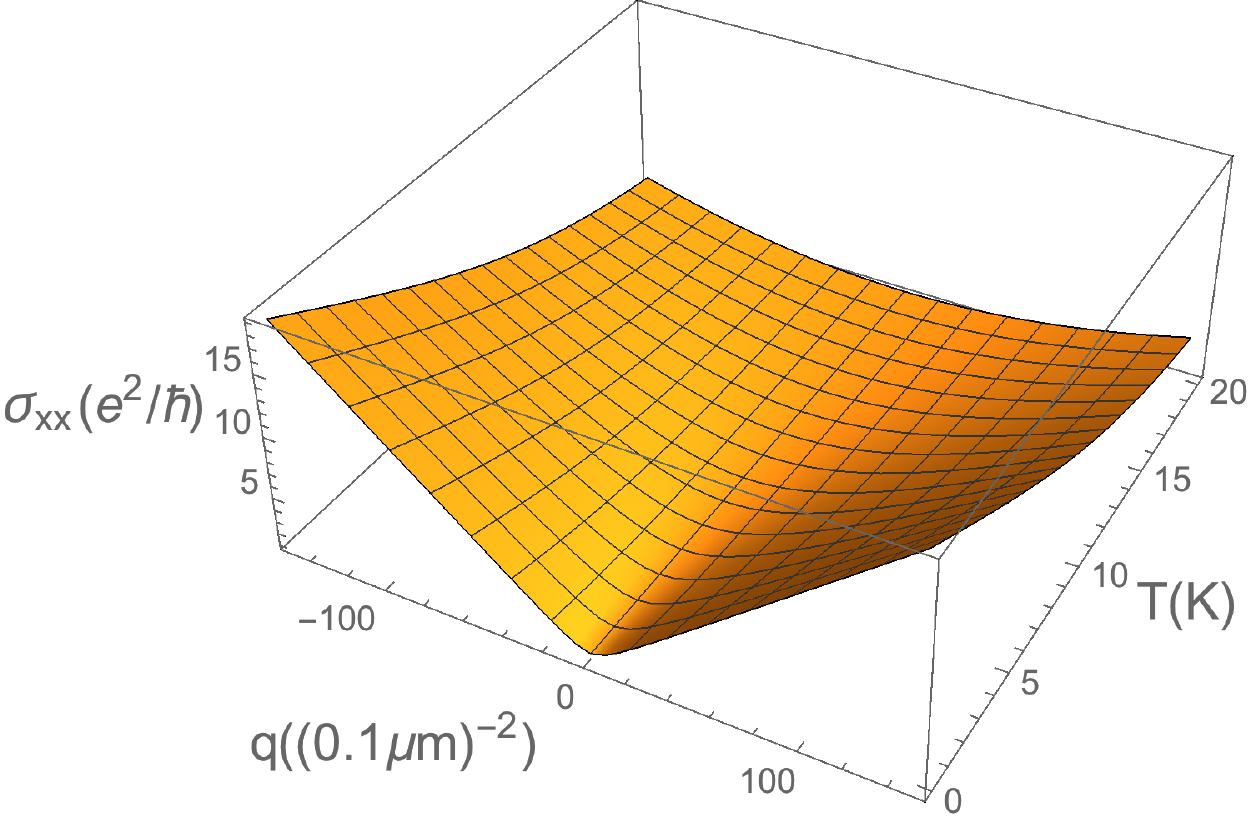} \label{}}
   \hspace{1cm}
       \subfigure[$\sigma_{xy}$]
   {\includegraphics[width=5.5cm]{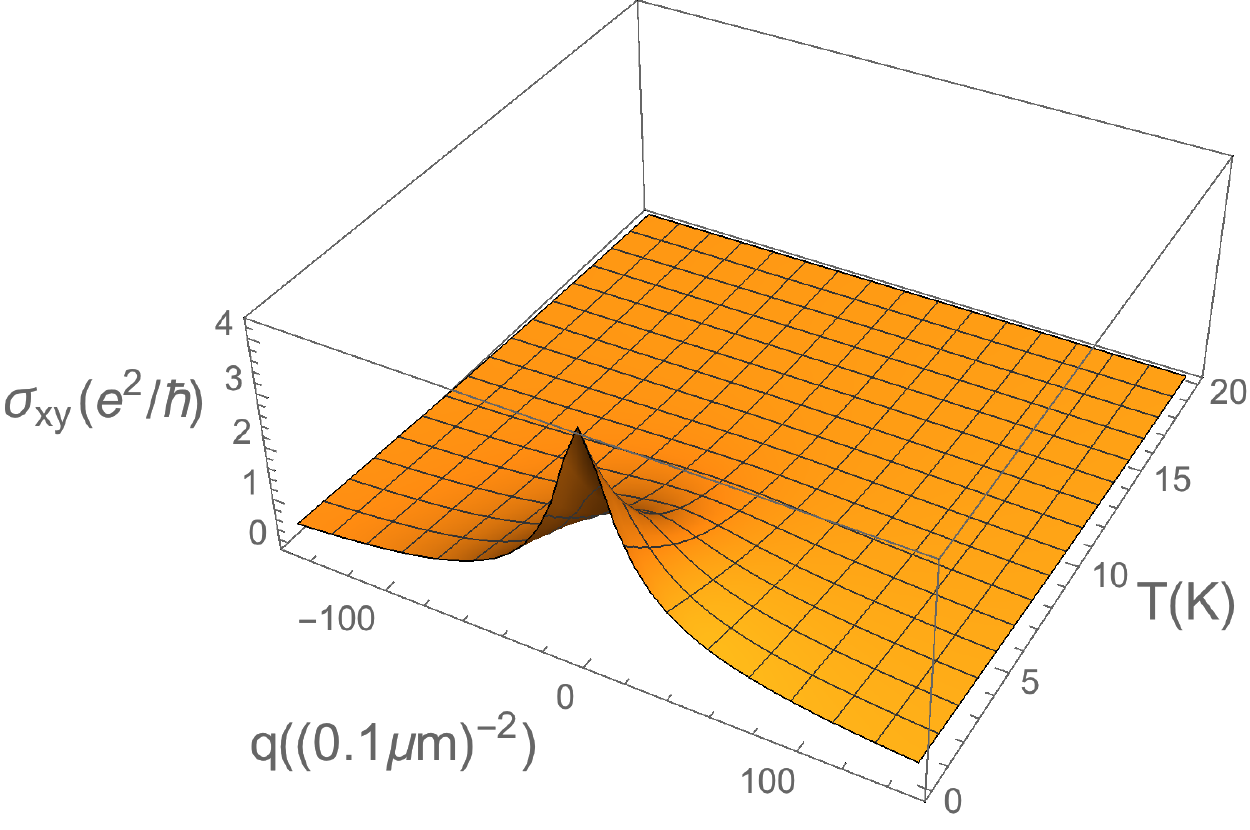} \label{}}
   
  \caption{The temperature and the charge density dependence of (a) the longitudinal conductivity $\sigma_{xx}$ and (b) the transverse conductivity $\sigma_{xy}$.   } \label{fig:QTS}
\end{figure}

The longitudinal and the transverse conductivity are drawn in Figure \ref{fig:QTS}. The longitudinal conductivity $\sigma_{xx}$ does not have $q_{\chi}$ dependence and hence is the same as one in RN-AdS with momentum relaxation. From (\ref{asymr0}), we can expand the longitudinal conductivity in small and large density region:
\begin{align}\label{asymSxx}
\sigma_{xx}\Big|_{q \ll 1} &\sim 1 + \frac{36 q^2}{(4 \pi T +\sqrt{(4\pi T)^2 +6 \beta^2})^2 \, \beta^2} + {\cal O}(q^3) \cr
\sigma_{xx} \Big|_{q \gg 1}   &\sim 2\sqrt{3} \frac{q}{\beta^2} -\frac{4 \sqrt{2} T}{3^{1/4} \beta^2} \sqrt{q} +{\cal O}(1/\sqrt{q}).
\end{align}
In the limit of $\beta \gg T$, (\ref{asymSxx}) becomes   simpler form
\begin{align}\label{longiS01}
 \sigma_{xx} \Big|_{q \ll 1} \sim 1 + \frac{6 q^2}{\beta^4}  , \quad \quad
 \sigma_{xx} \Big|_{q \gg 1} \sim 2 \sqrt{3} \frac{q}{\beta^2}.
\end{align}
In large carrier density limit, the longitudinal conductivity is linearly increasing in $q$, which is observed in the  graphene. By solving Boltzman equation,  it was shown to be 
\cite{Novoselov:2005aa, nomura2007quantum, Peres:2007aa}
\begin{align}
\sigma_{xx} = \frac{e^2}{\hbar} \, \frac{(\hbar v_{F})^2}{u_0^2} n,
\end{align}
where $n$ is the charge carrier density which is the same as $q$ in this paper and $u_0^2$ is proportional to the impurity density. If we identify $ \frac{(\hbar v_{F})^2}{u_0^2} \equiv \frac{2\sqrt{3}}{\beta^2}$, then the second line of (\ref{longiS01}) looks like  the result of Fermi liquid theory.
On the other hand, the longitudinal conductivity has quadratic behavior near charge neutrality point. If we introduce dimensionless quantity  $\tilde{Q} \equiv  2 \sqrt{3}\, q/\beta^2$, then
\begin{align}
\sigma_{xx} \Big|_{q\ll 1} \sim 1+ \frac{1}{2} \tilde{Q}^2,\hspace{1cm} \sigma_{xx} \Big|_{q \gg 1} \sim \tilde{Q}.
\end{align}

The transverse conductivity $\sigma_{xy}$ is shown in Figure \ref{fig:QTS}(b).   $\sigma_{xy}$ has maximum at zero temperature and charge neutrality point and the height is $6 q_{\chi} \gamma$. Large density and temperature behavior can be obtained from (\ref{asymr0})
\begin{align}
\sigma_{xy} \Big|_{q \gg 1}  \sim \frac{q_{\chi} \beta^2 \gamma}{q} ,\quad \quad
\sigma_{xy} \Big|_{T \gg 1}  \sim \frac{q_{\chi} \beta^2 \gamma}{T^2}.
\end{align}

\begin{figure}[ht!]
\centering
    \subfigure[$\rho_{xx}$, $q_{\chi}\gamma= 0$  ]
   {\includegraphics[width=5.5cm]{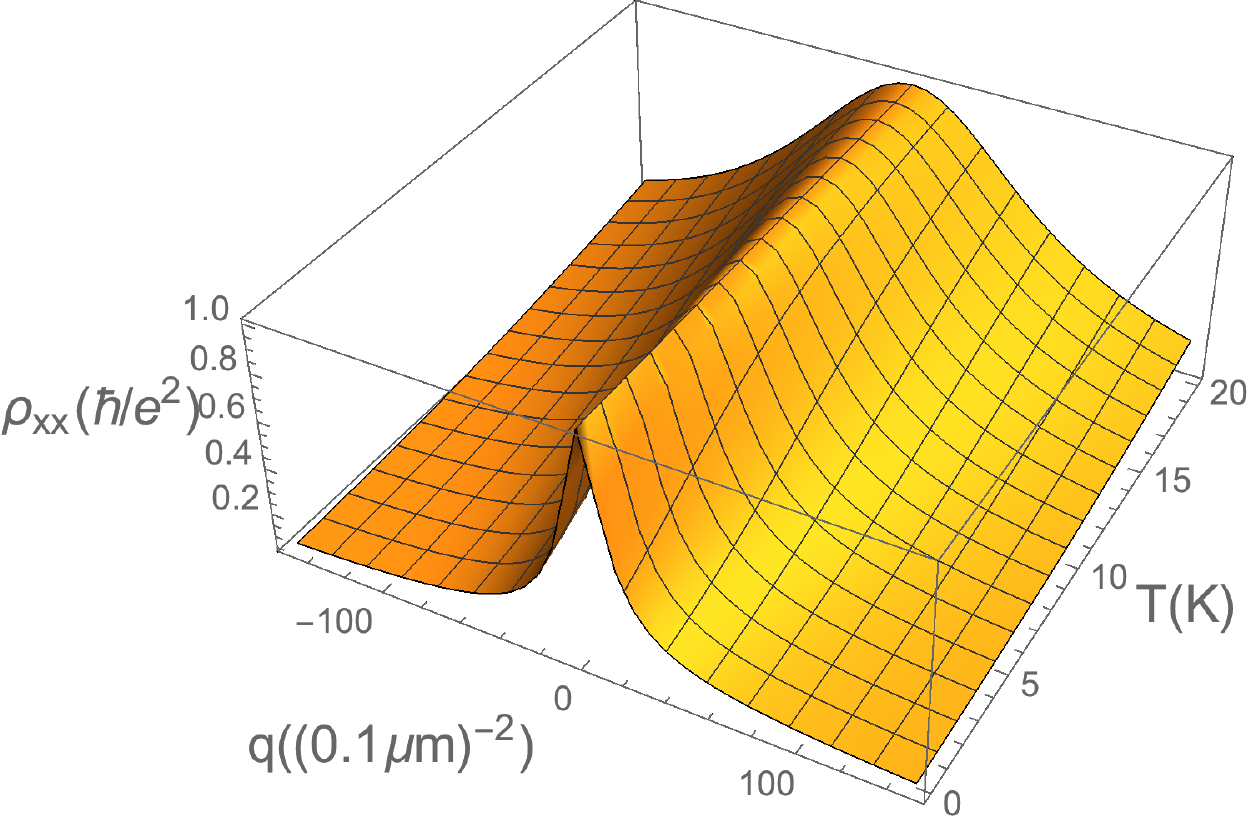} \label{}}
   \hspace{1cm}
       \subfigure[$\rho_{xx}$, $q_{\chi} \gamma = 0.7$]
   {\includegraphics[width=5.5cm]{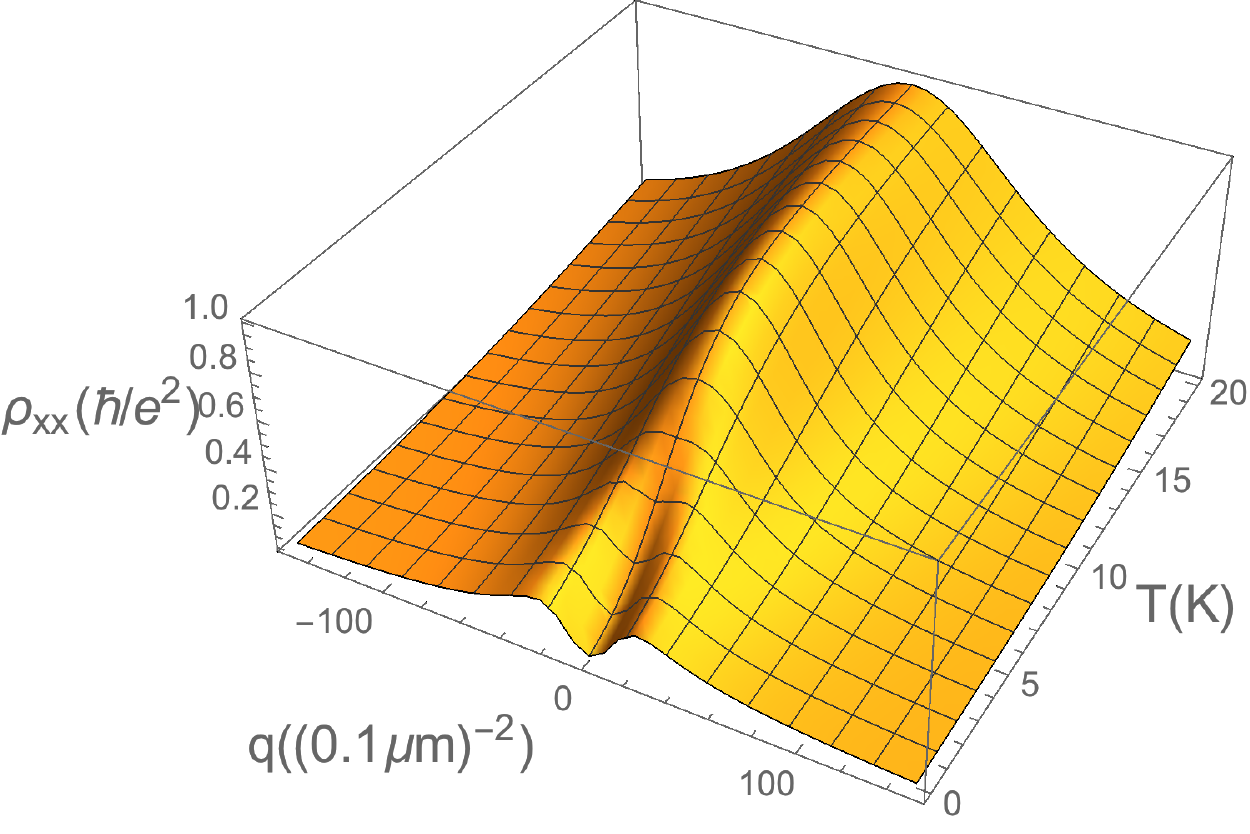} \label{}}
   
    \subfigure[$\rho_{xx}$, $q =0$ ]
   {\includegraphics[width=5.5cm]{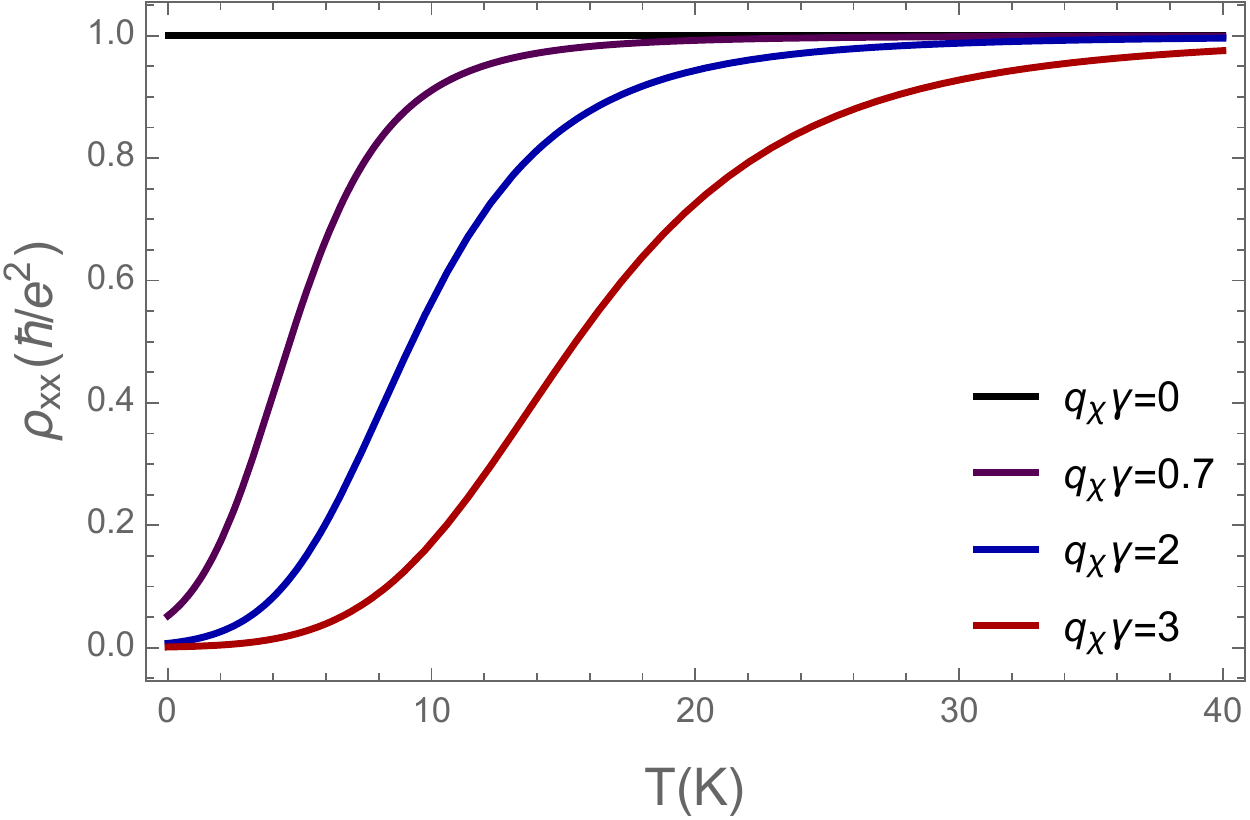} \label{}}
   \hspace{1cm}
       \subfigure[$\rho_{xx}$, $T = 2 K$]
   {\includegraphics[width=5.5cm]{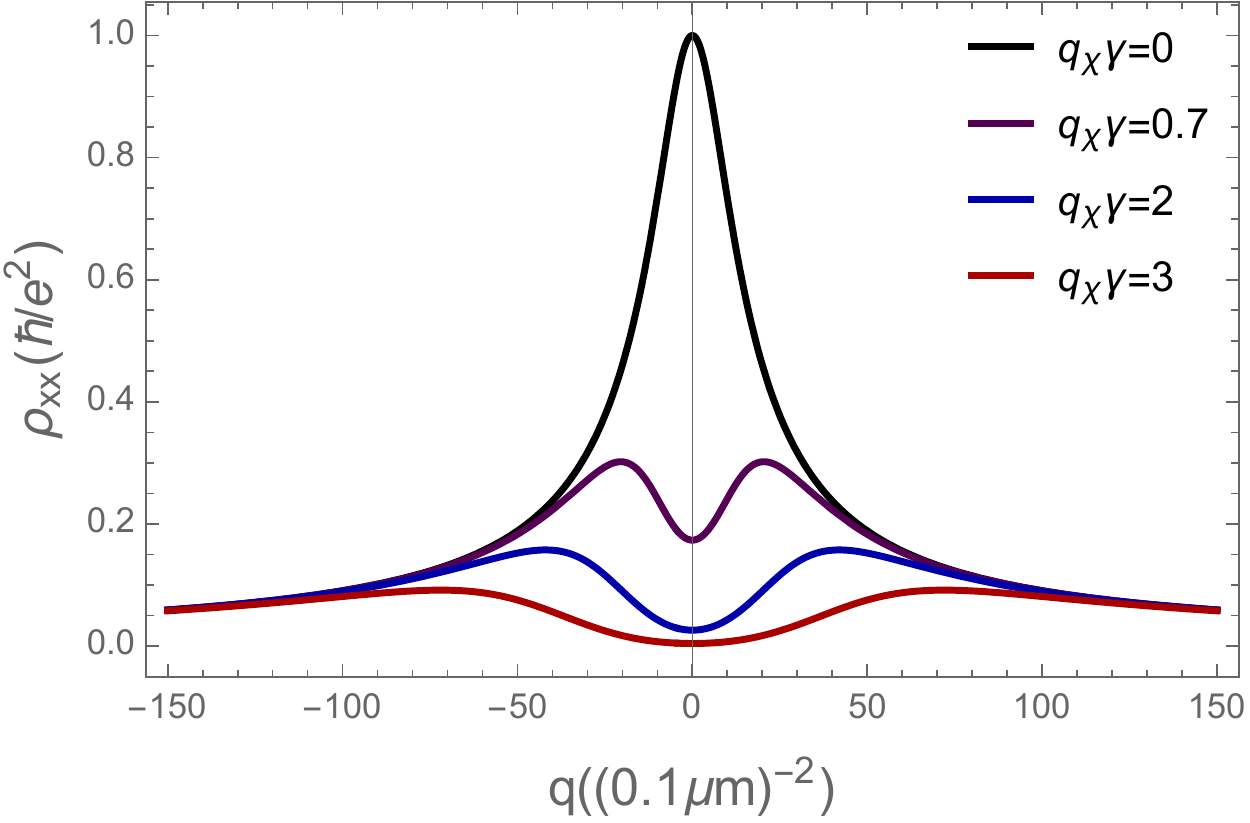} \label{}}

  \caption{The charge and the temperature dependence of the longitudinal resistivity without (a) and with (b) magnetic impurity. (c) The temperature dependence of $\rho_{xx}$ at the charge neutral point and (d) The charge density dependence of $\rho_{xx}$ at low temperature ($T=2 K$) with different value of $q_{\chi} \gamma$.      } \label{fig:RxxH0}
\end{figure}

The resistivity can be obtained by inverting the electric conductivity martix;
\begin{align}\label{RH0}
\rho_{xx}  = \frac{(1+\frac{q^2}{r_0^2 \beta^2})}{{\left(1+\frac{q^2}{r_0^2 \beta^2 }\right)^2 + \theta^2}} ,\quad \quad
\rho_{xy} = -\frac{\theta}{{\left(1+\frac{q^2}{r_0^2 \beta^2 }\right)^2 + \theta^2}} .
\end{align}
Notice $\rho_{xx}$ contains different information of $\sigma_{xx}$ due to the presence of $\theta$ even in the absence of the external magnetic field.
The transverse resistivity is proportional to $\theta$ which is related to the value of the magnetization. This phenomena is called anomalous Hall effect which comes from the intrinsic magnetic property of material.  The scaling property of the anomalous Hall effect was discussed in \cite{Kim:2015wba}. 
In this paper, we focus on the charge and the temperature dependence of the resistivity. The effect of the magnetic impurity on the longitudinal resistivity is drawn in Figure \ref{fig:RxxH0}.

The Figure \ref{fig:RxxH0} (a) and (b) show the density and the temperature dependence of the longitudinal resistivity. In the absence of the magnetic impurity, the resistivity has maximum at the charge neutrality point in all temperature range. It is natural  that due to the absence of charge carrier density  at the charge neutrality point, the resistivity has maximum. On the other hand, in the presence of the magnetic impurity,  the resistivity at the charge neutral point is suppressed at low temperature. As doping parameter $\gamma$ increases, the suppressed region becomes larger.  Figure \ref{fig:RxxH0} (c) shows the temperature dependence of the longitudinal resistivity at the charge neutrality point for different value of $\gamma$. The density dependence of the longitudinal resistivity at low temperature is more interesting, see Figure \ref{fig:RxxH0} (d). In this figure, the maximum of the resistivity is not located at the charge neutrality point as $\gamma$ increases. It can be understood as an effect of the magnetization which is proportional to $\theta$. The denominator of the longitudinal resistivity (\ref{RH0}) is maximized when $\theta$ has maximum value and it happens at the charge neutrality point as shown in Figure \ref{fig:QTS}(b). And the competition between the longitudinal and the transverse conductivity shift the maximum of the resistivity away from the charge neutrality point.

\begin{figure}[ht!]
\centering
     
    \subfigure[$\rho_{yx}$, $q_{\chi}\gamma =0$ ]
   {\includegraphics[width=5.5cm]{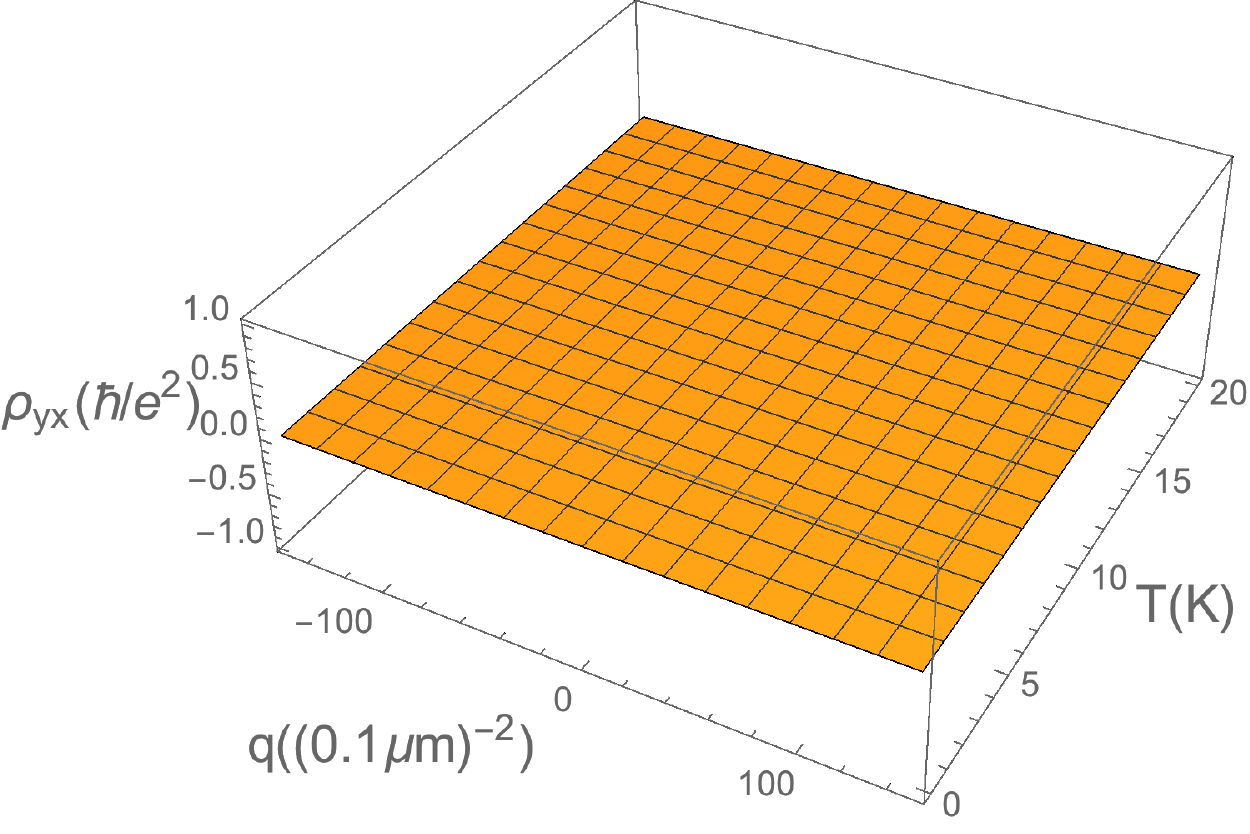} \label{}}
   \hspace{1cm}
       \subfigure[$\rho_{yx}$, $q_{\chi} \gamma= 0.7$]
   {\includegraphics[width=5.5cm]{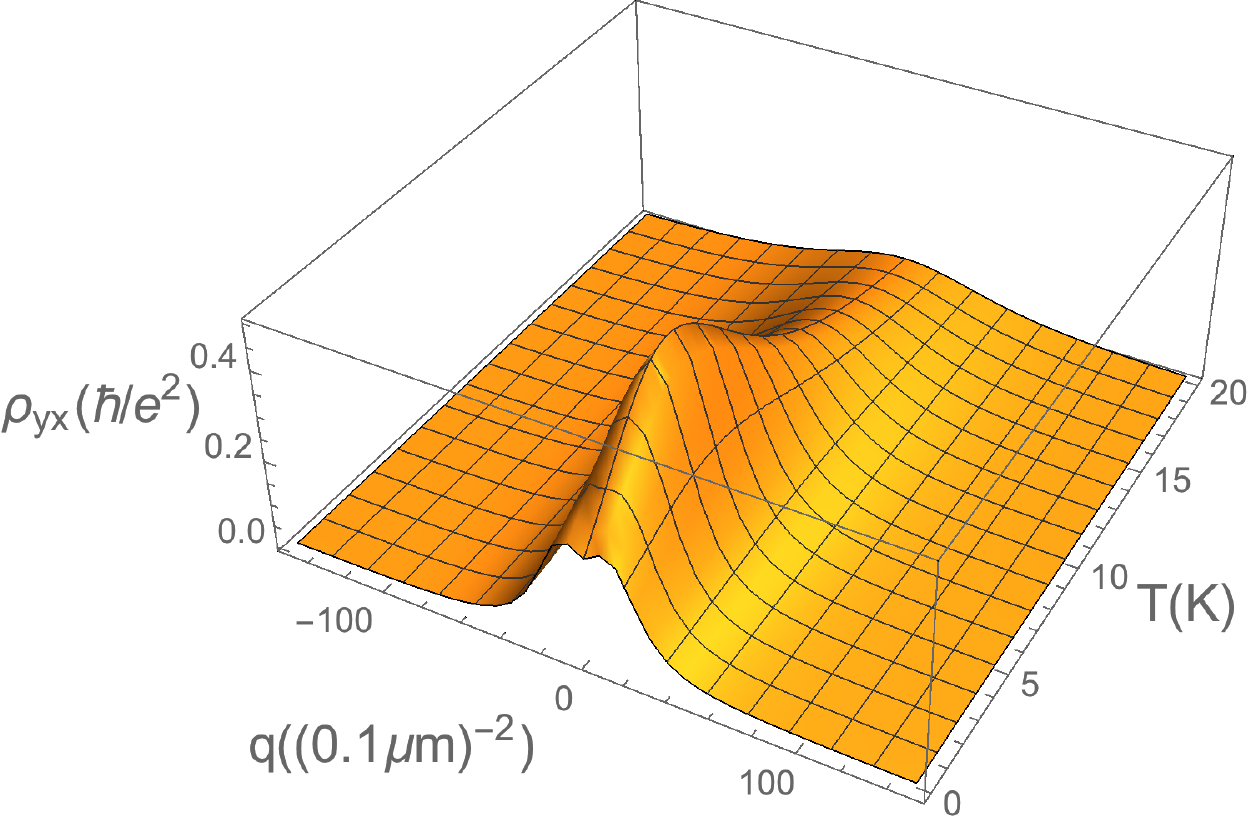} \label{}}
   
     \subfigure[$\rho_{yx}$, $q =0$ ]
   {\includegraphics[width=5.5cm]{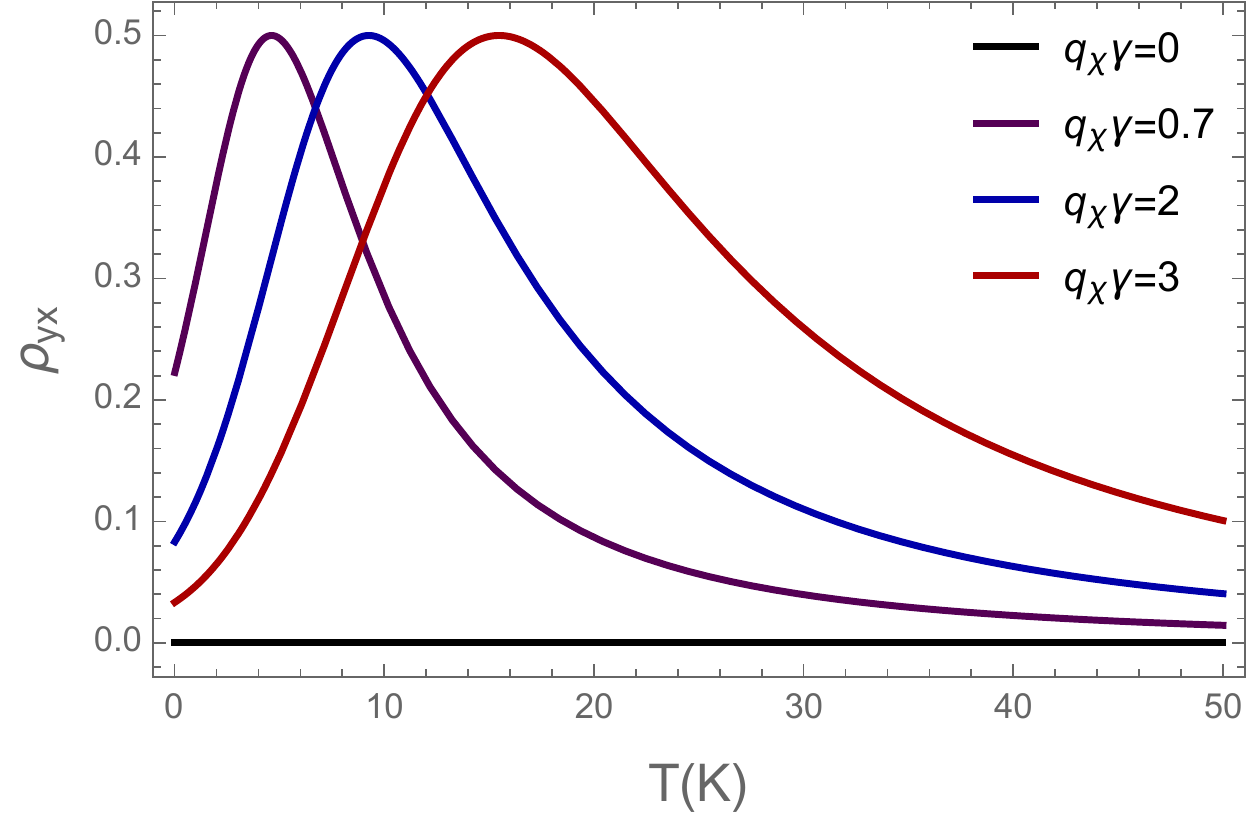} \label{}}
   \hspace{1cm}
       \subfigure[$\rho_{yx}$, $T = 2 K$]
   {\includegraphics[width=5.5cm]{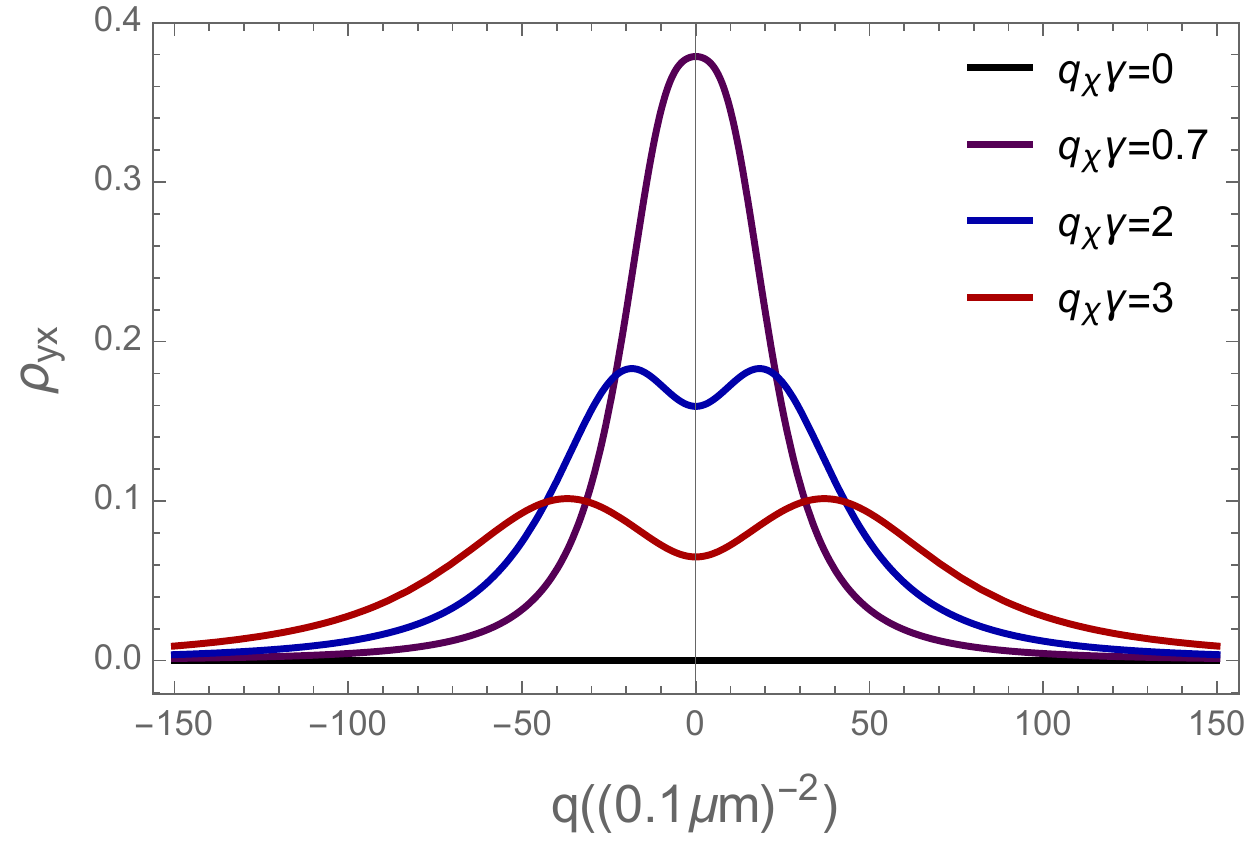} \label{}}
   
  \caption{The charge and the temperature dependence of the transverse resistivity without (a) and with (b) magnetic impurity. (c) The temperature dependence of $\rho_{yx}$ at the charge neutrality point. (d) The charge density dependence of $\rho_{yx}$ at low temperature ($T = 2 K$)  with different value of $q_{\chi} \gamma$.   } \label{fig:RyxH0}
\end{figure}

The effect of the magnetic impurity  on the transverse resistivity is drawn in Figure \ref{fig:RyxH0}. As shown in (\ref{RH0}), the transverse resistivity is proportional to $\theta$ and hence there is no transverse resistivity in the absence of the magnetic impurity. If we put magnetic impurity, there is maximum of the transverse resistivity at finite temperature and  charge neutrality point, see Figure \ref{fig:RyxH0} (b). As we increase impurity density, the peak of transverse resistivity moves to high temperature region. 

Non-zero value of $\sigma_{xy}$ gives non-trivial $\theta$ dependence to the thermal conductivity.  $\kappa$ is defined by the thermal current without electric current as it was given by Eq.(\ref{kappa}). 
From (\ref{H0trans}), the longitudinal and the transverse thermal conductivity are
\begin{align}\label{KappaH0}
\kappa_{xx} &=
\frac{s^2 T}{r_0^2 \beta^2} \cdot \frac{\sigma_{xx} +\sigma_{xy}^2}{\sigma_{xx}^2 +\sigma_{xy}^2} \cr
\kappa_{xy} &=
\frac{s^2 T q^2}{r_0^4 \beta^4} \cdot \frac{\sigma_{xy}}{\sigma_{xx}^2 +\sigma_{xy}^2}.
\end{align}

\begin{figure}[ht!]
\centering
    \subfigure[$\kappa_{xx}$, $q_{\chi} \gamma=0$ ]
   {\includegraphics[width=5.5cm]{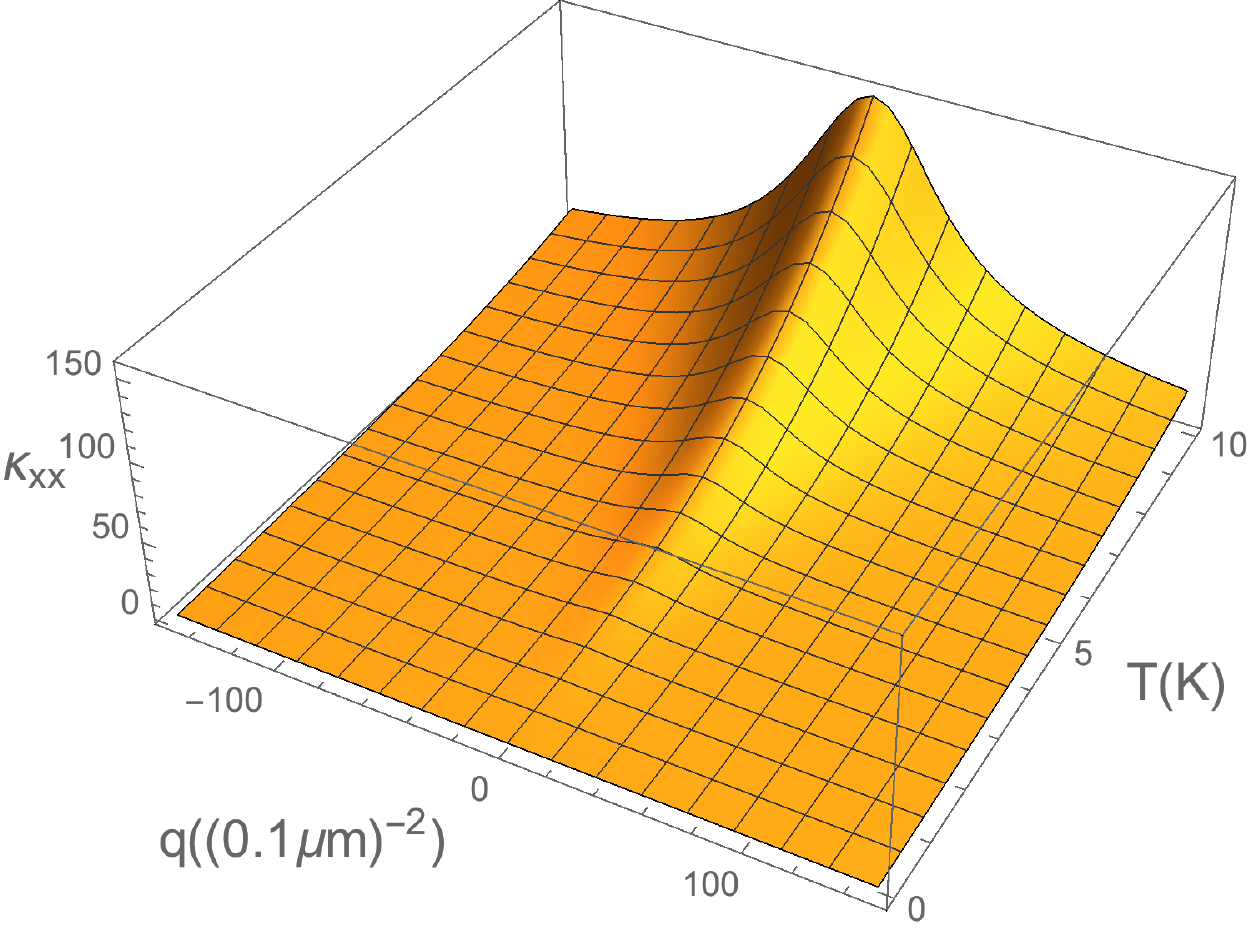} \label{}}
      \hspace{1cm}
    \subfigure[$\kappa_{xx}$, $q_{\chi} \gamma=5$ ]
   {\includegraphics[width=5.5cm]{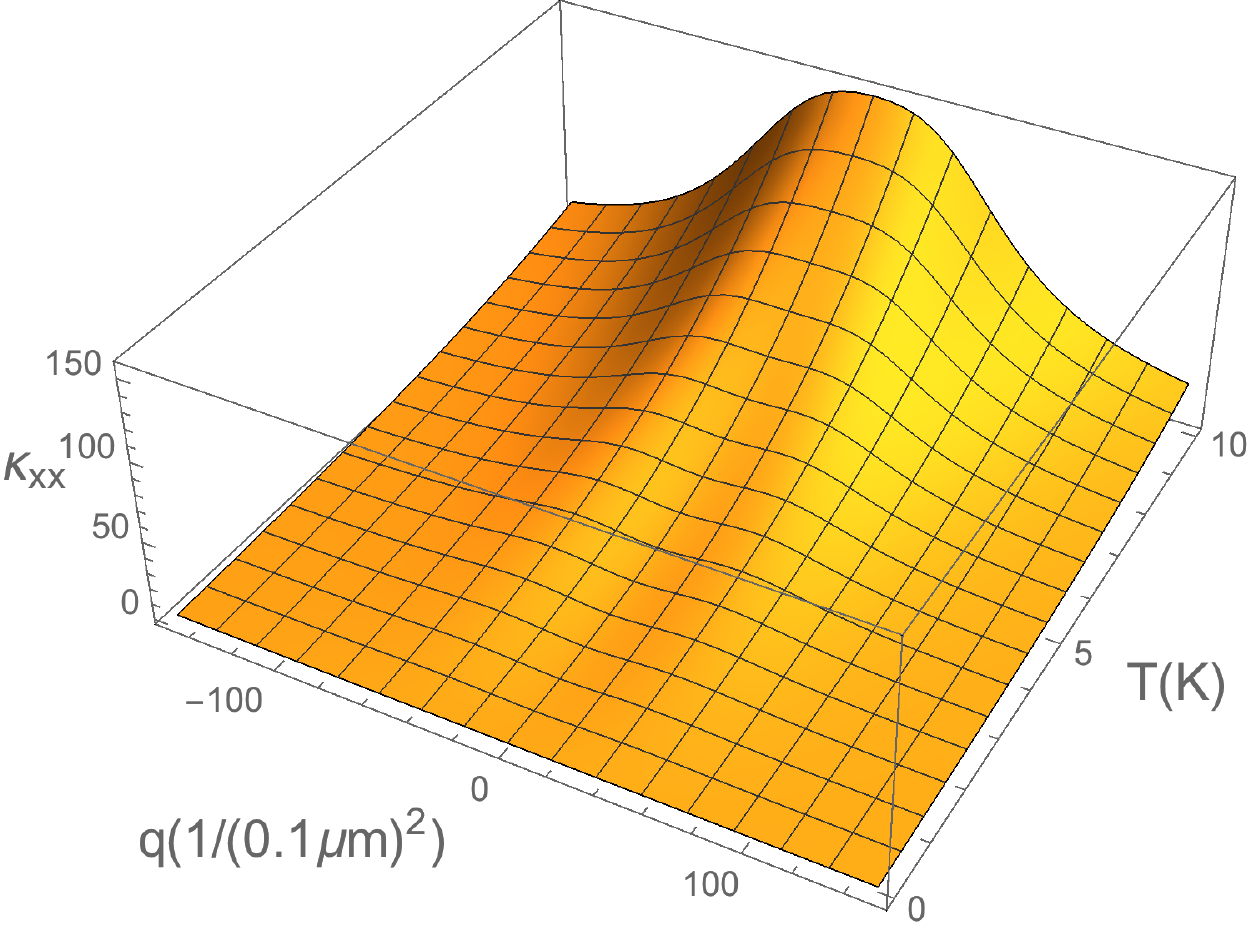} \label{}}

       \subfigure[$\kappa_{xy}$, $q_{\chi} \gamma=0$]
   {\includegraphics[width=5.5cm]{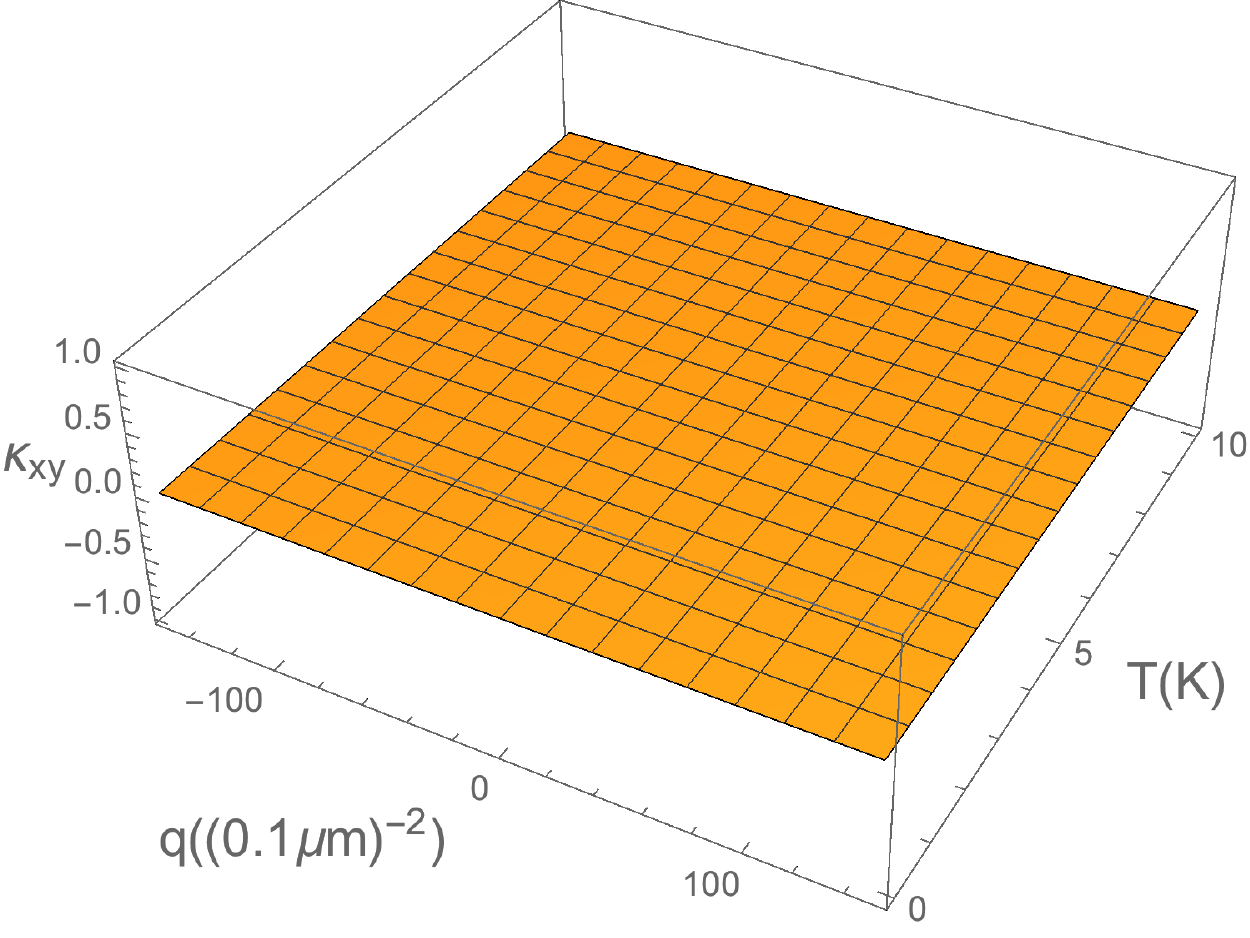} \label{}}
      \hspace{1cm}
     \subfigure[$\kappa_{xy}$, $q_{\chi} \gamma=0.7$]
   {\includegraphics[width=5.5cm]{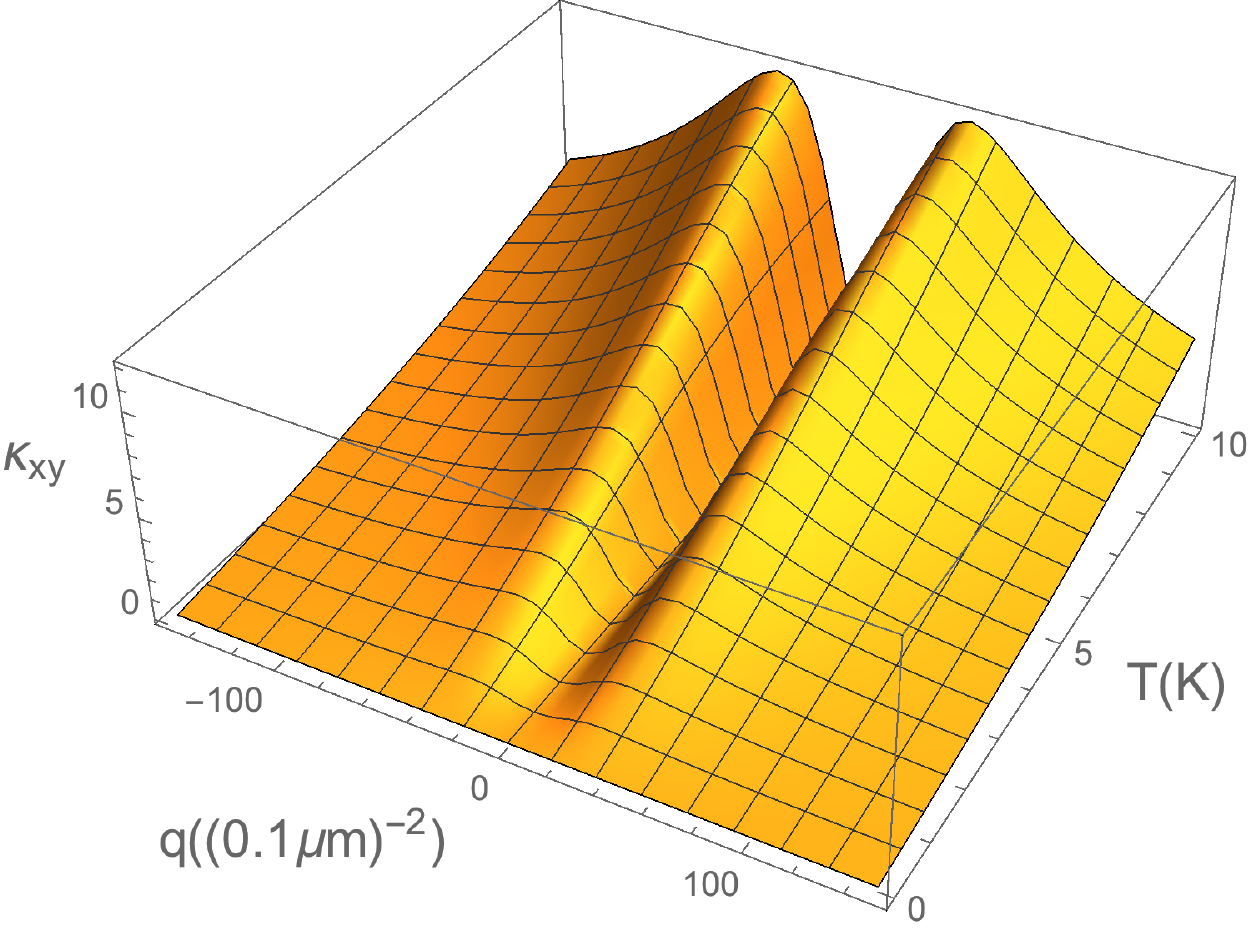} \label{}} 
  \caption{ The longitudinal thermal conductivity $\kappa_{xx}$  as a function of  the charge and the temperature    (a) without  and (b) with magnetic impurity.  (c) and (d) are  Transverse thermal conductivity $\kappa_{xy}$ without and with magnetic impurity respectively.   } \label{fig:QTKH0}
\end{figure} 
These results   are shown in Figure \ref{fig:QTKH0}.
Here,  figure   \ref{fig:QTKH0} (a), (b) show $\kappa_{xx}$  without and with magnetic impurity. In the absence of magnetic impurity, the longitudinal component of the thermal conductivity has maximum at charge neutrality point and grows as $T$ at low temperature and $T^3$ at high temperature. For finite magnetic impurity density, small dip appears at charge neutrality point. Figure  \ref{fig:QTKH0} (c) and (d) show the transverse thermal conductivity which shows $q^2$ behavior near charge neutrality point.

\begin{figure}[t]
\centering
    \subfigure[$\beta=\gamma=1$, $T\sim 0$ ]
   {\includegraphics[width=5.5cm]{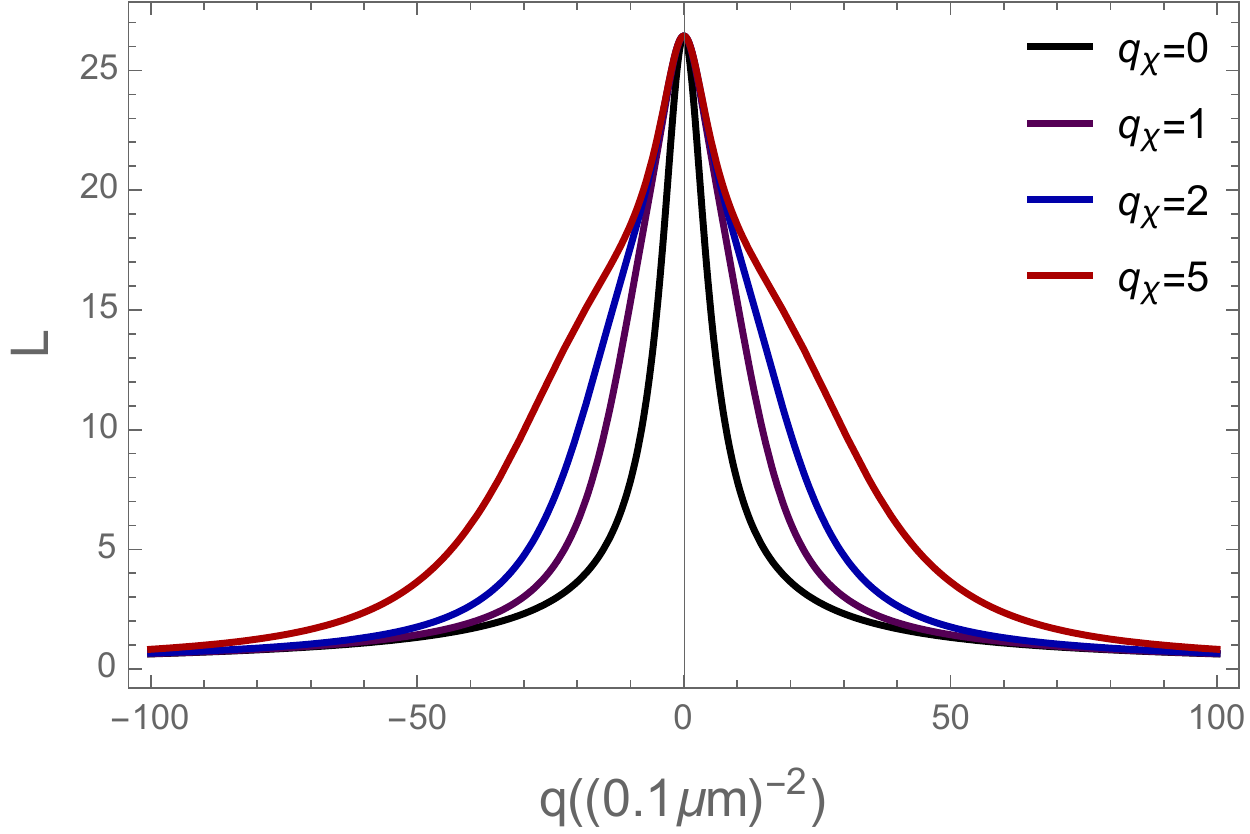} \label{}}
   \hspace{1cm}
       \subfigure[$q_{\chi} \gamma=2$, $T=50K$]
   {\includegraphics[width=5.5cm]{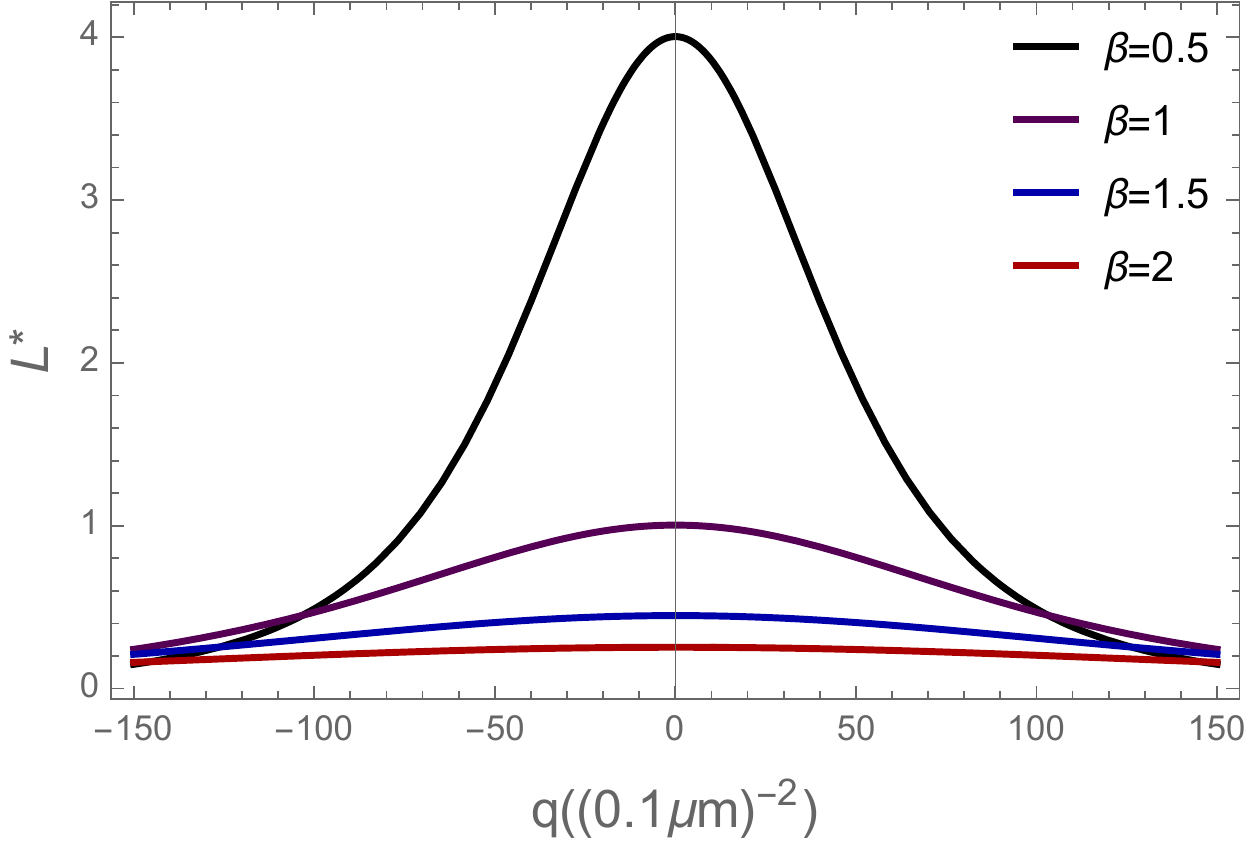} \label{}}
  
\caption{(a) Effect of interaction strength $q_{\chi}$ on Lorentz ratio  near zero temperature. (b) Effect of impurity density on Lorentz ratio, $L^* = L\cdot\left(\frac{9}{(4\pi)^4 T^2} \right)$,  at high temperature.
           } \label{fig:WF0}
\end{figure}

The Lorentz ratio is defined by the ratio between the longitudinal thermal conductivity and the longitudinal electric conductivity;
\begin{align}
L \equiv \frac{\kappa_{xx}}{\sigma_{xx} T}.
\end{align}
In the large $q$ limit, the transverse electric conductivity can be ignored and the Lorentz ratio is suppressed as $1/q$
\begin{align}
L \, \Big|_{q \gg 1} 
\sim \frac{2 \pi^2 }{3 \sqrt{3}}\cdot\frac{\beta^2}{q}.
\end{align} 
On the other hand, at the charge neutrality point($q=0$), the longitudinal electric conductivity becomes $1$ and the Lorentz ratio becomes
\begin{align}
L \, \Big|_{q=0} \sim \Big\{ \begin{array}{c} 8\cdot \frac{\pi^2}{3}~~~(T\ll 1) \\ \frac{(4\pi)^4}{9}\cdot \frac{T^2}{\beta^2}~~~(T \gg 1)  \end{array} ,
\end{align}
where we use 
\begin{align}
r_0 = \frac{2\pi}{3}\left( T +\sqrt{T^2 +6\left(\frac{\beta}{4\pi}\right)^2 } \right)
\end{align}
at the charge neutrality point. Figure \ref{fig:WF0} shows the charge dependence of Lorentz ration for various interaction strength $q_{\chi}$ and total impurity density $\beta$. In the figure, the violation of the Wiedemann-Frantz law is maximized at charge neutrality point and the violation region increases as interaction strength $q_{\chi}$ increases, Figure \ref{fig:WF0} (a) and the violation is suppressed as the total impurity density increases, Figure \ref{fig:WF0} (b). Notice that for large violation of the Wiedemann-Frantz law, one should use clean material($\beta \ll 1$).

\begin{figure}[ht!]
\centering
    \subfigure[$S$, $q_{\chi}\gamma=0$ ]
   {\includegraphics[width=5.5cm]{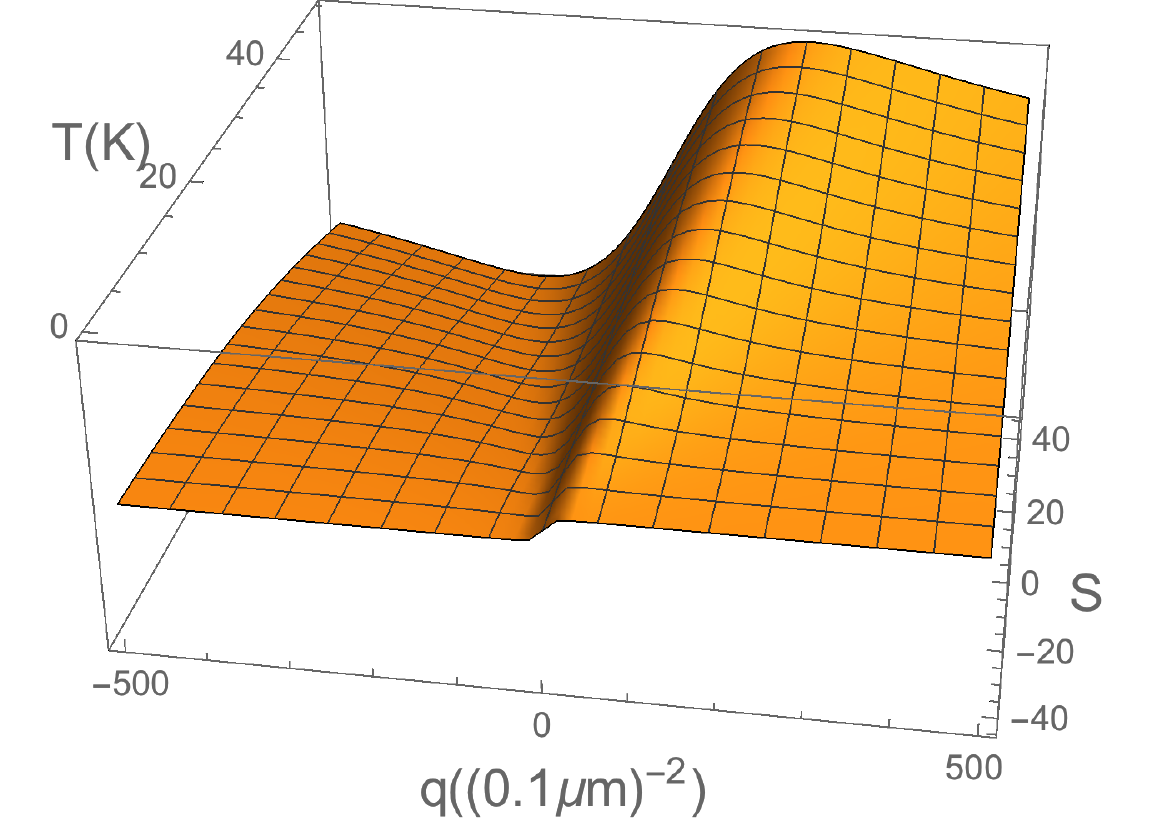} \label{}}
   \hspace{1cm}
       \subfigure[$S$, $q_{\chi}\gamma=3$]
   {\includegraphics[width=5.5cm]{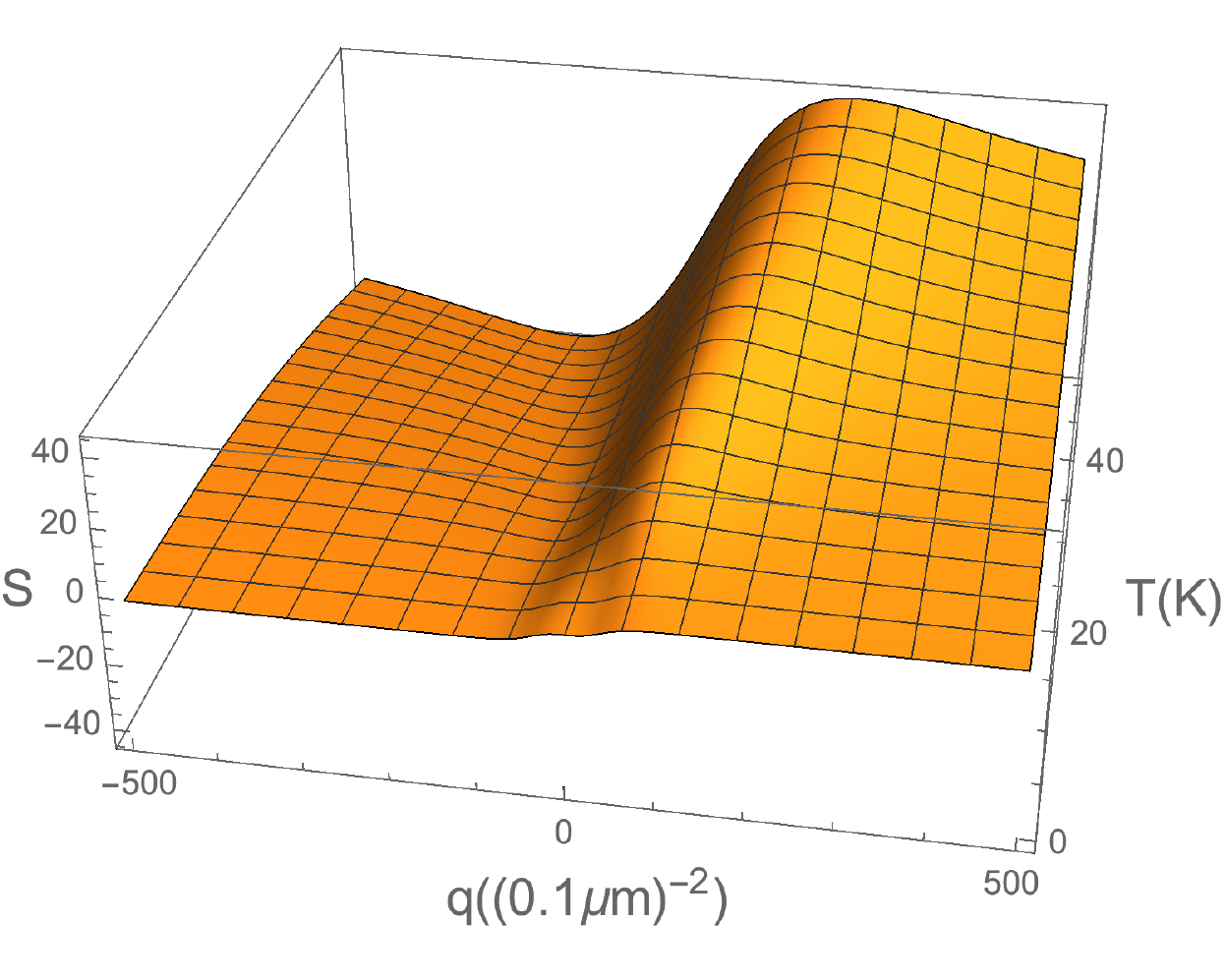} \label{}}
%
%
 \caption{The temperature and the charge density dependence of Seebeck coefficient without(a) with(b) magnetic impurity.     } \label{fig:SH0}
\end{figure}

 Seebeck coefficient($S$) and the Nernst signal(N) (\ref{SN01}) can be written as
\begin{align}\label{SN02}
S &= \frac{4\pi r_0^2 q(q^2 + r_0^2 \beta^2 )}{(q^2 + r_0^2 \beta^2 )^2 + (r_0^2 \beta^2)^2 \theta^2} \cr
N & = -\frac{4\pi r_0^4  q \beta^2 \theta}{(q^2 + r_0^2 \beta^2 )^2 + (r_0^2 \beta^2)^2 \theta^2},
\end{align}
where we use (\ref{H0trans}).
Notice that the Nernst signal is non zero in the absence of the external magnetic field because of the  non zero component of the transverse electric conductivity. Both of $S$ and $N$ are odd function of $q$ and it goes to zero for large $q$ limit. The temperature and the charge density dependence of Seebeck coefficient is drawn in Figure \ref{fig:SH0}.

In Figure \ref{fig:SH0} (a) and (b), the overall structure of the Seebeck coefficient is similar for both cases except near charge neutrality point.  In the absence of magnetic impurity, The behaviors of the Seebeck coefficient is linear in $q$ at charge neutrality point. But if we put magnetic impurity, step-like behavior appears near charge neutrality point for large value of $q_{\chi}$. The Seebeck coefficient near the charge neutrality point at zero temperature is
\begin{align}
S \Big|_{q \ll 1, T=0} \sim \frac{4\pi}{\beta^2 (1+36 q_{\chi}^2 \gamma^2)} \, q + {\cal O}(q^3).
\end{align}

\begin{figure}[ht!]
\centering
    \subfigure[$N$, $q_{\chi} \gamma=0$ ]
   {\includegraphics[width=5.5cm]{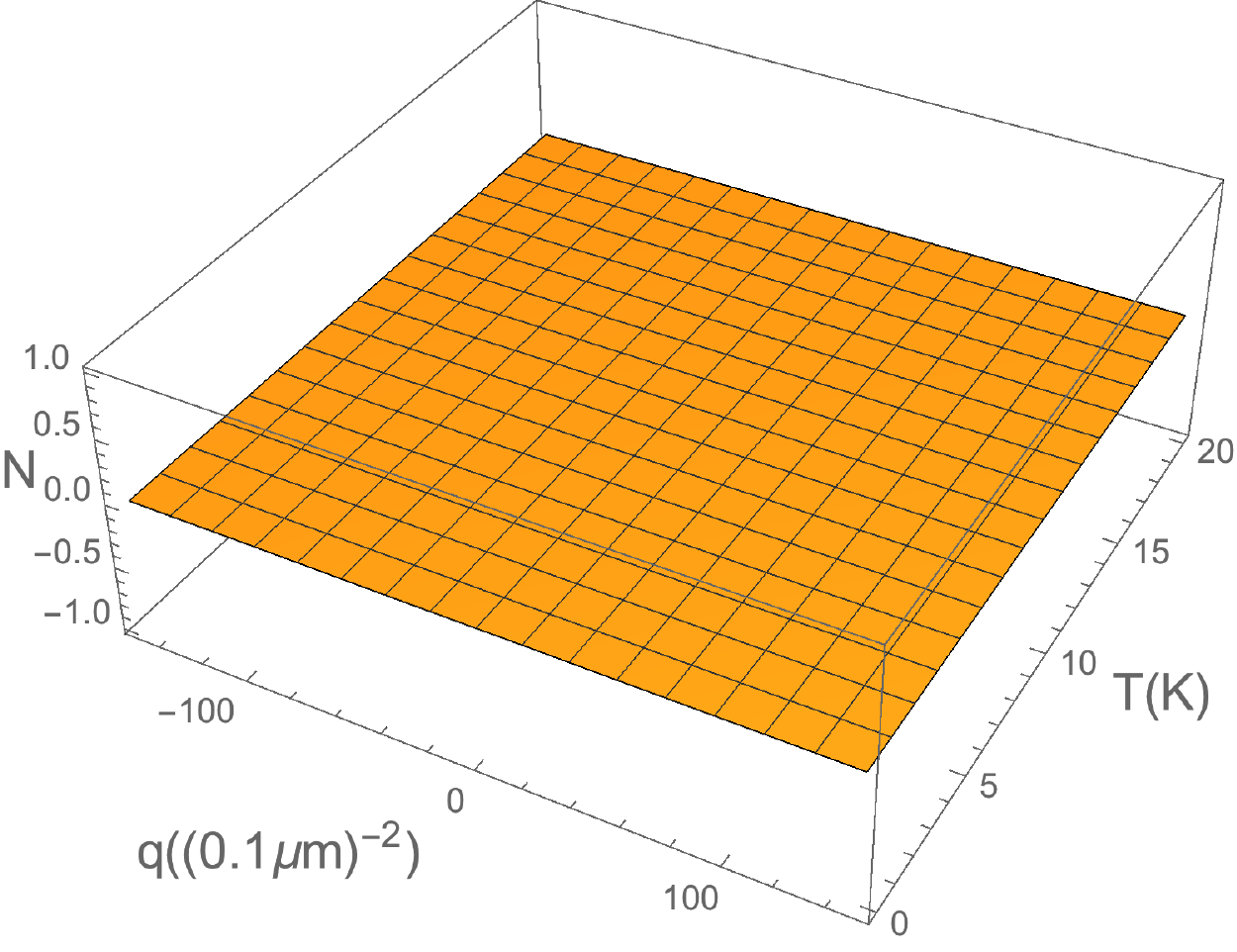} \label{}}
   \hspace{1cm}
       \subfigure[$N$, $q_{\chi} \gamma=0.7$]
   {\includegraphics[width=5.5cm]{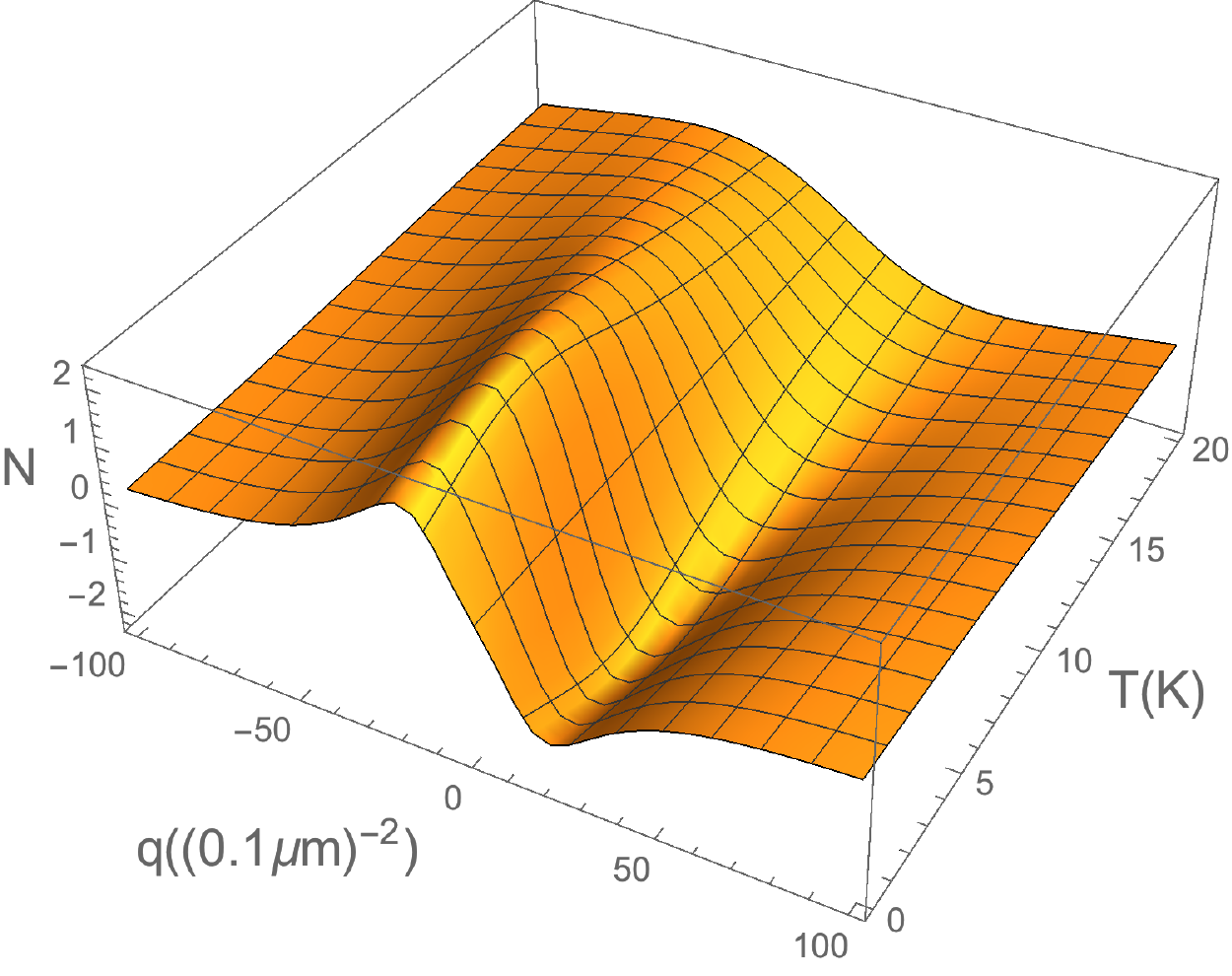} \label{}}
   
 \caption{The temperature and the charge density dependence of Seebeck coefficient without(a) with(b) magnetic impurity. We set $\beta^2=\frac{2747}{(\mu m)^2}$ and  $v_{F} =7.5 \times 10^4 m/s$. 
           } \label{fig:NH0}
\end{figure}

The charge density and the temperature dependences of the Nernst signal are drawn in Figure \ref{fig:NH0}.

\begin{figure}[ht!]
\centering
    \subfigure[ ]
   {\includegraphics[width=5.5cm]{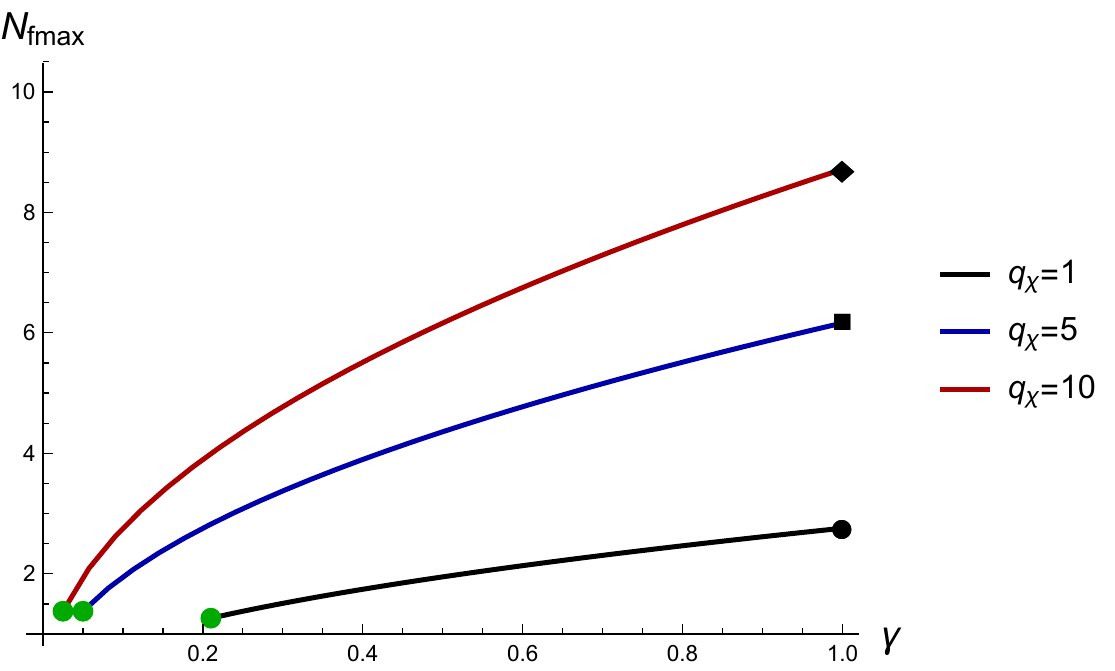} \label{}}
   \hspace{1cm}
       \subfigure[]
   {\includegraphics[width=5.5cm]{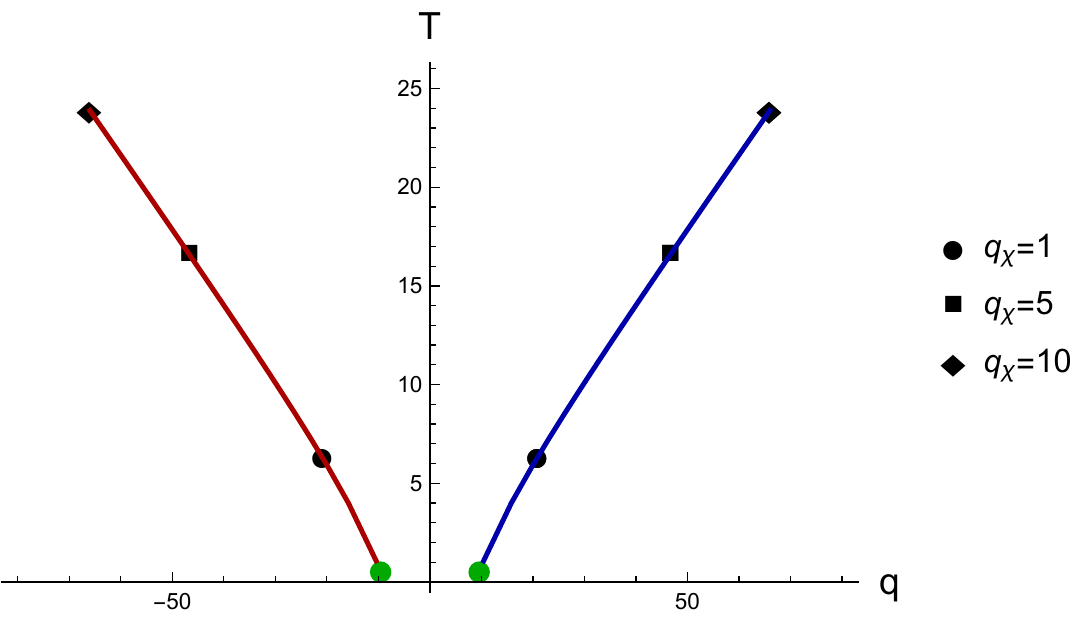} \label{}}
   
 \caption{(a) Maximum value Nernst signal for different $\gamma$ and $q_{\chi}$. (b) Position of maximum and minimum of Nernst signal.             } \label{fig:Nmax}
\end{figure}

Nernst signal is proportional to $\theta$ from (\ref{SN02}), therefore it vanishes at zero magnetic impurity case. But in the presence of magnetic impurity, there are maximum and minimum at finite density and temperature as shown in Figure \ref{fig:NH0} (b).  As we change the ratio of the magnetic impurity $\gamma$ or interaction strength $q_{\chi}$, the position of the maximum and minimum also changes. Figure \ref{fig:Nmax} shows $\gamma$ and $q_{\chi}$ dependence of the position of the maximum and minimum of Nernst signal.

Numerical study shows that as we increase $\beta$, overall shape of Seebeck coefficient 
 is broaden quickly while 
system does not depends on $q_{\chi}$ very much.  On the other hand, the Nernst signal has maximum value at finite density and temperature as shown in Figure \ref{fig:NH0}(b). Increasing $\beta$ makes the overall shape broaden similar to the Seebeck coefficient. If we increase $q_{\chi}$, overall shape is similar but the position of maximum moves to large $q$ and large $T$ direction and height  increases.  

\subsection{Graphical predictions for Bi$_2$Se$_3$}
\begin{figure}[ht!]
\centering
    \subfigure[ ]
   {\includegraphics[width=5.5cm]{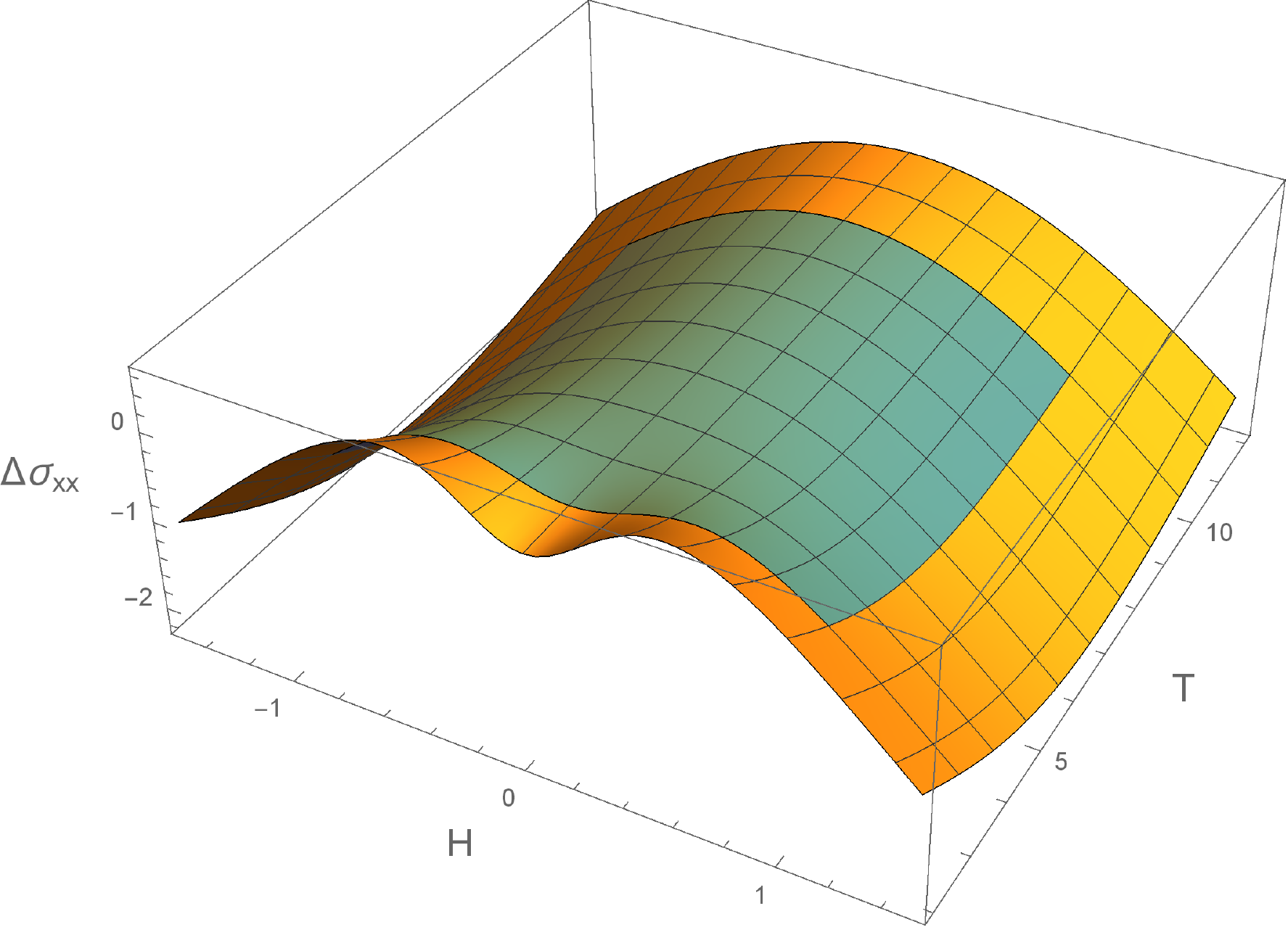} \label{}}
   \hspace{1cm}
       \subfigure[]
   {\includegraphics[width=5.5cm]{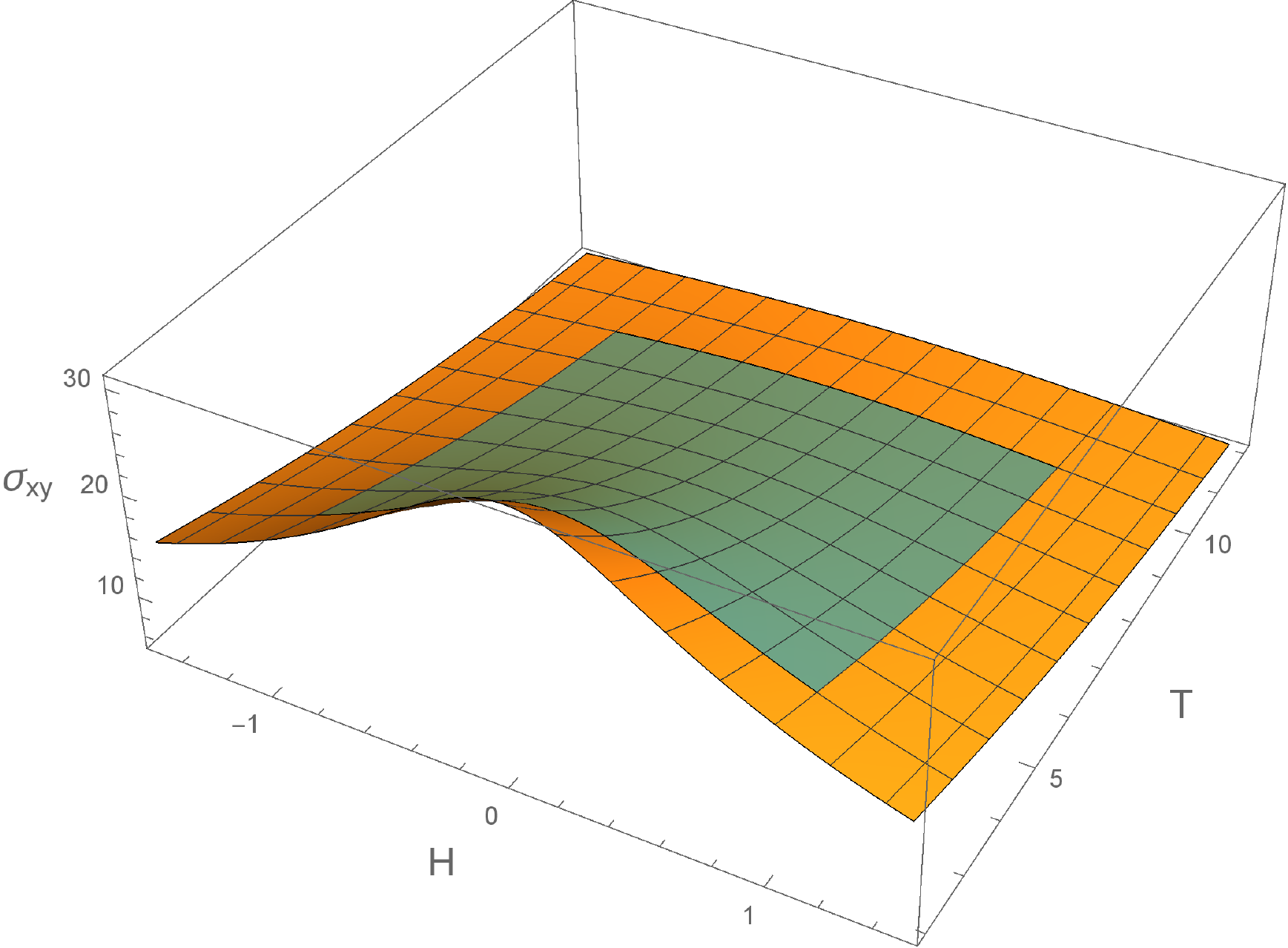} \label{}}
 \caption{The temperature and the magnetic field dependence of (a) parallel, (b) transverse magnetoconductivity. In this subsection we used  $v_F=c/3000,  \beta=1.82, q_{\chi}\gamma=5.6$ in all the figures. The green colored island corresponds to the region where it is expected that our theory might be valid.         } \label{fig:MCisland}
\end{figure}

\begin{figure}[h]
\centering
    \subfigure[ ]
   {\includegraphics[width=5.5cm]{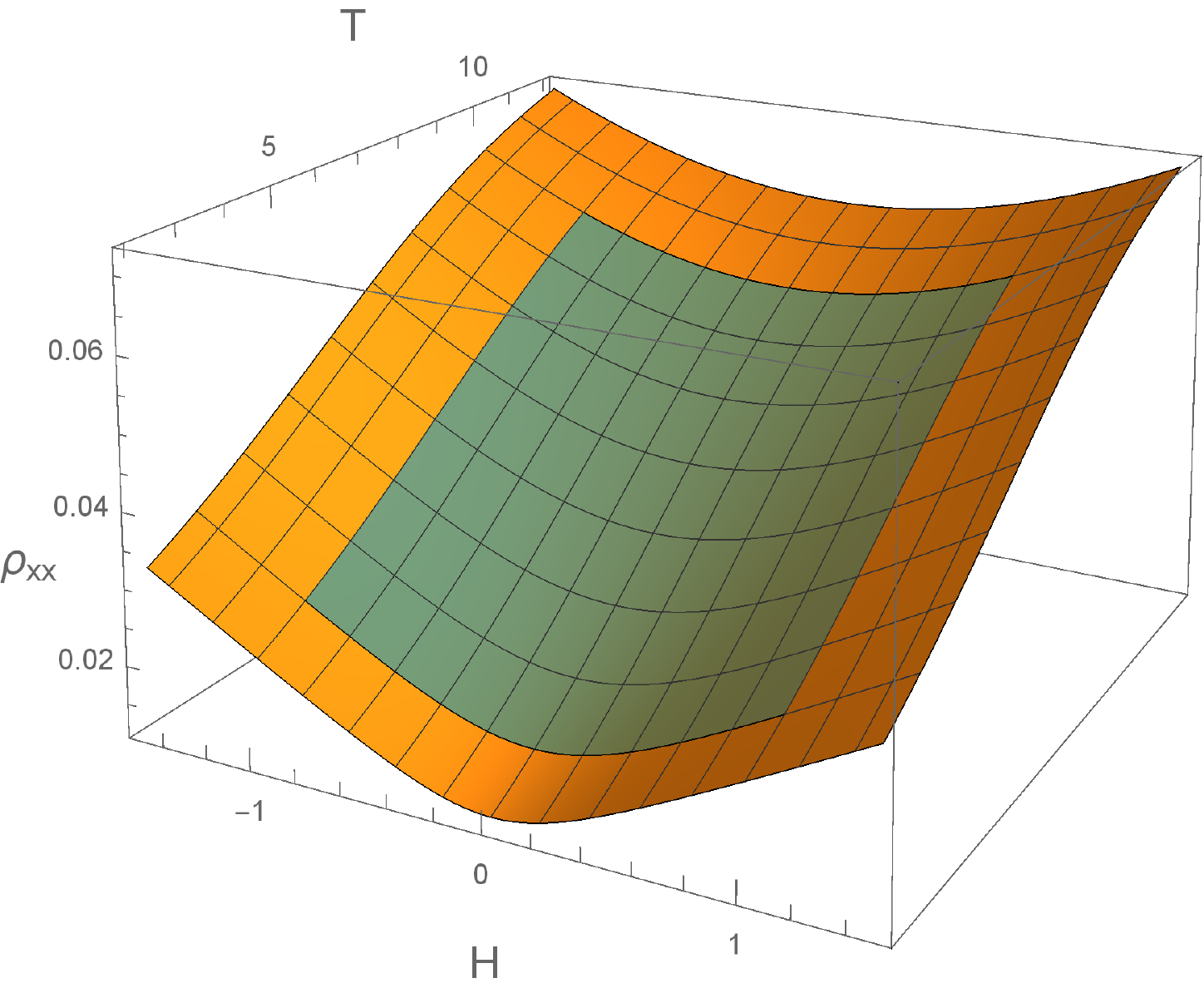} \label{}}
   \hspace{1cm}
       \subfigure[]
   {\includegraphics[width=5.5cm]{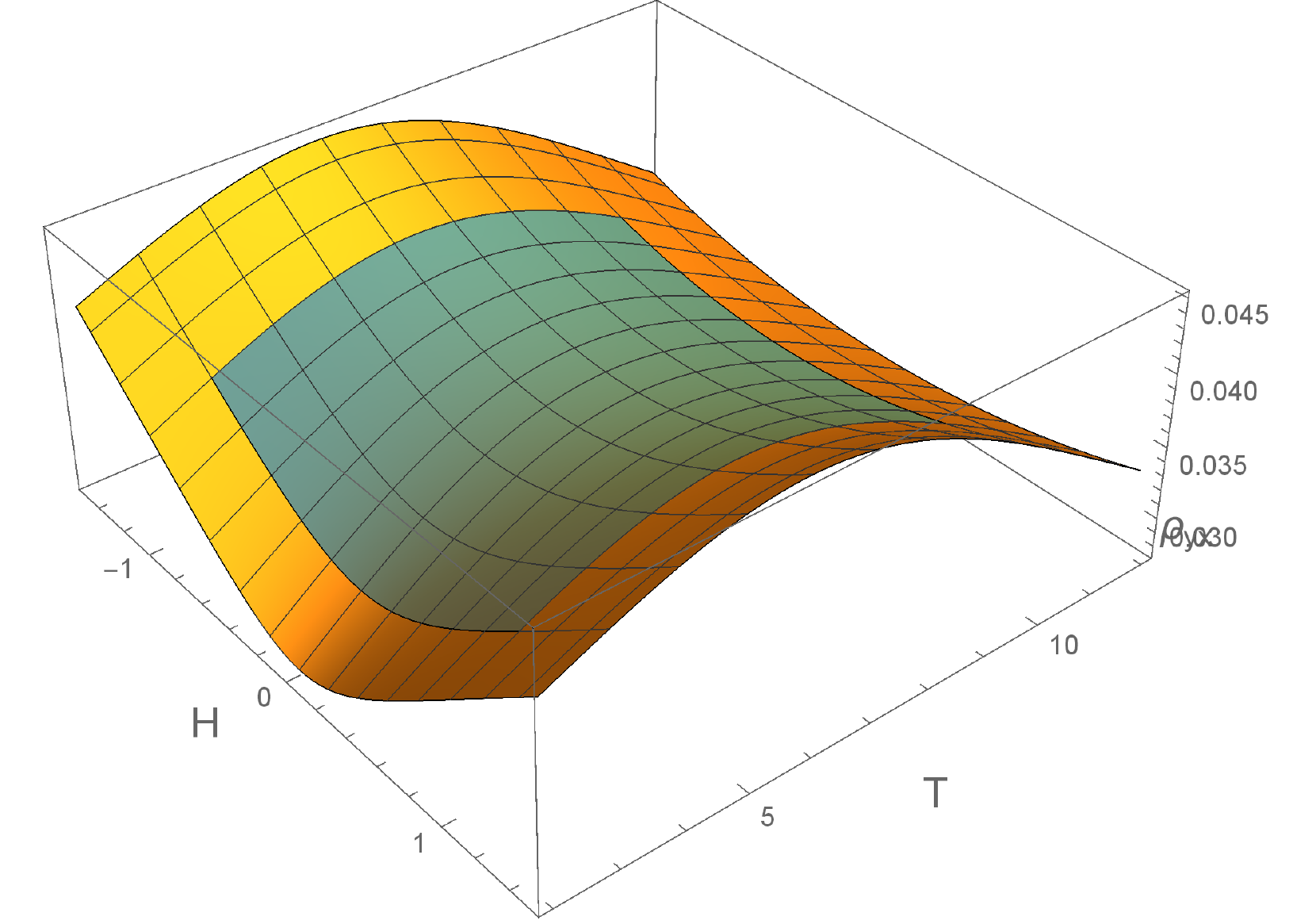} \label{}}
 \caption{The temperature and the magnetic field dependence of (a) parallel, (b) transverse magnetoresitivity.         } \label{fig:MRisland}
\end{figure}

\begin{figure}[h]
\centering
    \subfigure[ ]
   {\includegraphics[width=5.5cm]{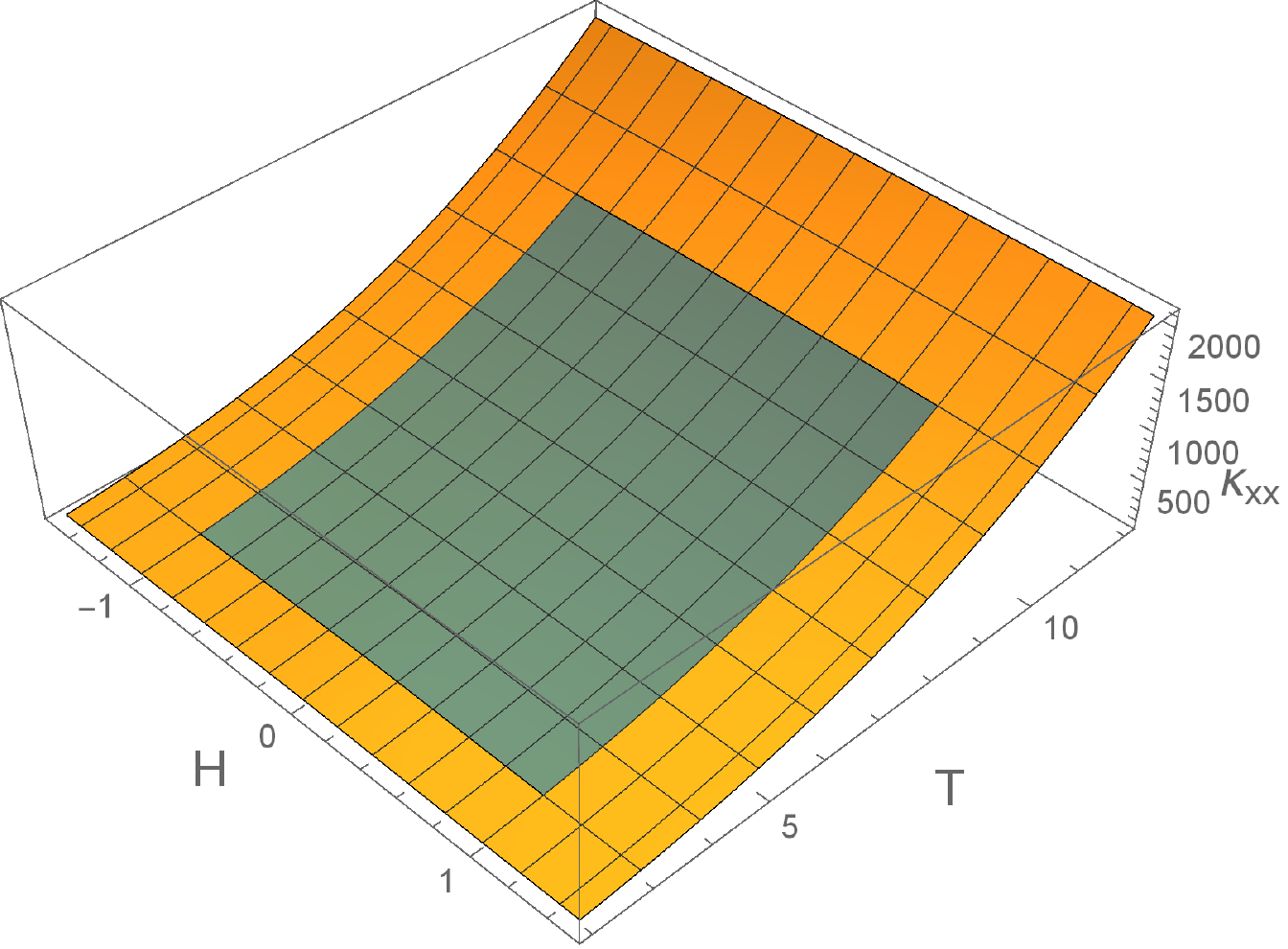} \label{}}
   \hspace{1cm}
       \subfigure[]
   {\includegraphics[width=5.5cm]{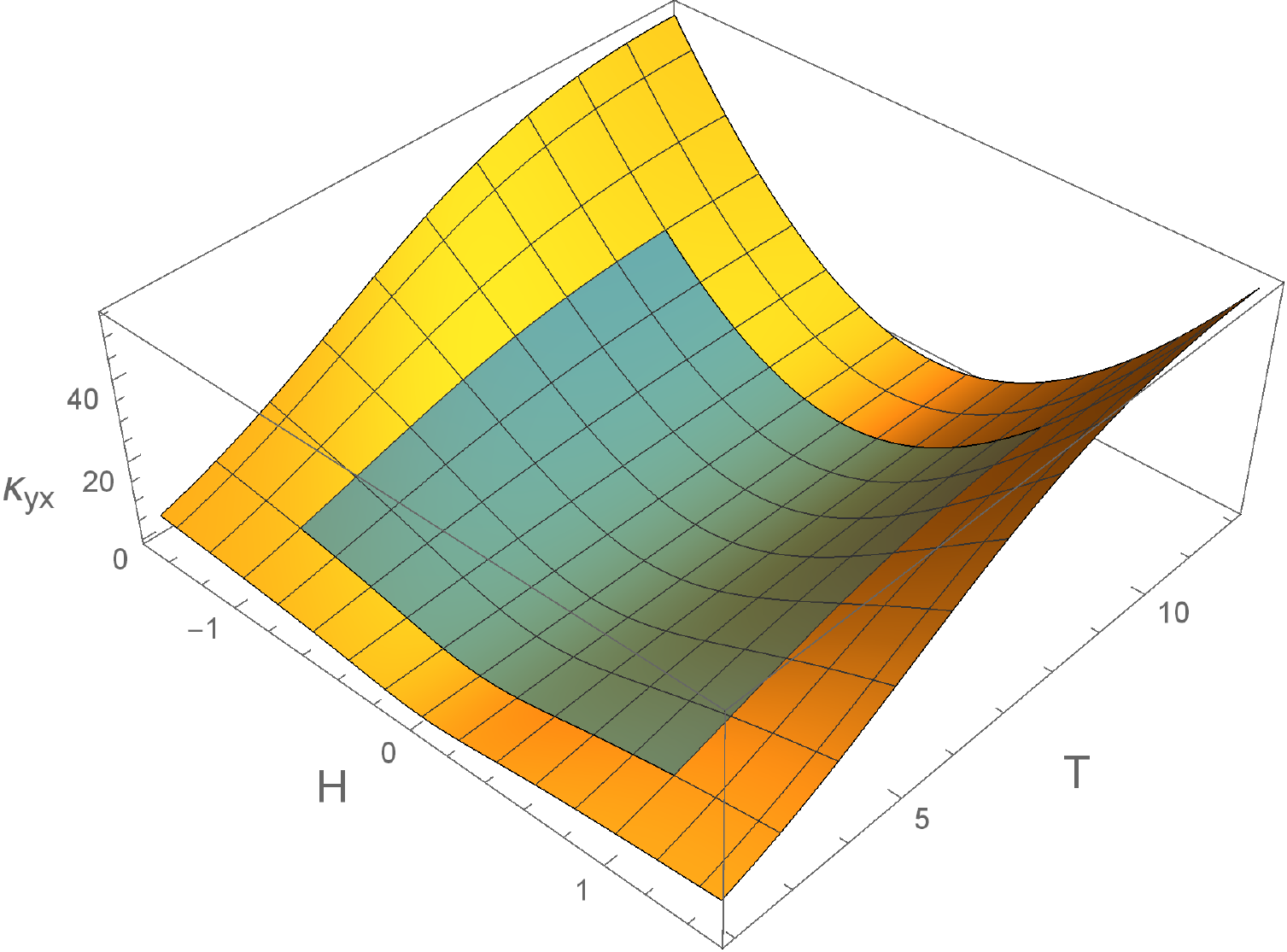} \label{}}
 \caption{The temperature and the magnetic field dependence of (a) parallel, (b) transverse thermal conductivity.   } \label{fig:Kisland}
\end{figure}

\begin{figure}[h]\
\centering
    \subfigure[ ]
   {\includegraphics[width=5.5cm]{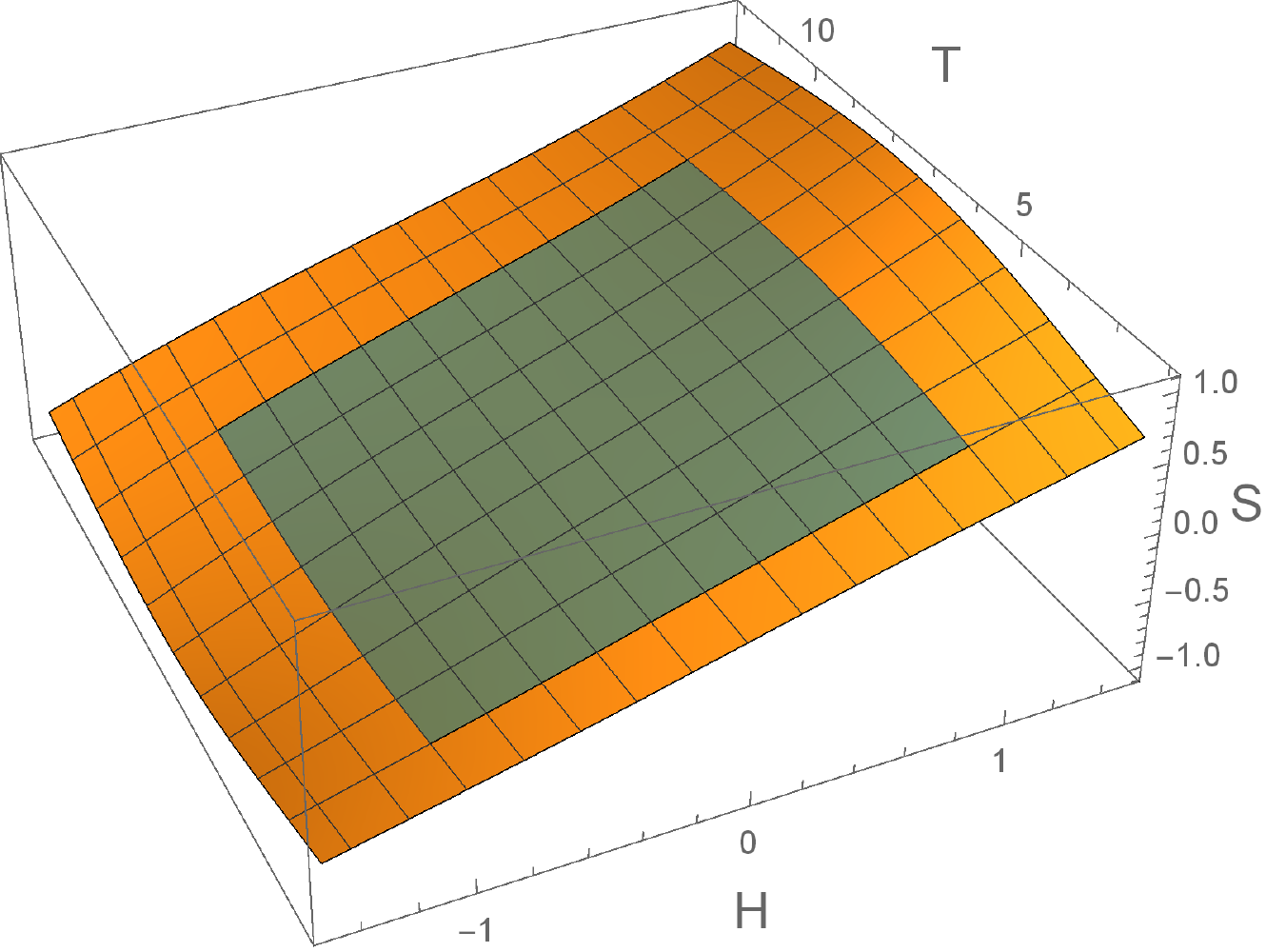} \label{}}
   \hspace{1cm}
       \subfigure[]
   {\includegraphics[width=5.5cm]{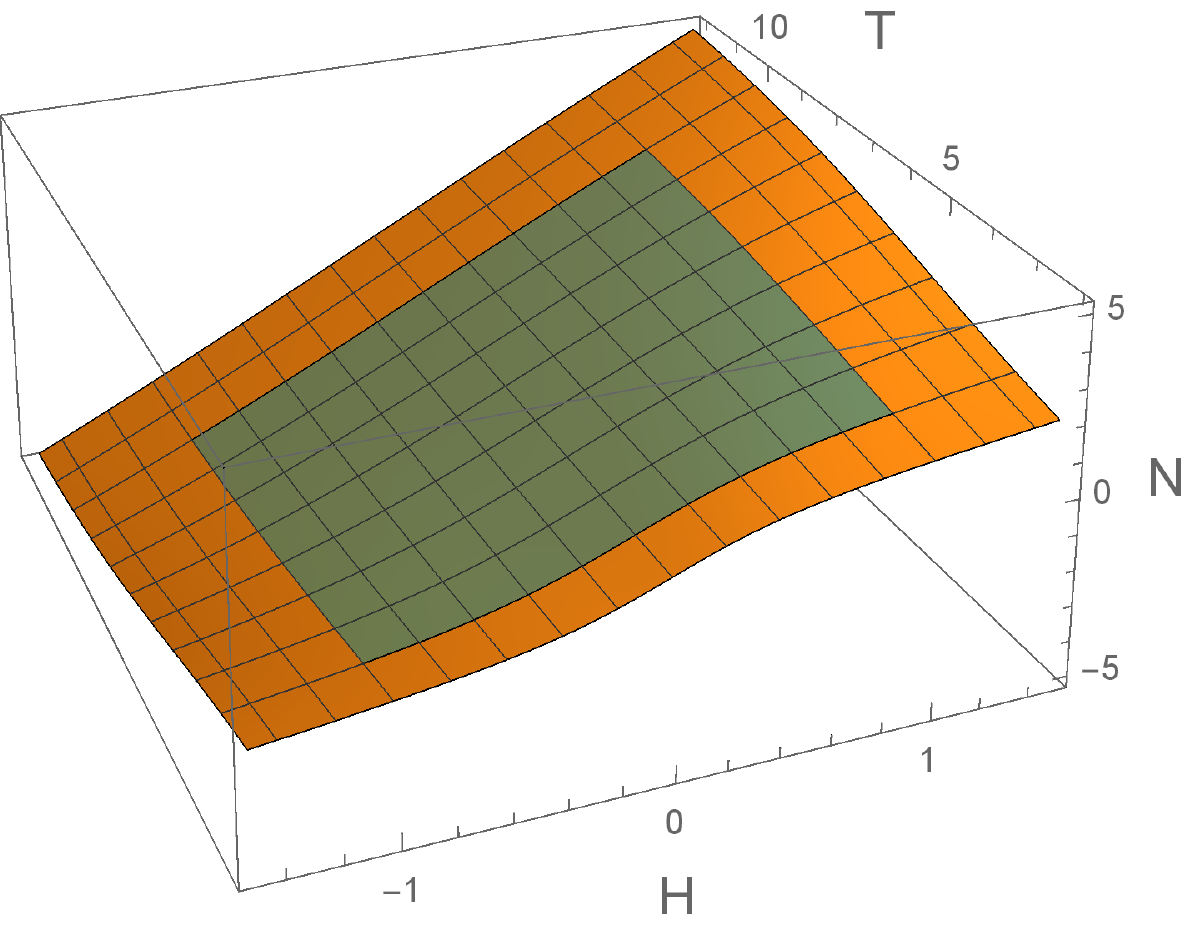} \label{}}
 \caption{The temperature and the magnetic field dependence of (a) Seebeck coefficient, (b) Nernst signal.          } \label{fig:SNisland}
\end{figure}

So far we plotted our results with set of parameters  such that interesting features appear.
For the future experiment, however, it will be more useful to use the parameters which was used 
to fit magneto conductivity. Here we redraw  all the figures such that all the 
figures can be compared with the data of $Bi_{2}Se_{3}$. See Figure \ref{fig:MCisland} - Figure \ref{fig:SNisland}.  
One caution is that we set  chemical potential zero. 
Individual material sample can have finite chemical potential 
for various reason.  Gating and impurity doping can bring finite charge and chemical density, for example.

\section{Conclusion}
In this paper we set up a model for a Dirac material where magnetic  impurity is coupled with massless degree of freedom in a time reversal symmetry breaking way.  
We targeted mainly for the surface states of TI with magnetic impurity doped. However we expect that 
the model may have more general validity. 

From the experience so far, we can mention  a few general aspects of Dirac materials. 
For undoped or   weakly doped  TI, one normally sees a sharp peak, which is the characteristic of weak  anti-localization. We, however,   expect that if we 
can set the  fermi surface    near Dirac point  by gating,  we  will see the disappearance of the sharp peak as we move down the fermi surface. We also expect  that  the transition behavior from WAL $\to$ WL in the medium doping is universal so that 
magneto-conductivity of all two dimensional Dirac material with broken TRS can be described by our formula, which is
independent of the detail of the system.   
For Cr$_{x}$Bi$_{2-x}$Te$_{3}$ with $x=0.1$ where  the system in our picture is  strongly interacting for $T\ge 2K$, we expect that ARPES data will show fuzzy density of state (DOS).   DOS will be non-zero in the region between dispersion curves,  
where quasi-particle case would show empty DOS leading to the gap or pseudo gap. Currently we are 
studying these effects using fermion two point functions.  
For general Dirac material, we can say  that 
near Dirac point there should be  large violation of Wiedemann-Franz Law just like graphene.  

Finally we mention some of the future projects. 
In this paper, we  studied  the  zero charge case mostly.  
Non-zero charge parameter will be discussed in follow-up paper in relation with the hysteresis of magnetization
and various other physical observables. 
The graphene has even number of Dirac cones,  weak spin-orbit interaction  and  therefore  mechanism for WL/WAL is different from the one analyzed here. 
Different interaction term  is necessary.  
Five dimensional extension of this work will be related to the study of Weyl semi-metal. 
It is also interesting to study coupling of  impurity density with  $R\wedge R$ as well as $F\wedge F$. 
Because of such differences, we need to find other  interaction term in holographic model for graphene.
It is also interesting to classify all possible patterns of interaction that provide the fermion surface gap in the presence of strong electron-electron correlation in our context. 

\acknowledgments
 We thank E.G. Moon, K.S. Kim, Y.B. Kim and K. Park  for  discussions. This  work is supported by Mid-career Researcher Program through the National Research Foundation of Korea grant No. NRF-2016R1A2B3007687.  YS is also supported in part by Basic Science Research Program through NRF grant No. NRF-2016R1D1A1B03931443. Work of CP is  supported by Basic Science Research Program through the National Research Foundation of Korea funded by the Ministry of Education (NRF-2016R1D1A1B03932371) and also supported by the Korea Ministry of Education, Science and Technology, Gyeongsangbuk-Do and Pohang City. We also thank APCTP for hospitality during the focus program ``Geometry and Holography of quantum criticality''.

\begin{appendices}
 \section{Magnetic induction and magnetization}
Here, 
we describe the problem in magnetization of 2+1  and possible resolution.
 
 \begin{enumerate}
\item 
 In 2+1 system,  
we can not  add M and H since two have different mass dimensions: $[M]=1, [H]=2$. 
Therefore although we can theoretically calculate it as a conjugacy of $H$, 
  we can not compare it with experiment. 

\item 
To fix the problem,  
we can either  uplift to or embed our 2+1 dimensional system into 3+1 one. 
Uplifting is considering a 3+1  system with translational invariance 
whose  dimensionally reduction is the original 2+1 system. 
Embedding is  to consider the 2+1  dimensional system as a thin film in  3+1   with a small thickness $L$. 
 
 \item 
 In any case we should   have $M_{3+1}=M_{2+1}/L$ for some length scale $L$. 
For the thin film case, 
$L$ is the thickness of the film. For uplifting case, $L$ is inverse temperature if that is the only scale. 
 If we turn   on $H$ and $q$,   $r_{0}$ is the most natural scale parameter to enter since 
the energy density $\sim r_{0}^{d}$ in any dimension. 

\item
For topological insulator's surface, we do not have a physical scale $L$ so its better to uplift it. 
We can argue that    there     is   a unique choice of $L$ used  in raising $M_{2+1}= L M_{3+1}$: 
$M_{2+1}$ begins with a term $-\frac{H}{r_{0}} +\cdots $, 
at the zero temperature and zero charge and zero impurity limit. That is, if we consider the dyonic black hole,   
 then $r_{0}= (12)^{-1/4} \sqrt{H}$ so that  $M_{2+1}=- ((12)^{1/4}) \sqrt{H}$. 
 To find magnetic induction, we need to uplift or embed it to 3+1 as we discussed it by multipying $1/L$ factor for some length scale $L$. Then for $r_{0}$ indepependent $L$, the magnetic induction becomes 
\begin{equation}
B=H+M=H- ((12)^{1/4}/L) \sqrt{H}<0, \quad {\rm for} \quad   H<\sqrt{12}/L.
\end{equation}
 This is more than  perfect magnetization, which is nonsense.  Similar but more complicated argument hold 
 for finite density and temperature.  
Therefore    $L$ should be  $r_{0}$ dependent. 

\item 
The only way to fix this problem 
is to choose $1/L= c. r_{0}$ with some constant $c$.  
 Now If we seek the energy density such that  $M_{3+1} =  M_{2+1}/L= - \frac{\partial\varepsilon_{3+1}}{\partial H}$, then we get $\varepsilon_{3+1}=\frac c 2 H^{2}+\cdots $. 
 Only when we take $c=1$,  the fist term  is identified with energy density of the magnetic field at the vacuum which 
 should be subtracted from the total energy  density  when we consider the response of the system 
 to  the applied magnetic field $H$ \cite{landau2013electrodynamics}. 
 Notice that the uplifting procedure give us a unique way to fix the problem by giving us a natural reasoning  
 to subtract unphysical part of magnetization by attributing it as the derivative of vacuum energy density.  
Similar description was done in completely condensed matter context \cite{nersesyan1989low,kkss}. 
   \end{enumerate}

In short,  in the absence of a canonical thickness of the surface of TI, we need to uplift 
to solve the dimensionality problem and  to make sure vacuum does not have any non-zero magnetization, 
we need to redefine the free energy by  subtracting 
 the energy   of background magnetic field in the absence of the matter,
\begin{align}
F = r_0 \epsilon -\frac{H^2}{2},
 \label{magnetization}
 \quad\quad 
M \equiv -\frac{\partial F}{ \partial H} = \frac{q \theta}{3} -\frac{\theta^2 H}{5}. 
\end{align}
The magnetic induction now is identified as 
\begin{align}
B = H + M = H +\frac{q \theta}{3} -\frac{\theta^2 H}{5}.
\end{align}
In the case of the dyonic black hole, $\theta =0$, and we have  $M=0$ and hence $B=H$.

Once we re-define magnetization $M$, we can study several properties of it.
  First,  the system is still ferromagnetic if charge density is nonzero, that is the magnetization (\ref{magnetization}) has  finite value $M_{0}$  even   in the absence of the external magnetic field:
\begin{align}\label{M0}
M_0 =\frac{q \theta}{3} = \frac{ q \lambda^2 q_{\chi}}{3 r_0^2},
\end{align}
\begin{figure}[ht!]
\centering
    \subfigure[ ]
   {\includegraphics[width=5.5cm]{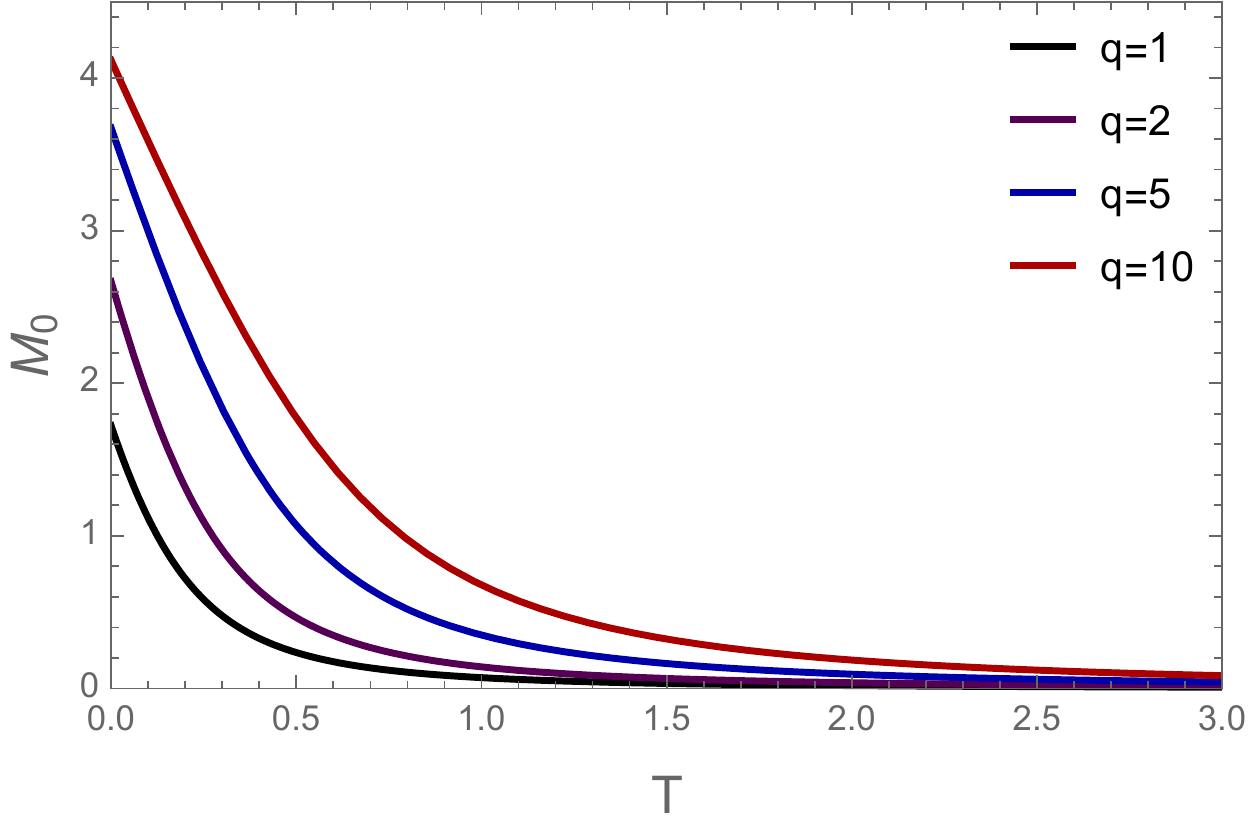} \label{}}
   \hspace{1cm}
       \subfigure[ ]
   {\includegraphics[width=5.5cm]{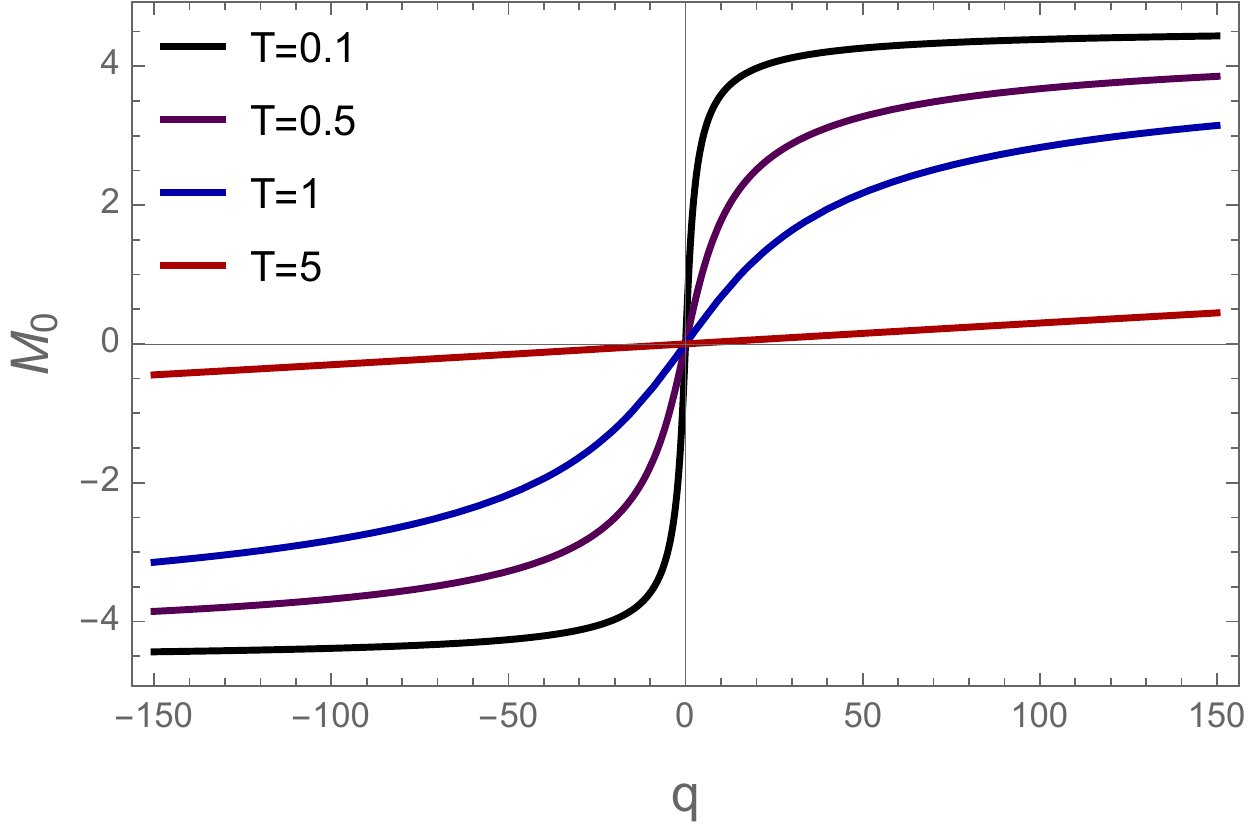} \label{}}
   
 \caption{(a) Temperature dependence of $M_0$ with for different density. (b) Density dependence of $M_0$ with different temperature. In both case, we set $q_{\chi}=1$, $\alpha =0$ and $\lambda=2$. 
           } \label{fig:M0}
\end{figure}

Figure \ref{fig:M0} (a) shows temperature dependence of $M_0$. At zero temperature 
it  becomes
\begin{align}
M_0 (T=0) = \frac{4 q \lambda^2 q_{\chi}}{\alpha^2+\lambda^2 + \sqrt{12 q^2 +\alpha^2+\lambda^2}},
\end{align}
which is proportional to $q \lambda^2 q_{\chi}$ for small $\lambda$ limit and to $q q_{\chi}$ for large $\lambda$ limit. 
On the other hand,  at high temperature,  it suppressed by  $1/T^2$ because $r_0\sim T$ in this regime.  

Figure \ref{fig:M0}(b) shows density dependence of $M_0$ for different temperature. For large value of $q$, $r_0\sim \sqrt{q}$ and the magnetization is saturated to a finite value 
\begin{align}\label{M02}
M_0(q =\infty) \sim  \lambda^2 q_{\chi}.
\end{align} 
We interpret $\lambda^2$  as the magnetic impurity density and $q_{\chi}$  as the strength of  coupling for each magnetic impurity. 

\begin{figure}[ht!]
\centering
    \subfigure[]
   {\includegraphics[width=6cm]{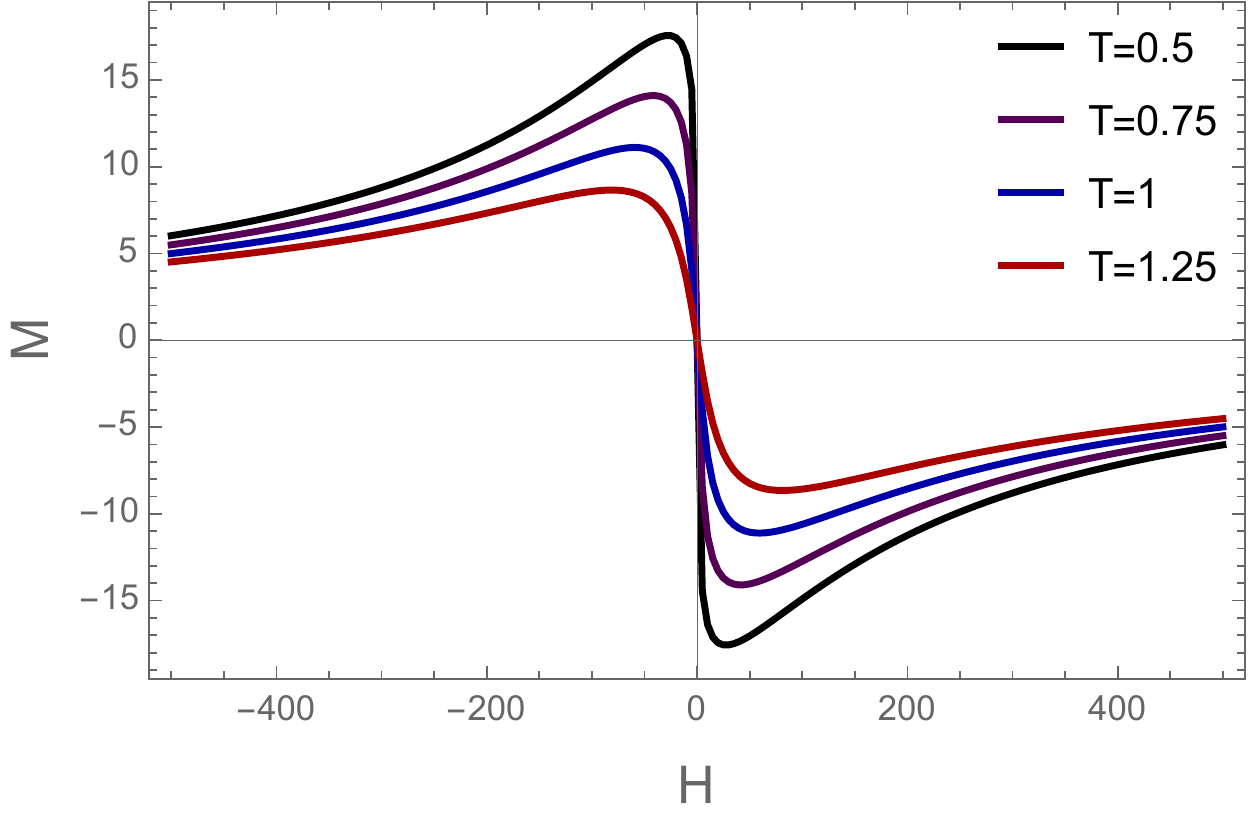} \label{}}
   \hspace{1cm}
       \subfigure[]
   {\includegraphics[width=6cm]{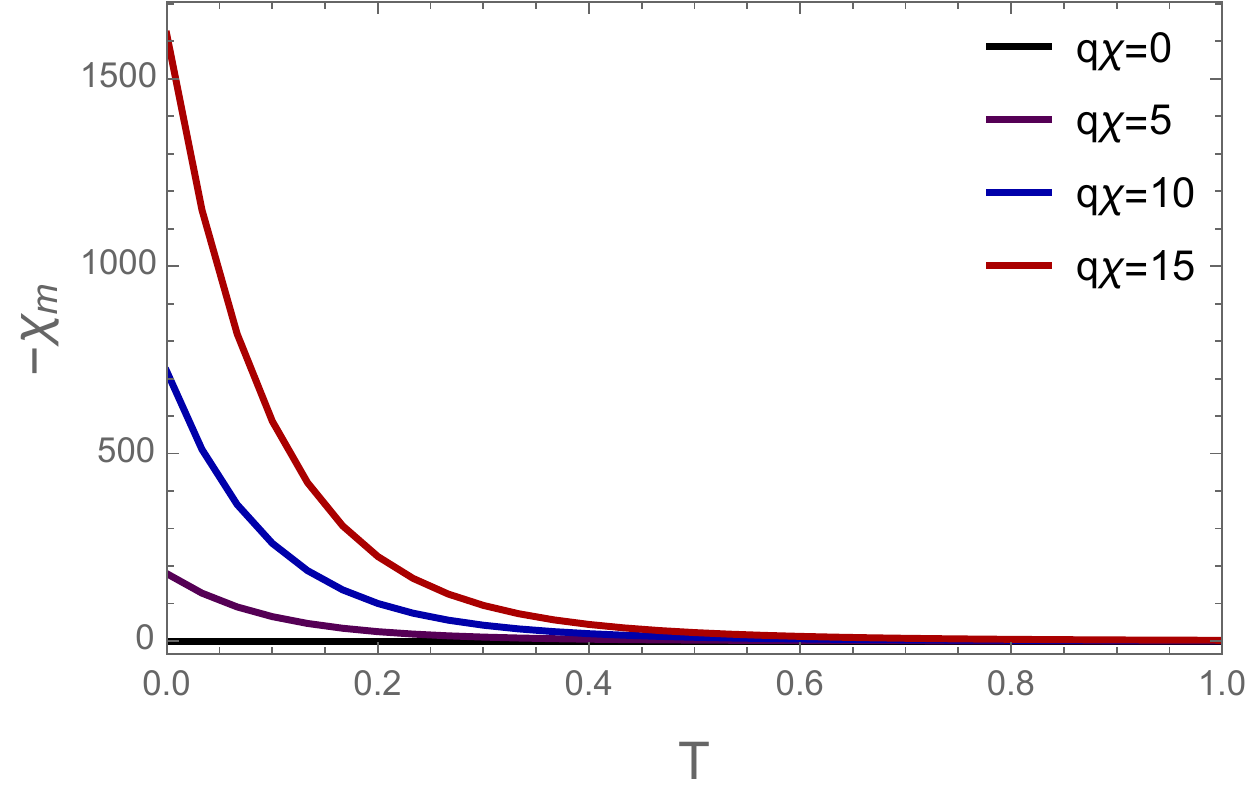} \label{}}
   
 \caption{(a) The external magnetic field dependence of the magnetization for different temperature. 
  We fix $\lambda=2$, $\alpha=0$, $q_{\chi}$=10 and $q=0$. (b) Temperature dependence of the magnetic susceptibility at $H=0$ with different value of $q_{\chi}$ with $\lambda=2$, $\alpha =0$ and $q=0$.
           } \label{fig:HM01}
\end{figure}

Second, in the presence of the external magnetic field, the coefficient of $H$ in the magnetization $M$ in (\ref{magnetization})  is alway negative, so that  the system is diamagnetic. At finite magnetic field, the horizon radius (\ref{temperature}) is a function of the external parameters $T$, $H$, $\beta$ and $q_{\chi}$.  The results are shown in Figure \ref{fig:HM01} (a). These diamagnetic behavior is similar to the experimental data of certain type of graphite sample \cite{Kopelevich:aa,suzuki2002magnetic}.

Third, once we obtain the magnetization, we can also calculate the magnetic susceptibility by taking derivative of the magnetization with respect to the external magnetic field;
\begin{align}
\chi_{m} \equiv \frac{\delta M}{\delta H}  
= -\frac{\theta^2}{5} + \frac{4\theta (6 H \theta -5 q)(H \theta^2 + H - q \theta)}{15\left\{ (3+ 7\theta^2) H^2 -10 H q \theta +3 q^2 +2 r_0^2 (6 r_0^2 +\alpha^2 +\lambda^2)    \right\}},
\end{align}
where second term comes from the variation of $r_0$ in (\ref{magnetization}). Figure \ref{fig:HM01} (b) shows temperature dependence of the magnetic susceptibility with different value of $q_{\chi}$ at zero magnetic field.

\end{appendices}
\vskip 1cm

\bibliographystyle{JHEP}
 
 \bibliography{Refs}

\providecommand{\href}[2]{#2}\begingroup\raggedright\begin{thebibliography}{10}

\bibitem{mott1937discussion}
N.~Mott and R.~Peierls, \emph{Discussion of the paper by de boer and verwey},
  {\emph{Proceedings of the Physical Society} {\bf 49} (1937) 72}.

\bibitem{Sachdev:2011mz}
S.~Sachdev, \emph{Quantum Phase Transitions.}
\newblock Cambridge University Press. (2nd ed.), 2011.

\bibitem{zaanen2015holographic}
J.~Zaanen, Y.~Liu, Y.-W. Sun and K.~Schalm, \emph{Holographic duality in
  condensed matter physics}.
\newblock Cambridge University Press, 2015.

\bibitem{Oh:2013qxn}
E.~Oh and S.-J. Sin, \emph{{Non-spherical collapse in AdS and Early
  Thermalization in RHIC}},
  \href{http://dx.doi.org/10.1016/j.physletb.2013.08.040}{\emph{Phys. Lett.}
  {\bf B726} (2013) 456--460}, [\href{http://arxiv.org/abs/1302.1277}{{\tt
  1302.1277}}].

\bibitem{Sin:2013yha}
S.-J. Sin, \emph{{Physical mechanism of AdS instability and universality of
  holographic thermalization}},
  \href{http://dx.doi.org/10.3938/jkps.66.151}{\emph{J. Korean Phys. Soc.} {\bf
  66} (2015) 151--157}, [\href{http://arxiv.org/abs/1310.7179}{{\tt
  1310.7179}}].

\bibitem{HKMS}
S.~A. {Hartnoll}, P.~K. {Kovtun}, M.~{M{\"u}ller} and S.~{Sachdev},
  \emph{{Theory of the Nernst effect near quantum phase transitions in
  condensed matter and in dyonic black holes}},
  \href{http://dx.doi.org/10.1103/PhysRevB.76.144502}{\emph{Phys. Rev. B} {\bf
  76} (Oct., 2007) 144502}, [\href{http://arxiv.org/abs/0706.3215}{{\tt
  0706.3215}}].

\bibitem{Lucas:2015sya}
A.~Lucas, J.~Crossno, K.~C. Fong, P.~Kim and S.~Sachdev, \emph{{Transport in
  inhomogeneous quantum critical fluids and in the Dirac fluid in graphene}},
  \href{http://dx.doi.org/10.1103/PhysRevB.93.075426}{\emph{Phys. Rev.} {\bf
  B93} (2016) 075426}, [\href{http://arxiv.org/abs/1510.01738}{{\tt
  1510.01738}}].

\bibitem{Maldacena:1997re}
J.~M. Maldacena, \emph{{The Large N limit of superconformal field theories and
  supergravity}},
  \href{http://dx.doi.org/10.1023/A:1026654312961}{\emph{Int.J.Theor.Phys.}
  {\bf 38} (1999) 1113--1133}, [\href{http://arxiv.org/abs/hep-th/9711200}{{\tt
  hep-th/9711200}}].

\bibitem{Witten:1998qj}
E.~Witten, \emph{{Anti-de Sitter space and holography}}, {\emph{Adv. Theor.
  Math. Phys.} {\bf 2} (1998) 253--291},
  [\href{http://arxiv.org/abs/hep-th/9802150}{{\tt hep-th/9802150}}].

\bibitem{Gubser:1998bc}
S.~S. Gubser, I.~R. Klebanov and A.~M. Polyakov, \emph{{Gauge theory
  correlators from noncritical string theory}},
  \href{http://dx.doi.org/10.1016/S0370-2693(98)00377-3}{\emph{Phys. Lett.}
  {\bf B428} (1998) 105--114}, [\href{http://arxiv.org/abs/hep-th/9802109}{{\tt
  hep-th/9802109}}].

\bibitem{pkim}
J.~{Crossno}, J.~K. {Shi}, K.~{Wang}, X.~{Liu}, A.~{Harzheim}, A.~{Lucas}
  et~al., \emph{{Observation of the Dirac fluid and the breakdown of the
  Wiedemann-Franz law in graphene}},
  \href{http://dx.doi.org/10.1126/science.aad0343}{\emph{Science} {\bf 351}
  (Mar., 2016) 1058--1061}, [\href{http://arxiv.org/abs/1509.04713}{{\tt
  1509.04713}}].

\bibitem{Seo:2016vks}
Y.~Seo, G.~Song, P.~Kim, S.~Sachdev and S.-J. Sin, \emph{{Holography of the
  Dirac Fluid in Graphene with two currents}},
  \href{http://dx.doi.org/10.1103/PhysRevLett.118.036601}{\emph{Phys. Rev.
  Lett.} {\bf 118} (2017) 036601}, [\href{http://arxiv.org/abs/1609.03582}{{\tt
  1609.03582}}].

\bibitem{hasan2010colloquium}
M.~Z. Hasan and C.~L. Kane, \emph{Colloquium: topological insulators},
  {\emph{Reviews of Modern Physics} {\bf 82} (2010) 3045}.

\bibitem{qi2011topological}
X.-L. Qi and S.-C. Zhang, \emph{Topological insulators and superconductors},
  {\emph{Reviews of Modern Physics} {\bf 83} (2011) 1057}.

\bibitem{Yu61}
R.~Yu, W.~Zhang, H.-J. Zhang, S.-C. Zhang, X.~Dai and Z.~Fang, \emph{Quantized
  anomalous hall effect in magnetic topological insulators},
  \href{http://dx.doi.org/10.1126/science.1187485}{\emph{Science} {\bf 329}
  (2010) 61--64}.

\bibitem{PhysRevLett.102.216403}
L.~Fu and C.~Kane, \emph{Probing neutral majorana fermion edge modes with
  charge transport},
  \href{http://dx.doi.org/10.1103/PhysRevLett.102.216403}{\emph{Phys.Rev.Lett.}
  {\bf 102} (May, 2009) 216403}.

\bibitem{PhysRevB.78.195424}
X.-L. Qi, T.~L. Hughes and S.-C. Zhang, \emph{Topological field theory of
  time-reversal invariant insulators},
  \href{http://dx.doi.org/10.1103/PhysRevB.78.195424}{\emph{Phys. Rev. B} {\bf
  78} (Nov, 2008) 195424}.

\bibitem{bergmann1982weak}
G.~Bergmann, \emph{Weak anti-localization---an experimental proof for the
  destructive interference of rotated spin 12}, {\emph{Solid State Commun.}
  {\bf 42} (1982) 815--817}.

\bibitem{liu2012crossover}
M.~Liu, J.~Zhang, C.-Z. Chang, Z.~Zhang, X.~Feng, K.~Li et~al., \emph{Crossover
  between weak antilocalization and weak localization in a magnetically doped
  topological insulator}, {\emph{Phys. Rev. Lett.} {\bf 108} (2012) 036805}.

\bibitem{zhang2012interplay}
D.~Zhang et~al., \emph{Interplay between ferromagnetism, surface states, and
  quantum corrections in a magnetically doped topological insulator},
  {\emph{Physical Review B} {\bf 86} (2012) 205127}.

\bibitem{bao2013quantum}
L.~Bao, W.~Wang, N.~Meyer, Y.~Liu, C.~Zhang, K.~Wang et~al., \emph{Quantum
  corrections crossover and ferromagnetism in magnetic topological insulators},
  {\emph{Scientific reports} {\bf 3} (2013) }.

\bibitem{hikami1980spin}
S.~Hikami, A.~I. Larkin and Y.~Nagaoka, \emph{Spin-orbit interaction and
  magnetoresistance in the two dimensional random system}, {\emph{Progress of
  Theoretical Physics} {\bf 63} (1980) 707--710}.

\bibitem{lu2011competition}
H.-Z. Lu, J.~Shi and S.-Q. Shen, \emph{Competition between weak localization
  and antilocalization in topological surface states}, {\emph{Phys. Rev. Lett.}
  {\bf 107} (2011) 076801}.

\bibitem{lang2012competing}
M.~Lang, L.~He, X.~Kou, P.~Upadhyaya, Y.~Fan, H.~Chu et~al., \emph{Competing
  weak localization and weak antilocalization in ultrathin topological
  insulators}, {\emph{Nano letters} {\bf 13} (2012) 48--53}.

\bibitem{mccann2006weak}
E.~McCann, K.~Kechedzhi, V.~I. Fal'ko, H.~Suzuura, T.~Ando and B.~Altshuler,
  \emph{Weak-localization magnetoresistance and valley symmetry in graphene},
  {\emph{Phys. Rev. Lett.} {\bf 97} (2006) 146805}.

\bibitem{tikhonenko2009transition}
F.~Tikhonenko, A.~Kozikov, A.~Savchenko and R.~Gorbachev, \emph{Transition
  between electron localization and antilocalization in graphene}, {\emph{Phys.
  Rev. Lett.} {\bf 103} (2009) 226801}.

\bibitem{Seo:2017oyh}
Y.~Seo, G.~Song and S.-J. Sin, \emph{{Strong Correlation Effects on Surfaces of
  Topological Insulators via Holography}},
  \href{http://dx.doi.org/10.1103/PhysRevB.96.041104}{\emph{Phys. Rev.} {\bf
  B96} (2017) 041104}, [\href{http://arxiv.org/abs/1703.07361}{{\tt
  1703.07361}}].

\bibitem{Witten:2015aoa}
E.~Witten, \emph{{Three Lectures On Topological Phases Of Matter}},
  \href{http://dx.doi.org/10.1393/ncr/i2016-10125-3}{\emph{Riv. Nuovo Cim.}
  {\bf 39} (2016) 313--370}, [\href{http://arxiv.org/abs/1510.07698}{{\tt
  1510.07698}}].

\bibitem{kkss}
Y.~Seo, K.-Y. Kim, K.~K. Kim and S.-J. Sin, \emph{{Character of matter in
  holography: Spin--orbit interaction}},
  \href{http://dx.doi.org/10.1016/j.physletb.2016.05.059}{\emph{Phys. Lett.}
  {\bf B759} (2016) 104--109}, [\href{http://arxiv.org/abs/1512.08916}{{\tt
  1512.08916}}].

\bibitem{Blake:2015ina}
M.~Blake, A.~Donos and N.~Lohitsiri, \emph{{Magnetothermoelectric Response from
  Holography}}, \href{http://dx.doi.org/10.1007/JHEP08(2015)124}{\emph{JHEP}
  {\bf 08} (2015) 124}, [\href{http://arxiv.org/abs/1502.03789}{{\tt
  1502.03789}}].

\bibitem{Andrade:2013gsa}
T.~Andrade and B.~Withers, \emph{{A simple holographic model of momentum
  relaxation}}, \href{http://dx.doi.org/10.1007/JHEP05(2014)101}{\emph{JHEP}
  {\bf 1405} (2014) 101}, [\href{http://arxiv.org/abs/1311.5157}{{\tt
  1311.5157}}].

\bibitem{Blake:2013bqa}
M.~Blake and D.~Tong, \emph{{Universal Resistivity from Holographic Massive
  Gravity}}, \href{http://dx.doi.org/10.1103/PhysRevD.88.106004}{\emph{Phys.
  Rev.} {\bf D88} (2013) 106004}, [\href{http://arxiv.org/abs/1308.4970}{{\tt
  1308.4970}}].

\bibitem{Blake:2013owa}
M.~Blake, D.~Tong and D.~Vegh, \emph{{Holographic Lattices Give the Graviton an
  Effective Mass}},
  \href{http://dx.doi.org/10.1103/PhysRevLett.112.071602}{\emph{Phys. Rev.
  Lett.} {\bf 112} (2014) 071602}, [\href{http://arxiv.org/abs/1310.3832}{{\tt
  1310.3832}}].

\bibitem{Donos:2014cya}
A.~Donos and J.~P. Gauntlett, \emph{{Thermoelectric DC conductivities from
  black hole horizons}},
  \href{http://dx.doi.org/10.1007/JHEP11(2014)081}{\emph{JHEP} {\bf 1411}
  (2014) 081}, [\href{http://arxiv.org/abs/1406.4742}{{\tt 1406.4742}}].

\bibitem{Blake:2014yla}
M.~Blake and A.~Donos, \emph{{Quantum Critical Transport and the Hall Angle}},
  \href{http://dx.doi.org/10.1103/PhysRevLett.114.021601}{\emph{Phys. Rev.
  Lett.} {\bf 114} (2015) 021601}, [\href{http://arxiv.org/abs/1406.1659}{{\tt
  1406.1659}}].

\bibitem{Kim:2015wba}
K.-Y. Kim, K.~K. Kim, Y.~Seo and S.-J. Sin, \emph{{Thermoelectric
  Conductivities at Finite Magnetic Field and the Nernst Effect}},
  \href{http://dx.doi.org/10.1007/JHEP07(2015)027}{\emph{JHEP} {\bf 07} (2015)
  027}, [\href{http://arxiv.org/abs/1502.05386}{{\tt 1502.05386}}].

\bibitem{Ghahari:2016aa}
F.~Ghahari, \emph{Enhanced thermoelectric power in graphene: Violation of the
  mott relation by inelastic scattering},
  \href{http://dx.doi.org/10.1103/PhysRevLett.116.136802}{\emph{Phys. Rev.
  Lett.} {\bf 116} (2016) }.

\bibitem{wang2006nernst}
Y.~Wang, L.~Li and N.~Ong, \emph{Nernst effect in high-t c superconductors},
  {\emph{Physical Review B} {\bf 73} (2006) 024510}.

\bibitem{Anderson:2007aa}
P.~W. Anderson, \emph{Two new vortex liquids}, {\emph{Nat Phys} {\bf 3} (03,
  2007) 160--162}.

\bibitem{Novoselov:2005aa}
K.~S. Novoselov, A.~K. Geim, S.~V. Morozov, D.~Jiang, M.~I. Katsnelson, I.~V.
  Grigorieva et~al., \emph{Two-dimensional gas of massless dirac fermions in
  graphene}, {\emph{Nature} {\bf 438} (11, 2005) 197--200}.

\bibitem{nomura2007quantum}
K.~Nomura and A.~MacDonald, \emph{Quantum transport of massless dirac
  fermions}, {\emph{Physical review letters} {\bf 98} (2007) 076602}.

\bibitem{Peres:2007aa}
N.~M.~R. Peres", \emph{Phenomenological study of the electronic transport
  coefficients of graphene},
  \href{http://dx.doi.org/10.1103/PhysRevB.76.073412}{\emph{Physical Review B}
  {\bf 76} (2007) 073412}.

\bibitem{landau2013electrodynamics}
L.~D. Landau, J.~Bell, M.~Kearsley, L.~Pitaevskii, E.~Lifshitz and J.~Sykes,
  \emph{Electrodynamics of continuous media}, vol.~8.
\newblock elsevier, 2013.

\bibitem{nersesyan1989low}
A.~Nersesyan and G.~Vachnadze, \emph{Low-temperature thermodynamics of the
  two-dimensional orbital antiferromagnet}, {\emph{Journal of Low Temperature
  Physics} {\bf 77} (1989) 293--303}.

\bibitem{Kopelevich:aa}
Y.~Kopelevich, P.~Esquinazi, J.~H.~S. Torres and S.~Moehlecke,
  \emph{Ferromagnetic- and superconducting-like behavior of graphite},
  {\emph{J. Low Temp. Phys.} {\bf 119} (2000) 691},
  [\href{http://arxiv.org/abs/cond-mat/9912413}{{\tt cond-mat/9912413}}].

\bibitem{suzuki2002magnetic}
M.~Suzuki, I.~S. Suzuki, R.~Lee and J.~Walter, \emph{Magnetic-field-induced
  superconductor--metal-insulator transitions in bismuth metal graphite},
  {\emph{Physical Review B} {\bf 66} (2002) 014533}.

\end{thebibliography}\endgroup

\end{document}